\documentclass[aip,jcp,reprint,twocolumn,superscriptaddress,showpacs,floatfix]{revtex4-1}

\usepackage{epsfig}
\usepackage{mathrsfs}
\usepackage{graphicx}
\usepackage{amssymb}
\usepackage{dcolumn}
\usepackage{bm}
\usepackage{textcomp}
\usepackage{gensymb}
\usepackage{physics}
\usepackage{setspace}
\usepackage{longtable}
\usepackage{multirow}
\usepackage{color}
\usepackage{bm}
\newcommand{\beq}{\begin{equation}}
\newcommand{\eeq}{\end{equation}}
\renewcommand{\thetable}{\arabic{table}}
\definecolor{red}{rgb}{1,0,0}
\begin{document}

\title{The Singlet--Triplet Gap of Cyclobutadiene: The CIPSI-Driven CC($\bm{P}$;$\bm{Q}$) Study}

\author{Swati S. Priyadarsini}
\affiliation{Department of Chemistry, Michigan State University, East Lansing, Michigan 48824, USA}

\author{Karthik Gururangan} 
\affiliation{Department of Chemistry, Michigan State University, East Lansing, Michigan 48824, USA}

\author{Jun Shen}
\affiliation{Department of Chemistry, Michigan State University, East Lansing, Michigan 48824, USA}

\author{Piotr Piecuch}
\thanks{Corresponding author}
\email[e-mail: ]{piecuch@chemistry.msu.edu.}
\affiliation{Department of Chemistry, Michigan State University, East Lansing, Michigan 48824, USA}
\affiliation{Department of Physics and Astronomy, Michigan State University, East Lansing, Michigan 48824, USA}

\date{\today}

\begin{abstract}
\textbf{ABSTRACT:} An accurate determination of singlet--triplet gaps in biradicals, including cyclobutadiene in the
automerization barrier region where one has to balance the substantial nondynamical many-electron
correlation effects characterizing the singlet ground state with the predominantly dynamical
correlations of the lowest-energy triplet, remains a challenge for many quantum chemistry methods.
High-level coupled-cluster (CC) approaches, such as the CC method with a full treatment of singly,
doubly, and triply excited clusters (CCSDT), are often capable of providing reliable results,
but the routine application of such methods is hindered by their high computational costs. We have
recently proposed a practical alternative to converging the CCSDT energetics at small fractions
of the computational effort, even when electron correlations become stronger and connected triply
excited clusters are larger and nonperturbative, by merging the CC($P$;$Q$) moment expansions
with the selected configuration interaction methodology abbreviated as CIPSI. We demonstrate that
one can accurately approximate the highly accurate CCSDT potential surfaces characterizing the
lowest singlet and triplet states of cyclobutadiene along the automerization coordinate and the gap
between them using tiny fractions of triply excited cluster amplitudes identified with the help of
relatively inexpensive CIPSI Hamiltonian diagonalizations.
\end{abstract}

\maketitle

%
\section{Introduction}
\label{sec1}

Biradicals play a key role in chemistry as reaction intermediates in 
thermal and photochemical pathways\cite{thermochem,sjs-ref1,photochem1,photochem2,photochem3,photochem4,abe2012} 
as well as functional materials used in molecular magnets,\cite{molmagnet1,molmagnet2,molmagnet3}
battery electrodes,\cite{battery2014} and organic photovoltaics.\cite{photovol1,photovol2,photovol3,singletfission} 
An important quantity characterizing the electronic structure of biradicals,
especially in the context of designing molecules
for magnetic, electrochemical, and photovoltaic applications,
is the energy gap $\Delta E_\text{S--T}$ separating the lowest-lying singlet and triplet states
(throughout this work,
we define $\Delta E_\text{S--T}$ as $E_{\rm S} - E_{\rm T}$, where $E_{\rm S}$ and
$E_{\rm T}$ are the electronic energies of the relevant singlet and triplet states, {\it i.e.},
when the singlet is lower than the triplet, $\Delta E_\text{S--T} < 0$).
Accurate computational determination of the
$\Delta E_{\text{S--T}}$ values in biradicals remains, however,
a difficult task because it requires balancing the strong nondynamical many-electron correlation
effects characterizing the low-spin singlet states with the predominantly dynamical 
correlations associated with their high-spin triplet counterparts.
\cite{ch2_krylov,crccl_jpc,ch2-rmrccsdt,%
icmrcc6,saito2011symmetry,ess2011ST,jspp-jctc2012,abe2013diradicals,garza2014electronic,%
ibeji2015singlet,ajala2017,stoneburner2017,zimmerman2017,jspp-dea-2021,gulania2021EOM,mazziotti2021,%
arnab-stgap-2022,demel-et-al-jcp-cbd-2023,szabados-et-al-molphys-2025}

This challenge is exemplified by the cyclobutadiene molecule, which
is the focus of the present study and which has fascinated
experimental and theoretical chemists for decades with questions
surrounding its low-lying electronic states, anti-aromaticity, and
reactivity in cycloaddition and isomerization reactions.
In the $D_{4\text{h}}$-symmetric square structure corresponding to
the barrier along the automerization coordinate or the minimum on
the lowest triplet potential, cyclobutadiene is a biradical with its
four valence $\pi$ orbitals arranged in a network consisting of
the nondegenerate $a_{2u}$ orbital, the doubly degenerate $e_g$
shell, and the nondegenerate $b_{1u}$ orbital. The distribution of two of the four
valence electrons among the pair of degenerate frontier $e_g$ orbitals gives
rise to three singlet states of the $B_{1g}(D_{4\text{h}})$, $A_{1g}(D_{4\text{h}})$,
and $B_{2g}(D_{4\text{h}})$ symmetries and a $A_{2g}(D_{4\text{h}})$-symmetric
triplet state, all involving the doubly occupied $a_{2u}$ and unoccupied $b_{1u}$
orbitals and the partially occupied $e_g$ shell in their zeroth-order description.
\cite{gulania2021EOM,craig1950,peyerimhoff1968,iwata1989,balkova1994,krylov2004,%
MR-AQCC,cbd-loos-2022}
As shown in the early {\it ab initio} calculations,
\cite{craig1950,peyerimhoff1968,borden1975,kollmar1977,borden1978}
and as confirmed in many subsequent theoretical studies, such as those reported in Refs.\
\onlinecite{gulania2021EOM,iwata1989,balkova1994,krylov2004,MR-AQCC,saito2011symmetry,%
ajala2017,stoneburner2017,zimmerman2017,mazziotti2021,cbd-loos-2022,arnab-stgap-2022,%
demel-et-al-jcp-cbd-2023,szabados-et-al-molphys-2025},
the lowest singlet of the $B_{1g}(D_{4\text{h}})$ symmetry, which has a substantial
multiconfigurational character, is the ground state, whereas the predominantly single-reference
$A_{2g}(D_{4\text{h}})$-symmetric triplet, in violation of Hund's rule, is the first excited state. 
To be more specific, if one orients cyclobutadiene
such that the two $C_{2}$ axes bisect the carbon--carbon
bonds, which is a convention adopted in the present study, the $^{1}B_{1g}(D_{4\text{h}})$
ground state of the square structure is dominated by
two closed-shell determinants in which one of the
two degenerate $e_g$ orbitals is occupied by two electrons and the other one is empty
(see, {\it e.g.},
Refs. \onlinecite{borden1978,iwata1989,krylov2004,cbd-loos-2022}). This should be
contrasted with the lowest-energy $^{3}A_{2g}(D_{4\text{h}})$ state, which is characterized by
single occupancy of each of the $e_g$ orbitals.

While the lowest $^{3}A_{2g}(D_\text{4h})$ state is stable in the square geometry, which
is a minimum on the corresponding triplet surface, the $^{1}B_{1g}(D_\text{4h})$ ground state
is unstable with respect to the rectangular distortion of the carbon--carbon bonds that
lifts the degeneracy of the valence $e_{g}$ orbitals and lowers its total electronic energy
due to the pseudo-Jahn--Teller effect. This results in the formation of the $D_{\text{2h}}$-symmetric
rectangular species characterized by the closed-shell, predominantly single-determinantal,
$^{1}A_{g}(D_\text{2h})$ ground state,
which represents a minimum on the lowest-energy singlet potential.
\cite{gulania2021EOM,borden1978,iwata1989,balkova1994,krylov2004,mazziotti2021,%
demel-et-al-jcp-cbd-2023,szabados-et-al-molphys-2025,cbd-loos-2022}
The distortion of the multiconfigurational 
${}^1B_{1g}(D_{4\text{h}})$ state into the ${}^1A_{g}(D_{2\text{h}})$ state coincides with
the automerization coordinate in cyclobutadiene, which describes
the conversion of the rectangular, $D_{2\text{h}}$-symmetric,
closed-shell reactant (R) into the equivalent product conformer by passing through
the square, $D_{4\text{h}}$-symmetric, biradical transition state (TS).
Obtaining an accurate description of the potential energy curves (PECs)
characterizing the lowest singlet [${}^1A_{g}(D_{2\text{h}})$] and
triplet [${}^3B_{1g}(D_{2\text{h}})$] states of cyclobutadiene
along its $D_{2\text{h}}$-symmetric automerization coordinate, and
the gap between them, particularly in the neighborhood of
the square biradical species, remains a significant challenge for modern \emph{ab initio} techniques
as it requires a high-level treatment of many-electron correlation effects in order to accurately
capture and balance the strong nondynamical correlations associated with the multiconfigurational 
singlet state with the largely dynamical correlations dominating the triplet state.

A traditional way of addressing this and similar challenges is to use multireference approaches,
\cite{Roos1987,ref:mrmpreview,chemrev-2012a,lindh-review-2012,sinha-review-2016,%
chemrev-2012b,succ5,evangelista-perspective-jcp-2018} but in this work we focus on
the single-reference coupled-cluster (CC) methodology,\cite{Coester:1958,Coester:1960,cizek1,cizek2,cizek4,%
paldus-li,bartlett-musial2007} which employs
the exponential wave function ansatz\cite{Hubbard:1957,Hugenholtz:1957}
\beq
|\Psi\rangle = e^T |\Phi\rangle,
\label{eq:srcc}
\eeq
where $|\Phi\rangle$ is the $N$-electron reference determinant that serves
as a Fermi vacuum and $T = \sum_{n=1}^N T_n$ is the cluster operator, with $T_n$ designating the $n$-body
component of $T$ responsible for generating connected $n$-particle--$n$-hole ($n$p-$n$h) excitations
out of $|\Phi\rangle$. It is well established that as long as the number of strongly correlated
electrons is not too large, the standard hierarchy of CC approximations, including the CC method
with singles and doubles (CCSD),\cite{ccsd,ccsd2,ccsdfritz,osaccsd}  
obtained by truncating $T$ at $T_{2}$, the CC approach with singles, doubles, and
triples (CCSDT),\cite{ccsdt1986,ccfullt,ccfullt2,ch2-bartlett2} in which $T$ is
truncated at $T_{3}$, the CC method with singles, doubles, triples, and quadruples
(CCSDTQ),\cite{ccsdtq0,ccsdtq1,ccsdtq2,ccsdtq3} where $T$ is truncated at $T_{4}$,
and so on, rapidly converges to the exact, full configuration interaction (CI) limit.
As a result, the single-reference CC approaches with a full treatment of higher-rank
$T_{n}$ clusters with $n > 2$, such as CCSDT or CCSDTQ, are often capable of accurately
describing multireference situations, including substantial bond rearrangements
in the course of chemical reactions and, what is especially important for this study,
singlet--triplet gaps in biradical species, by capturing the relevant dynamical and
nondynamical correlation effects via particle--hole excitations from a single determinant,
without having to involve genuine multireference concepts.

\begin{figure}[!ht]
\centering
\includegraphics[scale=0.5]{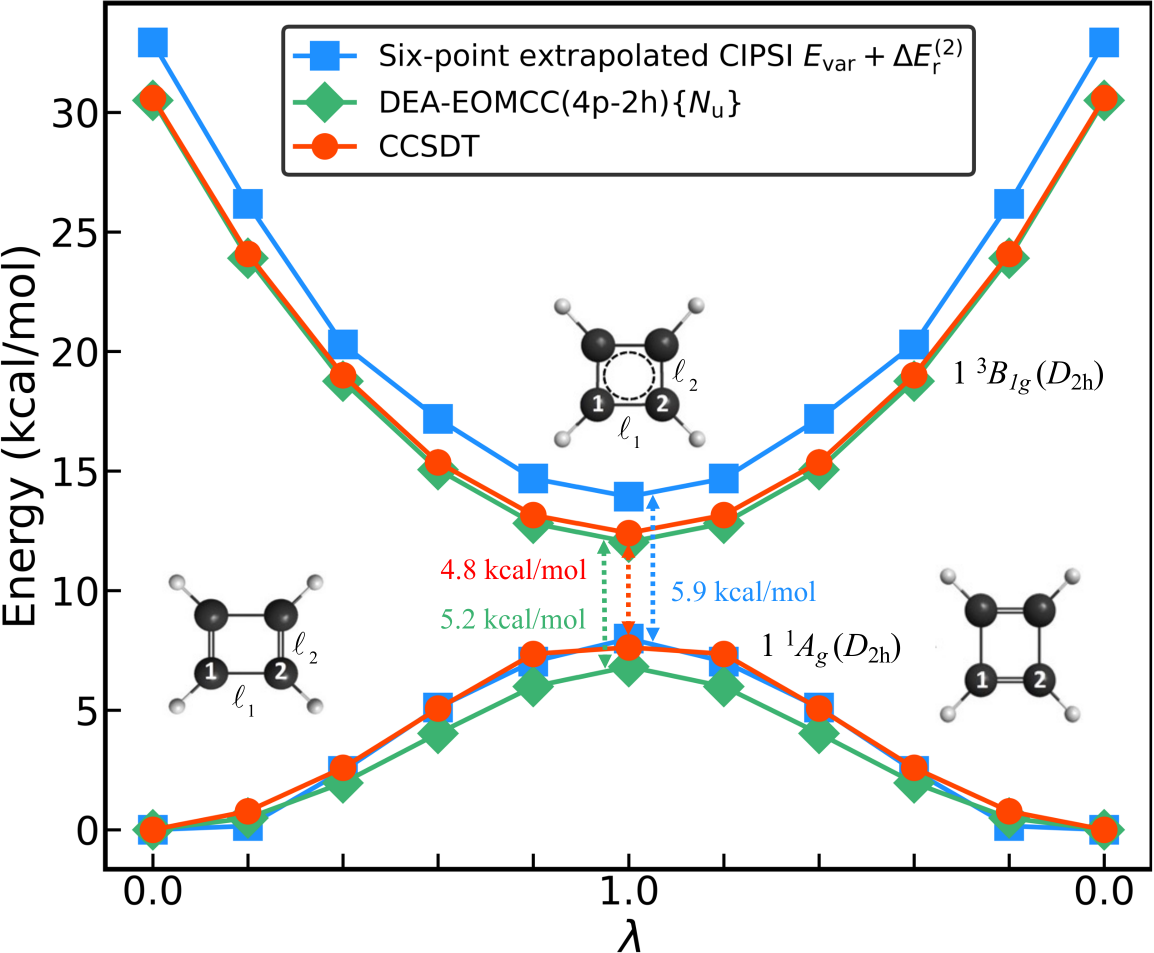}
\caption{
The PECs (in kcal/mol) characterizing the lowest-energy singlet and triplet states of cyclobutadiene
along the $D_{\rm 2h}$-symmetric automerization pathway, defined using the interpolation formula given
by Eq. (\ref{eq:ell}) and parameterized by dimensionless variable $\lambda$, resulting from the
full CCSDT (red solid circles and lines), active-space DEA-EOMCC(4p-2h)$\{N_\text{u}\}$
(green solid diamonds and lines), and perturbatively corrected and extrapolated CIPSI
(blue solid squares and lines) calculations using the cc-pVDZ basis set
described in the main text. For each of the three methods,
the energy of the singlet ground state at the reactant (R, $\lambda = 0$) geometry is set to 0.
The numbers in the middle, colored in the same way as the corresponding PECs, are the unsigned
values of the singlet--triplet gaps determined at the $\lambda= 1$ TS structure.}
\label{fig:figure1}
\end{figure}

In particular, the calculations reported in Refs.\ \onlinecite{balkova1994,arnab-stgap-2022,cbd-loos-2022}
show that the full treatment of $T_{1}$, $T_{2}$, and $T_{3}$ clusters provided by CCSDT offers
a highly accurate description of the total electronic energies of the lowest $^{1}B_{1g}(D_{\rm 4h})$
and $^{3}A_{2g}(D_{\rm 4h})$ states of the square cyclobutadiene and the gap between them.
As demonstrated in Figure \ref{fig:figure1}, this remains true when examining the PECs characterizing
the lowest-energy singlet and triplet states of cyclobutadiene along the entire $D_{\rm 2h}$-symmetric
automerization reaction path. In determining the lowest-energy
${}^1A_{g}(D_{2\text{h}})$ and ${}^3B_{1g}(D_{2\text{h}})$ PECs shown in Figure \ref{fig:figure1}
(for information about the electronic structure software used in our calculations, see Section \ref{sec2}), we 
followed the procedure described in Ref.\ \onlinecite{MR-AQCC} in which one constructs an approximate,
$D_{\rm 2h}$-symmetric, one-dimensional automerization pathway connecting the rectangular minima on the
lowest singlet potential
via the square TS species by linearly interpolating the carbon--carbon bond distances in cyclobutadiene
using the formula
\beq
\ell_i(\lambda) = (1-\lambda) \, \ell_i(\text{R}) + \lambda \, \ell_i(\text{TS}), \;\: i = 1, 2,
\label{eq:ell}
\eeq
where $\ell_1$ and $\ell_2$ are the C--C distances depicted in Figure \ref{fig:figure1}
and $\ell_i(\text{R})$ and $\ell_i(\text{TS})$ are the carbon--carbon bond lengths characterizing the
R and TS structures optimized (along with the C--H distances and H--C--C bond angles) in
Ref.\ \onlinecite{MR-AQCC} with the multireference average quadratic CC
(MR-AQCC) approach.\cite{mraqcc1,mraqcc2} The dimensionless parameter $\lambda$ defining the
automerization coordinate varies between 0, corresponding to the R species, and 1, corresponding
to the TS structure, going back to 0 (after replacing $\ell_1$ by $\ell_2$ and {\it vice versa})
when the automerization product equivalent to the R species is reached.
In the absence of information about the C--H bond lengths and H--C--C bond angles characterizing
the intermediate $\lambda = 0.2,$ 0.4, 0.6, and 0.8 geometries in Ref.\ \onlinecite{MR-AQCC},
in determining the lowest ${}^1A_{g}(D_{2\text{h}})$ and ${}^3B_{1g}(D_{2\text{h}})$
states of these structures, we fixed the C--H distances and H--C--C angles at their values corresponding
to the R species.

As shown in Figure \ref{fig:figure1}, the lowest singlet and triplet PECs computed as
functions of the $D_{\rm 2h}$-symmetric automerization coordinate $\lambda$ with full CCSDT
are in very good agreement with their counterparts obtained with the
CI method using perturbative selection made iteratively,
\cite{sci_3,cipsi_1,cipsi_2} abbreviated as CIPSI, extrapolated to the full CI limit
(see Section \ref{sec2} for further details).
The CCSDT energies are also very close to those determined using the double
electron-attachment (DEA) equation-of-motion (EOM) CC methodology\cite{dipea5,dipea6,%
jspp-dea-dip-2013,jspp-dea-dip-2014,ajala2017,gulania2021EOM,jspp-dea-2021}
with a full treatment of 2p and 3p-1h and an active-space treatment of 4p-2h correlations
on top of the CCSD description of the underlying closed-shell $({\rm C_{4}H_{4}})^{2+}$ core,
\cite{jspp-dea-dip-2013,jspp-dea-dip-2014,ajala2017,jspp-dea-2021}
denoted as DEA-EOMCC(4p-2h)$\{N_\text{u}\}$, where $N_{\text{u}}$ designates the number
of active unoccupied orbitals of $({\rm C_{4}H_{4}})^{2+}$ included in the calculations to capture
the leading 4p-2h effects in the target cyclobutadiene species
(to accurately describe the 4p-2h effects associated with the
valence orbitals of cyclobutadiene that correlate with the $e_g$ and $b_{1u}$ shells
of the square TS structure, we set $N_{\text{u}}$ to 3). To illustrate the agreement
between full CCSDT, perturbatively corrected and
extrapolated CIPSI, and DEA-EOMCC(4p-2h)$\{N_\text{u}\}$, which are three
independent {\it ab initio} methodologies, we compare the vertical $\Delta E_\text{S--T}$
values at the R and TS geometries. When using the cc-pVDZ\cite{ccpvnz} basis set, employed
in the calculations reported in Figure \ref{fig:figure1} and
in most of the computations discussed in
the rest of this article,
the CCSDT value of $\Delta E_\text{S--T}$
at the TS geometry is $-4.8$ kcal/mol. This is very close to $-5.2$ kcal/mol resulting from the
state-of-the-art DEA-EOMCC(4p-2h)$\{N_\text{u}\}$ calculations and $-5.9$ kcal/mol obtained with CIPSI.
The CCSDT, DEA-EOMCC(4p-2h)$\{N_\text{u}\}$, and perturbatively corrected and
extrapolated CIPSI values of $\Delta E_\text{S--T}$
at the R geometry are $-30.6$, $-30.5$, and $-32.9$ kcal/mol, respectively,
again in good agreement with one another.

Given the high accuracy of the lowest singlet and triplet potential surfaces of cyclobutadiene
and $\Delta E_\text{S--T}$ values along the $D_{\rm 2h}$-symmetric automerization
pathway offered by full CCSDT, it may be tempting to turn to the approximate treatments of
$T_{3}$ correlations that replace the expensive iterative $\mathscr{N}^8$ computational steps
of CCSDT, where $\mathscr{N}$ is a measure of the system size, by the more practical $\mathscr{N}^6$
operations of CCSD combined with the noniterative $\mathscr{N}^7$ steps needed to correct the
CCSD energetics for the leading $T_{3}$ correlations, as in the widely used CCSD(T) approach
\cite{ccsdpt,watts-gauss-bartlett-1993} or its more robust completely renormalized (CR) CC
counterpart abbreviated as CR-CC(2,3).\cite{crccl_jcp,crccl_cpl,crccl_molphys,crccl_jpc}
Unfortunately, neither CCSD(T) and CR-CC(2,3) nor any of the other
noniterative triples corrections to CCSD,
such as CCSD(T)$_{\Lambda}$,\cite{stanton1997,crawford1998,ref:26}
CCSD(2)$_{\rm T}$,\cite{eomccpt,ccsdpt2,gwaltney1,gwaltney3}
CR-CCSD(T)\cite{leszcz,ren1,irpc,PP:TCA}
and its locally renormalized extension,\cite{ndcmmcc}
$\Lambda$-CCSD(T),\cite{bartlett2008a,bartlett2008b}
and CCSD(${\rm T}\mbox{--}n$),\cite{eriksen1,eriksen2}
are capable of providing accurate results when the
coupling of the lower-order $T_{1}$ and $T_{2}$ components of the cluster operator with their
higher-rank $T_{3}$ counterpart becomes large. For example, even the most robust triples correction
to CCSD defining CR-CC(2,3), which improves CCSD(T) and other similar approaches in
situations involving electronic quasi-degeneracies, such as those present in single bond breaking,
\cite{jspp-chemphys2012,crccl_jcp,crccl_cpl,crccl_molphys,crccl_jpc,ptcp2007,ge1,ge2} struggles in
describing the PEC of the lowest ${}^1A_{g}(D_{2\text{h}})$ state of cyclobutadiene in
the neighborhood of the automerization barrier region, where $T_{3}$ clusters become large,
nonperturbative, and strongly coupled to $T_{1}$ and $T_{2}$.\cite{jspp-jcp2012}  As a result,
as shown in Ref.\ \onlinecite{arnab-stgap-2022} and this study,
the CR-CC(2,3) $\Delta E_{\text{S--T}}$ value
at the singlet TS structure,
of 4.4 kcal/mol when the cc-pVDZ basis
is used,
is in
large error (including incorrect sign)
relative to its CCSDT $-4.8$ kcal/mol counterpart.
CCSD(T) gives 3.9 kcal/mol, which is similarly inaccurate.
This is because neither CR-CC(2,3) nor CCSD(T), nor any other noniterative triples
correction to CCSD, can correctly describe $T_{3}$ contributions to $\Delta E_{\text{S--T}}$
that at the square TS geometry of cyclobutadiene are a few times larger, in absolute value,
than the CCSDT singlet--triplet gap itself. For example, the $T_{3}$ effects on $\Delta E_{\text{S--T}}$
at the TS structure, estimated by subtracting the singlet--triplet gap obtained in the
CCSD/cc-pVDZ calculations from its CCSDT/cc-pVDZ counterpart, are $-15.1$ kcal/mol, {\it i.e.},
they are more than three times larger than the value of $\Delta E_{\text{S--T}}$
obtained in the CCSDT/cc-pVDZ calculations.
The fact that the $T_{3}$ contributions to the singlet--triplet gap of cyclobutadiene
in the vicinity of the automerization barrier region become so massive and difficult to
describe by the noniterative triples corrections to CCSD is closely related to the dramatic
increase of $T_{1}$, $T_{2}$, and $T_{3}$ cluster amplitudes characterizing the lowest-energy
${}^1A_{g}(D_{2\text{h}})$ state, especially the amplitudes defining $T_{2}$ and $T_{3}$
operators that engage valence orbitals around the Fermi level, reflecting on the increasingly
strongly correlated character of this state as one transitions
from the R to TS structures. For example, the largest $T_{1}$, $T_{2}$, and $T_{3}$
amplitudes obtained in the CCSDT/cc-pVDZ calculations for the $A_{g}(D_{2\text{h}})$-symmetric
singlet ground state of cyclobutadiene at its R geometry are $-0.032682$, $-0.206225$, and
$0.003414$, respectively. At the TS geometry, they become $-0.054928$, $-0.895785$, and
$0.014836$, respectively, {\it i.e.}, we observe a four-fold increase in the largest $T_{2}$
and $T_{3}$ amplitudes compared to the R structure. This should be contrasted with the behavior
of the lowest-energy triplet state, where the largest $T_{1}$, $T_{2}$, and $T_{3}$ cluster
amplitudes resulting from the CCSDT/cc-pVDZ computations remain relatively small and barely change
when the ${\rm R} \rightarrow {\rm TS}$ ($\lambda = 0 \rightarrow 1$) geometrical transformation
is examined. They are $0.031220$ for $T_{1}$, $-0.163848$ for $T_{2}$, and $0.003285$
for $T_{3}$ at $\lambda = 0$ and 0.030476, $-0.137065$, and $0.002555$, respectively,
when $\lambda$ becomes 1. It is this drastically different behavior of the lowest
${}^1A_{g}(D_{2\text{h}})$ and ${}^3B_{1g}(D_{2\text{h}})$ states of cyclobutadiene
and the rapid growth of the cluster amplitudes characterizing the former state that engage
valence orbitals, especially those associated with $T_{2}$ and $T_{3}$, as
the automerization barrier region is approached, which result in failures of methods
such as CCSD(T) and CR-CC(2,3) in describing the corresponding singlet--triplet gap.

Problems with applying noniterative corrections to CCSD in situations where $T_{n}$ components
with $n > 2$, such as $T_{3}$, are not only large and nonperturbative, but also strongly coupled to
their lower-rank $T_{1}$ and $T_{2}$ counterparts, have motivated us to develop the generalization
of the biorthogonal moment expansions, which in the past resulted in the CR-CC approaches,
such as CR-CC(2,3) and its excited-state and higher-order extensions,
\cite{crccl_jcp,crccl_cpl,crccl_molphys,crccl_jpc,ptcp2006,crccl_ijqc2,7hq,%
ptcp2007,msg65,nuclei8,nbjspp-molphys2017,ccpq-be2-jpca-2018}
to unconventional truncations in the cluster and EOMCC\cite{emrich,emrich2,eomcc1,eomcc3}
excitation operators, designated as CC($P$;$Q$).
\cite{jspp-chemphys2012,jspp-jcp2012,jspp-jctc2012,nbjspp-molphys2017,%
ccpq-be2-jpca-2018,ccpq-mg2-mp-2019,stochastic-ccpq-prl-2017,stochastic-ccpq-molphys-2020,%
stochastic-ccpq-jcp-2021,arnab-stgap-2022,cipsi-ccpq-2021,adaptiveccpq2023,adaptive-active-ccpq-2025}
By incorporating the dominant contributions to the higher--than--two-body clusters into
the iterative steps, so that $T_{1}$ and $T_{2}$ amplitudes can
relax compared to their CCSD values when $T_{n}$ components with $n > 2$ become
more substantial, and correcting the results for the remaining
many-electron correlation effects of interest using suitably defined moment expansions,
the CC($P$;$Q$) formalism provides us with the opportunity to converge or accurately
approximate the parent high-level CCSDT, CCSDTQ, and similar energetics at small fractions
of the computational costs, even when noniterative corrections to CCSD fail or struggle.
Focusing on full CCSDT, which provides the parent data for the lowest singlet and triplet PECs of
cyclobutadiene examined in this work, a few different CC($P$;$Q$) approaches designed to converge
CCSDT energetics have been developed. In the initial, active-orbital-based, variant of CC($P$;$Q$),
abbreviated as CC(t;3), which is part of the larger CC(t;3), CC(t,q;3), CC(t,q;3,4), CC(q;4), {\it etc.}
hierarchy,\cite{jspp-chemphys2012,jspp-jcp2012,jspp-jctc2012,nbjspp-molphys2017,%
ccpq-be2-jpca-2018,ccpq-mg2-mp-2019,adaptive-active-ccpq-2025} the leading $T_{3}$ amplitudes that enter the iterative
steps preceding the determination of the CC($P$;$Q$) corrections are obtained using the
active-space CCSDt approach.\cite{semi0b,semi1,semi2,ccsdtq3,semih2o,ghose,semi3c,%
semi4,semi4new,piecuch-qtp} In the more black-box semi-stochastic
\cite{stochastic-ccpq-prl-2017,stochastic-ccpq-molphys-2020,stochastic-ccpq-jcp-2021,arnab-stgap-2022}
and CIPSI-driven CC($P$;$Q$)\cite{cipsi-ccpq-2021} approaches aimed at converging CCSDT, the dominant 
triply excited cluster amplitudes are identified with the help of CI
\cite{Booth2009,Cleland2010,fciqmc-uga-2019,ghanem_alavi_fciqmc_jcp_2019,ghanem_alavi_fciqmc_2020}
or CC\cite{Thom2010,Franklin2016,Spencer2016,Scott2017} Quantum Monte Carlo wave function
propagations in the many-electron Hilbert space, in the former case, and the sequences of
Hamiltonian diagonalizations constructed in the CIPSI algorithm\cite{sci_3,cipsi_1,cipsi_2}
in the case of the latter method. In the recently introduced adaptive CC($P$;$Q$) formalism,
\cite{adaptiveccpq2023,adaptive-active-ccpq-2025} the triply excited determinants and amplitudes defining the leading
$T_{3}$ contributions in the iterative steps of CC($P$;$Q$) calculations are identified
using the intrinsic structure of the CC($P$;$Q$) energy corrections.

In this study, we focus on the CIPSI-driven CC($P$;$Q$) methodology of
Ref.\ \onlinecite{cipsi-ccpq-2021}. Our main goal is to answer the question how 
efficient this methodology is in converging the lowest singlet and triplet
potentials of cyclobutadiene along the $D_{\rm 2h}$-symmetric automerization coordinate
resulting from
the high-level CCSDT
computations shown in Figure \ref{fig:figure1}.
The ability of the CC($P$;$Q$) framework using the CIPSI algorithm
to identify the leading triply excited determinants for the iterative steps
of the CC($P$;$Q$) procedure to accurately approximate the CCSDT energies of the
${}^1A_{g}(D_{2\text{h}})$ and ${}^3B_{1g}(D_{2\text{h}})$ states of cyclobutadiene
and the gap between, especially at the most challenging TS geometry, in calculations
using basis sets larger than cc-pVDZ
(represented in this work by cc-pVTZ\cite{ccpvnz}) is examined as well. As shown
in our initial study,\cite{cipsi-ccpq-2021} the CIPSI-driven CC($P$;$Q$)
approach is capable of producing the near-CCSDT energetics for singlet electronic states
using tiny fractions of triply excited cluster amplitudes in the iterative parts of
the CC($P$;$Q$) algorithm that
are smaller
than those used in the analogous
semi-stochastic and active-orbital-based CC($P$;$Q$) considerations. One of
the objectives
of this work is to determine if similar observations apply to the
CIPSI-driven CC($P$;$Q$) calculations for the lowest singlet and triplet potentials
of cyclobutadiene along its automerization coordinate and the gap between them.
The role of the CC($P$;$Q$) moment corrections in
accelerating convergence toward CCSDT and the key elements of our improved
implementation of the CIPSI-based CC($P$;$Q$) method, capable of efficiently handling small but generally
spotty subsets of triply excited determinants in the underlying CC iterations, illustrated by
computational timings, are discussed too.


%
\section{Theory and Computational Details}
\label{sec2}

We begin by summarizing the key ingredients of the CC($P$;$Q$) formalism, as applied to
the ground-state problem or, in general, to the lowest state of a given symmetry for which
a suitable single-determinantal reference can be found. Each CC($P$;$Q$) calculation
requires defining two disjoint subspaces of the many-electron Hilbert space, called the
$P$ and $Q$ spaces, designated as $\mathscr{H}^{(P)}$ and $\mathscr{H}^{(Q)}$, respectively.
The former space consists of the excited determinants $|\Phi_K\rangle = E_K|\Phi\rangle$
which, together with the reference function $|\Phi\rangle$, dominate the electronic state
of interest ($E_K$ is the elementary particle--hole excitation operator that generates
$|\Phi_K\rangle$ from $|\Phi\rangle$). The determinants spanning the complementary $Q$
space $\mathscr{H}^{(Q)}$ are used to form the noniterative correction $\delta(\mbox{$P$;$Q$})$
which captures higher-order correlation effects the CC calculations in the $P$ space do not
describe.

All CC($P$;$Q$) computations consist of two stages.
In the first, iterative, stage, denoted as CC($P$), we solve
the CC amplitude equations in $\mathscr{H}^{(P)}$ to determine amplitudes
$t_K$ that define the $P$-space cluster operator
\begin{equation}
T^{(P)} = \sum_{|\Phi_K\rangle \in \mathscr{H}^{(P)}}t_K E_K .
\label{ccpq-1}
\end{equation}
This is done by employing the conventional projective technique adopted in the majority of
single-reference CC calculations, {\it i.e.}, by solving the system
\begin{equation}
\mathfrak{M}_K(P) = 0, \; |\Phi_K\rangle \in \mathscr{H}^{(P)} ,
\label{ccpq-2}
\end{equation}
where
\begin{equation}
\mathfrak{M}_K(P) = \langle\Phi_K|\overline{H}^{(P)}|\Phi\rangle,
\label{ccpq-3}
\end{equation}
with
$\overline{H}^{(P)} = e^{-T^{(P)}} H e^{T^{(P)}}$
representing the similarity-transformed Hamiltonian,
are generalized moments of the CC($P$) equations.\cite{leszcz,ren1,moments} Once the cluster
amplitudes $t_K$ defining $T^{(P)}$ are determined, the CC($P$) energy is calculated in a usual
way as
\begin{equation}
E^{(P)} = \langle\Phi|\overline{H}^{(P)}|\Phi\rangle .
\label{ccpq-5}
\end{equation}
In the second stage of the CC($P$;$Q$) procedure,
we construct the aforementioned noniterative correction $\delta(\mbox{$P$;$Q$})$
using the expression
\begin{equation}
\delta(\mbox{$P$;$Q$}) = \sum_{|\Phi_K\rangle \in \mathscr{H}^{(Q)}} \ell_K(P) \, \mathfrak{M}_K(P),
\label{ccpq-6}
\end{equation}
where coefficients $\ell_K(P)$ multiplying moments $\mathfrak{M}_K(P)$ are defined as
\begin{equation}
\ell_K(P) = \langle\Phi|(1+\Lambda^{(P)})\overline{H}^{(P)}|\Phi_K\rangle / D_K^{(P)},
\label{ccpq-7}
\end{equation}
with
$D_K^{(P)} = E^{(P)}-\langle\Phi_K|\overline{H}^{(P)}|\Phi_K\rangle$ designating
the relevant Epstein--Nesbet-like denominators. 
The hole--particle deexcitation operator
\begin{equation}
\Lambda^{(P)} = \sum_{|\Phi_K\rangle \in \mathscr{H}^{(P)}}\lambda_K(E_K)^\dagger
\label{ccpq-9}
\end{equation}
in Eq. (\ref{ccpq-7}), which defines the bra state
$\langle\tilde{\Psi}^{(P)}| = \langle\Phi|(1+\Lambda^{(P)})e^{-T^{(P)}}$ matching
the CC($P$) ket state $|\Psi^{(P)}\rangle = e^{T^{(P)}}|\Phi\rangle$, is obtained by solving the
linear system
\begin{equation}
\langle\Phi|(1+\Lambda^{(P)})\overline{H}^{(P)}|\Phi_K\rangle = E^{(P)}\lambda_K, \;
|\Phi_K\rangle \in \mathscr{H}^{(P)} .
\label{ccpq-10}
\end{equation}
The final CC($P$;$Q$) energy is obtained using the formula
\begin{equation}
E^{(P+Q)} = E^{(P)} + \delta(\mbox{$P$;$Q$}) .
\label{ccpq-11}
\end{equation}

One of the main advantages of the CC($P$;$Q$) methodology is its flexibility. In particular, we can
make a wide variety of conventional as well as unconventional choices of the $P$ and $Q$ spaces,
adjusting them to the nature of the electronic states of interest and adopting different numerical
procedures in their construction.
Conventional choices for the $P$ and $Q$ spaces, based on the many-body ranks of the determinants
included in them, result in the left-eigenstate CR-CC methods, such as the CR-CC(2,3)
approach discussed in the Introduction, in which the former space consists of all singly and
doubly excited determinants and the latter space is spanned by all triples. We can, however,
also make unconventional choices, including those adopted in the
CC(t;3), CC(t,q;3), CC(t,q;3,4), CC(q;4), {\it etc.} hierarchy
\cite{jspp-chemphys2012,jspp-jcp2012,jspp-jctc2012,nbjspp-molphys2017,%
ccpq-be2-jpca-2018,ccpq-mg2-mp-2019,adaptive-active-ccpq-2025} and the semi-stochastic,
\cite{stochastic-ccpq-prl-2017,stochastic-ccpq-molphys-2020,stochastic-ccpq-jcp-2021,arnab-stgap-2022}
adaptive,\cite{adaptiveccpq2023,adaptive-active-ccpq-2025} and CIPSI-driven\cite{cipsi-ccpq-2021} CC($P$;$Q$) methods,
mentioned in the Introduction as well, in which the suitably
chosen subsets of higher--than--doubly excited determinants are incorporated into the underlying $P$
spaces, in addition to all singles and doubles, to relax the lower-rank $T_{1}$ and $T_{2}$ clusters
in the presence of their higher-rank counterparts, such as the leading $T_{3}$ contributions, which
the CCSD(T), CR-CC(2,3), $\Lambda$-CCSD(T), and similar approaches are not designed to do. Having some
higher--than--doubly excited determinants in the $P$ space
provides us with a straightforward and computationally efficient 
mechanism to account for the coupling between the lower- and higher-order components of the cluster
operator, which cannot be neglected when $T_{n}$ contributions with $n>2$, such as $T_{3}$,
become large and nonperturbative, as is the case when the automerization barrier region
of the lowest-energy singlet potential of cyclobutadiene is examined.
This, in turn, allows us to recover the full CCSDT, CCSDTQ, and similar energetics
without running into the very expensive, often prohibitive, computational costs associated with the
high-level CC methods of this type.

In the case of the CIPSI-driven CC($P$;$Q$) approach, introduced in Ref.\ \onlinecite{cipsi-ccpq-2021}
and investigated in this study, the desired subsets of higher--than--doubly excited determinants incorporated
into the underlying $P$ spaces are identified with the help of sequences of relatively inexpensive
Hamiltonian diagonalizations in systematically grown, recursively defined, subspaces of the many-electron
Hilbert space, denoted as $\mathscr{V}_\text{int}^{(k)}$,  where $k = 0, 1, 2, \ldots$ enumerates the
consecutive CIPSI iterations. In doing so, we follow the CIPSI algorithm, originally proposed in
Ref.\ \onlinecite{sci_3}, further developed in Refs.\ \onlinecite{cipsi_1,cipsi_2}, and available in
the Quantum Package 2.0 software.\cite{cipsi_2} Given our interest in using CIPSI, which is one of
the selected CI approaches\cite{sci_1,sci_2,sci_3,sci_4} (see Refs.\
\onlinecite{adaptive_ci_1,adaptive_ci_2,asci_1,asci_2,ici_1,ici_2,shci_1,shci_2,shci_3}
for other examples), within the
single-reference CC($P$;$Q$) framework, the initial subspaces $\mathscr{V}_\text{int}^{(0)}$ adopted in our
work are always spanned by the restricted Hartree--Fock (RHF) or restricted open-shell Hartree--Fock
(ROHF) determinants. Once $\mathscr{V}_{\text{int}}^{(0)}$ is defined, each subsequent subspace
$\mathscr{V}_{\text{int}}^{(k+1)}$ with $k \ge 0$ is constructed by enlarging its $\mathscr{V}_{\text{int}}^{(k)}$
predecessor with the subset of the leading singly and doubly exited determinants generated out of it,
identified with the help of the many-body perturbation theory (MBPT). Thus, if
$|\Psi_k^{(\text{CIPSI})}\rangle = \sum_{|\Phi_I\rangle \in \mathscr{V}_{\text{int}}^{(k)}} c_I |\Phi_I\rangle$
and $E_{\text{var},k}$ are the CI wave function and energy obtained in $\mathscr{V}_{\text{int}}^{(k)}$,
and if the space of all singles and doubles out of $|\Psi_k^{(\text{CIPSI})}\rangle$ is designated as
$\mathscr{V}_{\text{ext}}^{(k)}$, the subset of determinants
$|\Phi_\alpha\rangle \in \mathscr{V}_{\text{ext}}^{(k)}$ selected for inclusion in
$\mathscr{V}_{\text{int}}^{(k+1)}$ consists of those that have the largest
$e_{\alpha,k}^{(2)} = |\langle\Phi_\alpha|H|\Psi_k^\text{(CIPSI)}\rangle|^2 /
(E_{\text{var},k}-\langle\Phi_\alpha|H|\Phi_\alpha\rangle)$ contributions to the perturbative correction
$\Delta E_{k}^{(2)} = \sum_{|\Phi_\alpha\rangle \in \mathscr{V}_{\text{ext}}^{(k)}} e_{\alpha,k}^{(2)}$
to $E_{\text{var},k}$. Their selection is accomplished by arranging the sampled determinants
$|\Phi_\alpha\rangle \in \mathscr{V}_{\text{ext}}^{(k)}$ in descending order according to their
$|e_{\alpha,k}^{(2)}|$ values and enlarging $\mathscr{V}_{\text{int}}^{(k)}$, determinant by determinant,
starting with the $|\Phi_\alpha\rangle$s associated with the largest $|e_{\alpha,k}^{(2)}|$ contributions
and moving toward those characterized by smaller values of $|e_{\alpha,k}^{(2)}|$, until the dimension of
$\mathscr{V}_{\text{int}}^{(k+1)}$ exceeds that of its $\mathscr{V}_{\text{int}}^{(k)}$ predecessor
by a user-defined factor $f > 1$, which in all the calculations performed in this study was set to its
default value of 2 [the actual number of determinants included in $\mathscr{V}_{\text{int}}^{(k+1)}$
is usually slightly larger than $f$ times the dimension of $\mathscr{V}_{\text{int}}^{(k)}$ since one may
have to add extra determinants in $\mathscr{V}_{\text{int}}^{(k+1)}$ to make sure that the corresponding CI
wave function $|\Psi_{k+1}^{(\text{CIPSI})}\rangle$ is an eigenstate of the total spin $S^2$ and $S_z$
operators]. To reduce the computational costs associated with the above procedure of enlarging
the $\mathscr{V}_{\text{int}}^{(k)}$ space to obtain $\mathscr{V}_{\text{int}}^{(k+1)}$, in all the
CIPSI-driven CC($P$;$Q$) computations reported in this work, we relied on a semi-stochastic version
of the above determinant selection algorithm implemented in Quantum Package 2.0, in which one
stochastically filters out the most important singly and doubly excited determinants out of
$|\Psi_k^{(\text{CIPSI})}\rangle$, so that only a small subset of singles and doubles ends up
in the $\mathscr{V}_{\text{ext}}^{(k)}$ space prior to determining and analyzing the
$e_{\alpha,k}^{(2)}$ contributions. The $e_{\alpha,k}^{(2)}$ values, in addition to guiding
the process of enlarging diagonalization spaces $\mathscr{V}_{\text{int}}^{(k)}$
and allowing us to evaluate the perturbatively corrected CIPSI energies
$E_{\text{var},k}+\Delta E_{k}^{(2)}$, can be used to calculate the renormalized
second-order corrections $\Delta E_{\text{r},k}^{(2)}$ introduced in  Ref.\ \onlinecite{cipsi_2}
and the $E_{\text{var},k}+\Delta E_{\text{r},k}^{(2)}$ energies.

To produce the final wave function $|\Psi^{(\text{CIPSI})}\rangle$,
needed to construct the list of higher--than--doubly excited determinants to be
included in the $P$ space of a given CIPSI-driven CC($P$;$Q$) calculation, and
determine the associated variational ($E_{\text{var}}$) and perturbatively corrected
[$E_{\text{var}}+\Delta E^{(2)}$ or $E_{\text{var}}+\Delta E_{\text{r}}^{(2)}$]
CIPSI energies, the sequence of Hamiltonian diagonalizations defining the underlying
CIPSI run must be terminated. This could be done by stopping at the first iteration
$k$ for which the absolute value of the second-order MBPT correction $\Delta E_k^{(2)}$
falls below a user-defined threshold $\eta$, but, given our interest in examining the
convergence of the CIPSI-based CC($P$;$Q$) calculations toward the desired high-level CC
energetics, represented in this study by CCSDT, using systematically grown $P$ spaces
obtained with the help of CIPSI, in this work we follow Ref.\ \onlinecite{cipsi-ccpq-2021}
and stop when the number of determinants in the diagonalization space equalizes or exceeds
the user-defined parameter $N_{\text{det(in)}}$. To ensure that the CIPSI sequences
preceding our CC($P$;$Q$) calculations did not terminate too soon, before the dimensions
of terminal diagonalization spaces became greater than or equal to $N_{\text{det(in)}}$,
we set the aforementioned parameter $\eta$ to 1 microhartree. As a result
[putting aside the convergence threshold used in the CC($P$) iterations, which we set
to $10^{-7}$ hartree], all CIPSI-driven CC($P$;$Q$) computations reported in this article,
along with the underlying $P$ spaces, were controlled by a single input variable $N_{\text{det(in)}}$.
In addition to $N_{\text{det(in)}}$, in presenting our CC($P$) and CC($P$;$Q$)
results for the lowest singlet and triplet potentials of cyclobutadiene, we also
provide information about the numbers of determinants included in the terminal
CIPSI wave functions $|\Psi^{(\text{CIPSI})}\rangle$ obtained for various values of
$N_{\text{det(in)}}$, designated as $N_{\text{det(out)}}$.
Given our choice of the subspace enlargement parameter $f$, the $N_{\text{det(out)}}$ values
characterizing the $|\Psi^{(\text{CIPSI})}\rangle$ states used to identify the
triply excited determinants for inclusion in the $P$ spaces employed in our CIPSI-driven
CC($P$;$Q$) computations were
always between $N_{\text{det(in)}}$ and $2N_{\text{det(in)}}$. With all of this
in mind, the algorithm used in the CIPSI-enabled CC($P$;$Q$) calculations reported in
this study, aimed at converging the CCSDT energetics, can be summarized as follows:
\cite{cipsi-ccpq-2021}
\begin{enumerate}
\item Choose a wave function termination parameter $N_\text{det(in)}$ and execute a CIPSI 
diagonalization sequence starting from the one-dimensional subspace $\mathscr{V}_\text{int}^{(0)}$
spanned by the RHF or ROHF reference determinant $|\Phi\rangle$ to obtain the
$|\Psi^\text{(CIPSI)}\rangle$ state.
\item Extract the list of triply excited determinants included in $|\Psi^\text{(CIPSI)}\rangle$
and combine it with all singly and doubly excited determinants relative to $|\Phi\rangle$ to
obtain the $P$ space for CC($P$;$Q$) calculations.
\item Solve the CC($P$) amplitude equations, Eq. (\ref{ccpq-2}), to determine the cluster operator
$T^{(P)} = T_1 + T_2 + T_3^\text{(CIPSI)}$, where $T_3^\text{(CIPSI)}$ is the three-body component
of $T^{(P)}$ defined using the list of triply excited determinants extracted from $|\Psi^\text{(CIPSI)}\rangle$,
and energy $E^{(P)}$, Eq. (\ref{ccpq-5}).
Solve the left-eigenstate CC($P$) system given by Eq. (\ref{ccpq-10}) to obtain the companion
hole--particle deexcitation operator $\Lambda^{(P)} = \Lambda_1 + \Lambda_2 + \Lambda_3^\text{(CIPSI)}$
in which the triples entering $\Lambda_3^\text{(CIPSI)}$ are the same as those included in
$T_3^\text{(CIPSI)}$.
\item Calculate the noniterative correction $\delta(\mbox{$P$;$Q$})$, Eq. (\ref{ccpq-6}), in which the $Q$ space
is defined as the remaining triply excited determinants absent in $|\Psi^\text{(CIPSI)}\rangle$, and
add it to $E^{(P)}$ to obtain the CC($P$;$Q$) energy $E^{(P+Q)}$, Eq. (\ref{ccpq-11}).
\end{enumerate}
The above steps 1--4 can be repeated by increasing $N_\text{det(in)}$. The entire process can be stopped
when the difference between consecutive $E^{(P+Q)}$ values falls below some small, user-specified,
convergence threshold.

In order to perform the CIPSI-driven CC($P$) and CC($P$;$Q$) calculations for the singlet and triplet
PECs of cyclobutadiene along the automerization coordinate investigated in this work and examine their
convergence toward the
parent CCSDT potentials, we
used the computer programs described in Ref.\ \onlinecite{cipsi-ccpq-2021}
and the newer implementation of the same methods in our open-source CCpy package available on
GitHub.\cite{CCpy-GitHub} The former codes
take advantage of our highly efficient, automatically generated, Fortran CC
routines
that were previously exploited in implementing the
active-orbital-based\cite{jspp-chemphys2012,jspp-jcp2012,jspp-jctc2012,nbjspp-molphys2017,ccpq-be2-jpca-2018}
and semi-stochastic\cite{stochastic-ccpq-prl-2017,stochastic-ccpq-molphys-2020,stochastic-ccpq-jcp-2021}
CC($P$;$Q$) approaches.
The latter codes, available in CCpy, use a hybrid Python--Fortran programming approach.
All of our CIPSI-driven CC($P$) and CC($P$;$Q$) computer programs targeting CCSDT and our group's
computer-generated CC codes used to produce the parent CCSDT data are
interfaced with the RHF, ROHF, and integral transformation routines in
GAMESS.\cite{gamess,gamess2020,gamess2023} As already alluded to above,
the lists of triply excited determinants used to construct the
$P$ spaces for the CC($P$) and CC($P$;$Q$) calculations corresponding to the various choices of
the input parameter $N_{\text{det(in)}}$ were extracted from the terminal CIPSI wave functions
$|\Psi^{(\text{CIPSI})}\rangle$ obtained with Quantum Package 2.0, whereas the complementary $Q$
spaces, needed to determine the $\delta(\mbox{$P$;$Q$})$ corrections, consisted of the remaining
triples absent in the $|\Psi^{(\text{CIPSI})}\rangle$ states. We also used Quantum Package 2.0 to
obtain the $E_{\text{var}}$, $E_{\text{var}}+\Delta E^{(2)}$, and $E_{\text{var}}+\Delta E_{\text{r}}^{(2)}$
energies associated with the CIPSI runs that provided the lists of triples for the CC($P$)
iterations. The results
of the DEA-EOMCC(4p-2h)$\{N_\text{u}\}$ calculations discussed in the
Introduction were
carried out with the highly efficient DEA-EOMCC routines developed in Ref.\ \onlinecite{jspp-dea-2021},
which became part of the official GAMESS distribution
in 2023.

To obtain
the desired insights into the convergence of the singlet and triplet potentials of cyclobutadiene
toward their CCSDT counterparts, we adopted the strategy used in Ref.\ \onlinecite{cipsi-ccpq-2021}.
Thus, for each nuclear geometry along the automerization pathway considered in this work, we carried out
a series of CIPSI-driven CC($P$) and CC($P$;$Q$) calculations using the $P$ and $Q$ spaces
derived from the increasingly large CIPSI wave functions obtained by varying $N_{\text{det(in)}}$
in an approximately semi-logarithmic manner. We started with $N_{\text{det(in)}} = 1$, where
the CIPSI wave functions $|\Psi^{(\text{CIPSI})}\rangle$ are the single determinants defining
the reference states $|\Phi\rangle$ used in our CC computations (RHF in the case of the singlet and ROHF
in the triplet case) and the resulting CC($P$) and CC($P$;$Q$) energies become identical to those
obtained with CCSD and CR-CC(2,3), respectively, and went all the way to $N_{\text{det(in)}} = 10,000,000$,
to reflect on the fact that as the input variable $N_{\text{det(in)}}$ becomes increasingly large and
the CIPSI wave functions capture more and more triply excited determinants, the CC($P$;$Q$) energies
$E^{(P+Q)}$ approach their CCSDT parents (becoming identical to them when all triply
excited determinants are captured by the $|\Psi^{(\text{CIPSI})}\rangle$ states). The energies obtained
in the CC($P$) computations approach their CCSDT counterparts as well, but, as
discussed in the next section, their convergence toward CCSDT is much slower than that observed
in the CC($P$;$Q$) 
runs.

The ability
of the CIPSI-driven CC($P$;$Q$) methodology to generate the CCSDT-level
energetics using tiny fractions of triply excited determinants in the underlying $P$ spaces captured
by the relatively small Hamiltonian diagonalizations, observed in the calculations for the lowest
${}^1A_{g}(D_{2\text{h}})$ and ${}^3B_{1g}(D_{2\text{h}})$ potentials of cyclobutadiene reported in
this study, results in enormous savings in the computational effort compared to CCSDT. Indeed, as
explained in Ref.\ \onlinecite{cipsi-ccpq-2021}, the CIPSI runs using smaller $N_{\text{det(in)}}$
values are much faster than those needed to reach convergence, the CC($P$) calculations using tiny
fractions of triples in the $P$ space are
one or more
orders of magnitude faster than the corresponding CCSDT computations,
and the effort involved in obtaining the noniterative $\delta(\mbox{$P$;$Q$})$ corrections
is similar to the determination of the triples corrections of CR-CC(2,3) or CCSD(T).
As pointed out in Refs.\
\onlinecite{stochastic-ccpq-molphys-2020,stochastic-ccpq-jcp-2021}, the key to obtaining the
desired computational efficiency in the CC($P$;$Q$) calculations lies in the development of an
algorithm capable of offering significant speedups compared to the parent CC approach, such as
CCSDT, when the lists of higher--than--doubly excited determinants included in the CC($P$)
iterations do not necessarily form continuous manifolds, as is the case when these lists are
created by the sequences of Hamiltonian diagonalizations of CIPSI adopted in this study
and Ref.\ \onlinecite{cipsi-ccpq-2021}, the previously employed CIQMC/CCMC propagations,
\cite{stochastic-ccpq-prl-2017,stochastic-ccpq-molphys-2020,stochastic-ccpq-jcp-2021,%
stochastic-ccpq-jcp-2021,arnab-stgap-2022} or the moment expansions defining the
$\delta(\mbox{$P$;$Q$})$ corrections.\cite{adaptiveccpq2023,adaptive-active-ccpq-2025} In all these cases,
conventional diagrammatic (or algebraic) techniques assuming continuous excitation
manifolds labeled by occupied and unoccupied orbitals from the respective ranges
of indices, used to implement the standard CC methods employing rank-based truncations,
no longer apply. A different programming approach is needed. The key elements of
our algorithm used to
implement
the CC($P$) equations in which the $P$ space consists
of all singly and doubly excited determinants and a generally spotty subset of triply
excited determinants identified in this work by CIPSI, along with illustrative computational
timings obtained using our CCpy codes,\cite{CCpy-GitHub} are summarized in the Appendix
(for the analogous timings information obtained in the context of the adaptive CC($P$;$Q$)
calculations for cyclobutadiene, see Ref.\ \onlinecite{adaptiveccpq2023}).

Two different basis sets were employed in this work, namely, cc-pVDZ, which we also
used to construct the lowest-energy ${}^1A_{g}(D_{2\text{h}})$ and ${}^3B_{1g}(D_{2\text{h}})$ potentials
of cyclobutadiene shown in Figure \ref{fig:figure1}, and cc-pVTZ, used to discuss the effect of the
basis set on our conclusions regarding the performance of the CIPSI-driven CC($P$;$Q$) approach.
In generating the results of the CIPSI-based CC($P$;$Q$) computations for the lowest singlet and triplet
states of cyclobutadiene along its automerization coordinate at different values of the wave function
terminating parameter $N_{\text{det(in)}}$
using a smaller cc-pVDZ basis,
we adopted the same philosophy as that
exploited
in the case of the DEA-EOMCC(4p-2h)$\{N_\text{u}\}$, extrapolated $E_\text{var}+\Delta E_r^{(2)}$,
and parent CCSDT potentials shown in Figure \ref{fig:figure1}. Thus, we used Eq. (\ref{eq:ell}),
in which the geometries of the R and TS structures on the singlet potential obtained
in the MR-AQCC/cc-pVDZ optimizations
were taken from Ref.\ \onlinecite{MR-AQCC}, to set up a one-dimensional, $D_{\rm 2h}$-symmetric,
automerization pathway parameterized by dimensionless variable $\lambda \in [0,1]$,
and then, in
analogy to the PECs shown in Figure \ref{fig:figure1},
we executed our CIPSI-based CC($P$) and CC($P$;$Q$)
calculations for the lowest ${}^1A_{g}(D_{2\text{h}})$ and ${}^3B_{1g}(D_{2\text{h}})$ states
of cyclobutadiene and determined the corresponding $E_{\text{var}}$, $E_{\text{var}}+\Delta E^{(2)}$,
and $E_{\text{var}}+\Delta E_{\text{r}}^{(2)}$ CIPSI energies for $\lambda = 0$, 0.2, 0.4, 0.6, 0.8,
and 1, reflecting the resulting PEC segments about $\lambda = 1$ to obtain the potentials that connect
the rectangular reactant and product minima via the square TS.
We used a similar strategy in the CIPSI-driven CC($P$) and CC($P$;$Q$) computations
and the preceding CIPSI runs employing the cc-pVTZ basis set, but in this case, in reporting our results,
we limited ourselves to the key R ($\lambda = 0$) and TS ($\lambda = 1$) structures optimized at the
MR-AQCC/cc-pVTZ level in Ref.\ \onlinecite{MR-AQCC}.
Consistent with the overall
symmetry of the automerization pathway examined in this work, the $D_{\rm 2h}$ point group was adopted
throughout. In particular, the $P$ and $Q$ spaces used in our CC($P$;$Q$) calculations for the lowest-energy
singlet PEC consisted of the $S_z = 0$ determinants of the $A_{g}(D_{\rm 2h})$ symmetry. In the case of the
lowest-energy triplet potential, they consisted of the $S_z = 1$ $B_{1g}(D_{\rm 2h})$ determinants.
In all post-RHF/ROHF computations reported in this article, the four core molecular orbitals
correlating with the 1s shells of the carbon atoms were frozen.

While the numerical evidence discussed in the next section
clearly demonstrates that the CIPSI computations characterized by $N_{\text{det(in)}} = 10,000,000$
result in the $P$ spaces which are unnecessarily large for accurately approximating the CCSDT energetics
using the CIPSI-driven CC($P$;$Q$) approach
when the cc-pVDZ and cc-pVTZ basis sets are employed,
we include them in our analysis since they also allowed us to extrapolate the near-full-CI
$E_\text{var}+\Delta E_r^{(2)}$ potentials,
such as those presented in Figure \ref{fig:figure1} for a cc-pVDZ basis,
by following the procedure described in Refs.\ \onlinecite{cipsi_2,cipsi_benzene,cbd-loos-2022} (see, also,
Ref.\ \onlinecite{cipsi-ccpq-2021}). In this procedure, the $E_{\text{var},k}+\Delta E_{r,k}^{(2)}$ energies
extracted from the last four to six Hamiltonian diagonalizations of the CIPSI sequence leading to the
final $|\Psi^{(\text{CIPSI})}\rangle$ state are plotted against the corresponding $\Delta E_{r,k}^{(2)}$
corrections and the resulting data, fit to a line, are extrapolated to the $\Delta E_r^{(2)}=0$ limit.
In the case of the CIPSI calculations
employing the cc-pVDZ basis set,
to extrapolate the reasonably smooth
PECs for the lowest singlet and triplet states of cyclobutadiene out of our largest CIPSI runs
corresponding to $N_{\text{det(in)}}=10,000,000$, shown in Figure \ref{fig:figure1}, we
used
the last six $E_{\text{var},k}+\Delta E_{r,k}^{(2)}$ and $\Delta E_{r,k}^{(2)}$ values of each of these runs,
since using
fewer than six values resulted in unphysical bumps in the extrapolated potentials.
The determination of the extrapolated $E_\text{var}+\Delta E_r^{(2)}$ energies of the
lowest ${}^1A_{g}(D_{2\text{h}})$ and ${}^3B_{1g}(D_{2\text{h}})$ states of cyclobutadiene at the R and TS
geometries using the cc-pVTZ basis had to be handled differently. In this case, we relied on the last four
$E_{\text{var},k}+\Delta E_{r,k}^{(2)}$ and $\Delta E_{r,k}^{(2)}$ values obtained in the
$N_{\text{det(in)}}=10,000,000$ CIPSI runs. Using more values than four
led to problems with producing a sensible result for the extrapolated $E_\text{var}+\Delta E_r^{(2)}$
energy of the lowest singlet state at the TS geometry.
In addition to allowing us to comment on the quality of the CCSDT energetics in the Introduction,
the extrapolated
$E_\text{var}+\Delta E_r^{(2)}$ energies
serve in this study
as the reference data
for assessing the accuracy of the $E_\text{var}$, $E_\text{var}+\Delta E^{(2)}$, and $E_\text{var}+\Delta E_r^{(2)}$
values
obtained in the CIPSI calculations using various
choices
of $N_{\text{det(in)}}$ from a 1--10,000,000
range.

%
\section{Results and Discussion}
\label{sec3}

As explained in the Introduction, the primary objective of this study is to examine
efficiency of the CIPSI-driven CC($P$;$Q$) methodology in converging the full CCSDT data for
the lowest singlet and triplet potentials of cyclobutadiene along its automerization coordinate
and the gap between them. We are especially interested in investigating how effective the
CIPSI-driven CC($P$;$Q$) approach is in balancing the substantial nondynamical many-electron
correlation effects characterizing the lowest $^{1}A_{g}(D_{\rm 2h})$ state in the neighborhood
of the automerization barrier region, where $T_{3}$ clusters become large, nonperturbative,
and strongly coupled to their lower-rank $T_{1}$ and $T_{2}$ counterparts, with the predominantly
dynamical correlations characterizing the lowest $^{3}B_{1g}(D_{\rm 2h})$ state. As pointed out
in the Introduction, using comparisons with the DEA-EOMCC(4p-2h)$\{N_\text{u}\}$ and
perturbatively corrected and extrapolated CIPSI results, the CCSDT approach, despite its
intrinsically single-reference character, captures essentially all relevant many-electron
correlation effects needed to accurately describe the lowest singlet and triplet potentials
of cyclobutadiene and the separation between them. The question is if the CIPSI-driven CC($P$;$Q$)
computations using compact wave functions $|\Psi^{(\text{CIPSI})}\rangle$, resulting from the
relatively inexpensive CIPSI diagonalization sequences characterized by the $N_{\text{det(out)}}$
values that are much smaller than the numbers of all $T_{3}$ amplitudes, and tiny fractions
of the triply excited determinants in the underlying $P$ spaces are capable of accomplishing
the same. We also examine how effective the noniterative
$\delta(\mbox{$P$;$Q$})$ corrections are in accelerating convergence of the CC($P$) energetics 
toward their CCSDT parents and how the rate of convergence of the CIPSI-driven CC($P$;$Q$)
calculations toward CCSDT with $N_{\text{det(in)}}$ compares to the analogous convergence
of the perturbatively corrected CIPSI energies toward their extrapolated
$E_{\text{var}}+\Delta E_{\text{r}}^{(2)}$
values. Comparisons with the CCSDt and CC(t;3) methods, which belong to the
active-orbital-based CC($P$) and CC($P$;$Q$) hierarchies, and the effect of the basis set on
the performance of the CIPSI-based CC($P$;$Q$) methodology in calculations of the lowest
singlet and triplet states of cyclobutadiene are discussed as well.

\begin{figure}[!ht]
\centering
\includegraphics[scale=0.5]{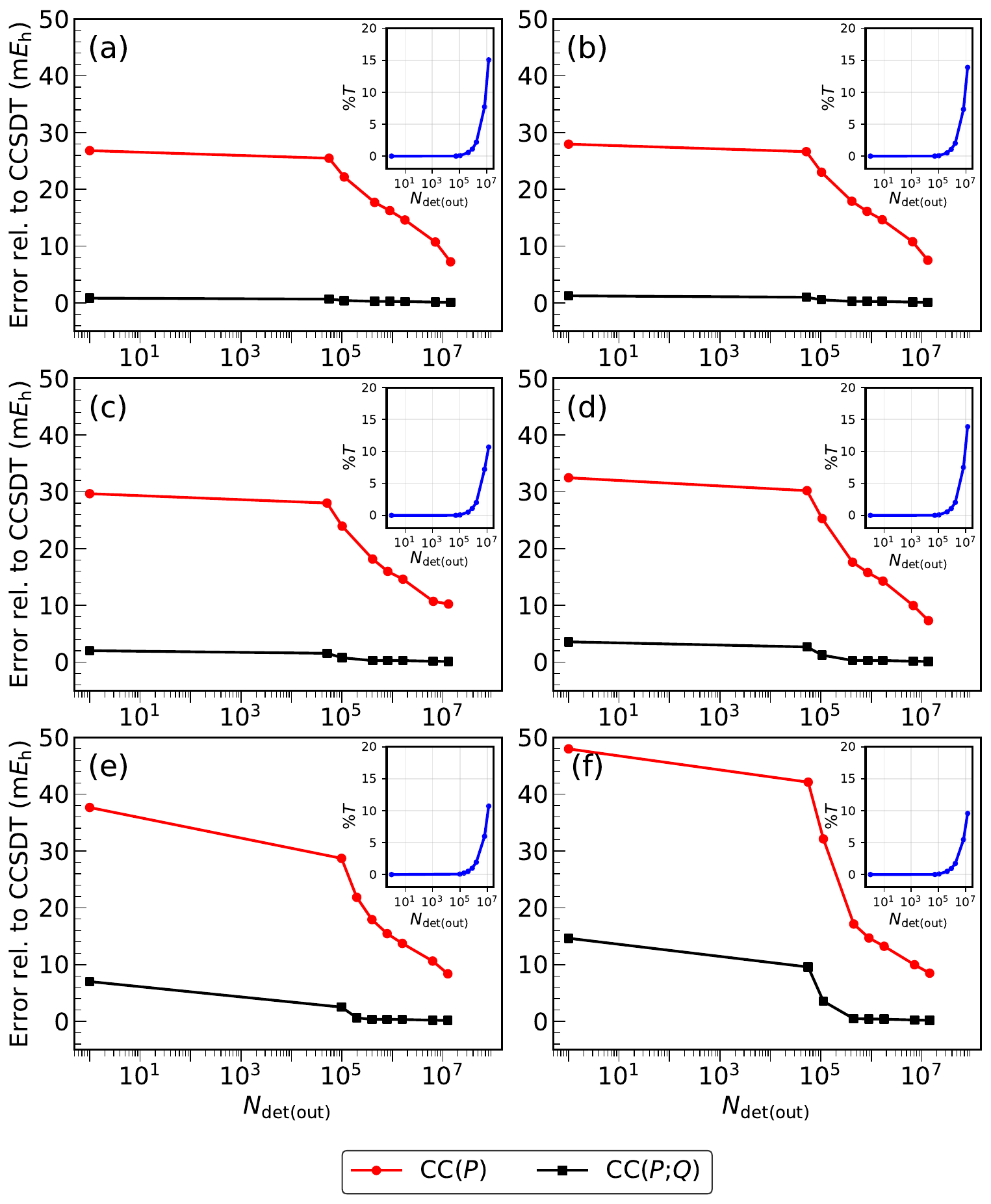}
\caption{
Graphical illustration of the convergence of the CC($P$) (red lines and circles) and
CC($P$;$Q$) (black lines and squares) energies characterizing the lowest singlet state of
cyclobutadiene, as described by the cc-pVDZ basis set, toward their CCSDT parents as functions
of the actual numbers of determinants $N_\mathrm{det(out)}$ that define the sizes of the terminal
wave functions $|\Psi^{(\text{CIPSI})}\rangle$ generated in the underlying CIPSI runs at (a)
$\lambda = 0$, (b) $\lambda = 0.2$, (c) $\lambda = 0.4$, (d) $\lambda = 0.6$, (e) $\lambda = 0.8$,
and (f) $\lambda = 1$. The insets show the percentages of the $S_z=0$ $A_{g}(D_{2\text{h}})$-symmetric
triply excited determinants captured by CIPSI as functions of $N_\text{det(out)}$.
}
\label{fig:figure2}
\end{figure}

\begin{figure}[!ht]
\centering
\includegraphics[scale=0.5]{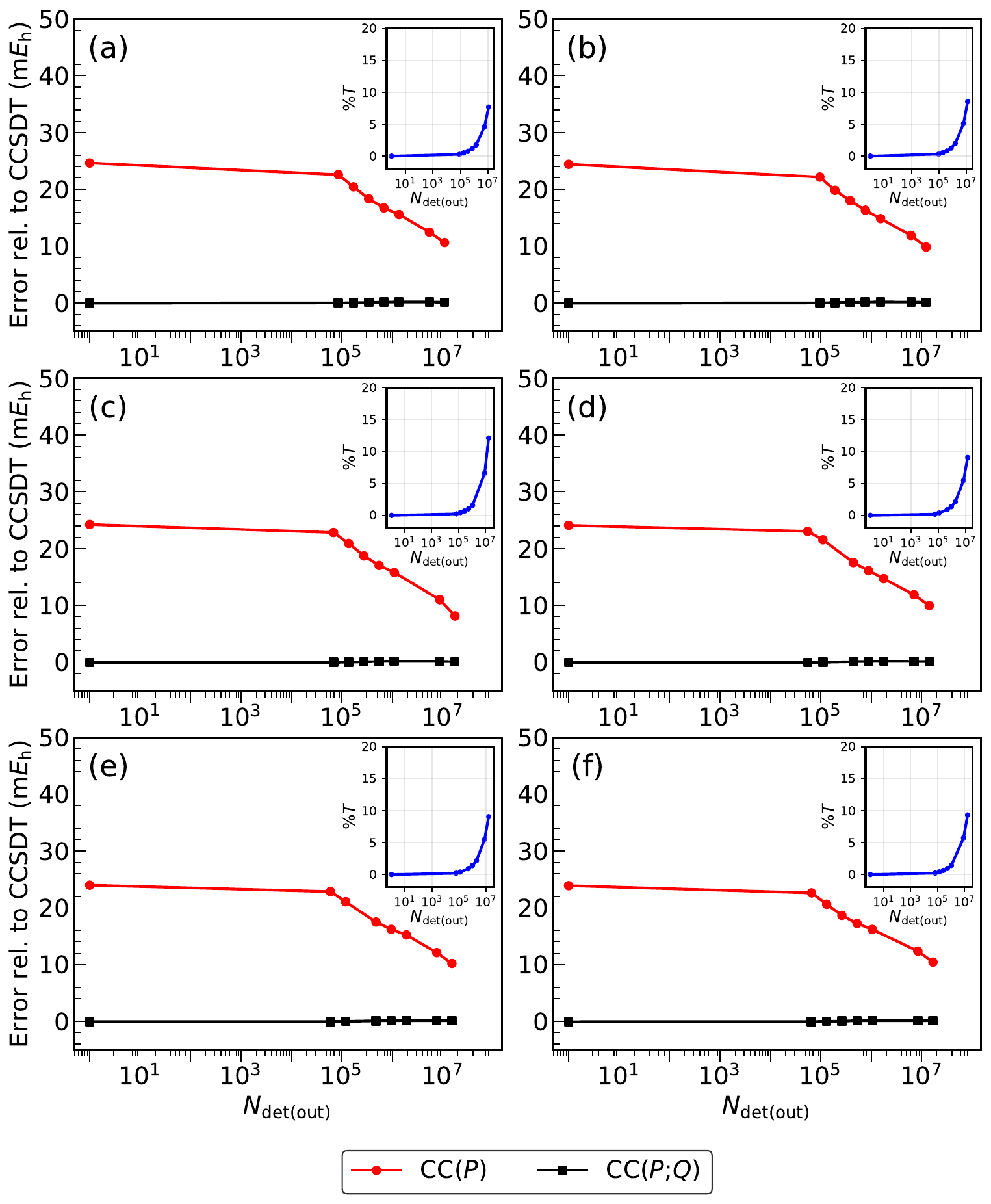}
\caption{
Graphical illustration of the convergence of the CC($P$) (red lines and circles) and
CC($P$;$Q$) (black lines and squares) energies characterizing the lowest triplet state of
cyclobutadiene, as described by the cc-pVDZ basis set, toward their CCSDT parents as functions
of the actual numbers of determinants $N_\mathrm{det(out)}$ that define the sizes of the terminal
wave functions $|\Psi^{(\text{CIPSI})}\rangle$ generated in the underlying CIPSI runs at (a)
$\lambda = 0$, (b) $\lambda = 0.2$, (c) $\lambda = 0.4$, (d) $\lambda = 0.6$, (e) $\lambda = 0.8$,
and (f) $\lambda = 1$. The insets show the percentages of the $S_z=1$ $B_{1g}(D_{2\text{h}})$-symmetric
triply excited determinants captured by CIPSI as functions of $N_\text{det(out)}$.
}
\label{fig:figure3}
\end{figure}

\begin{figure}[!ht]
\centering
\includegraphics[scale=0.5]{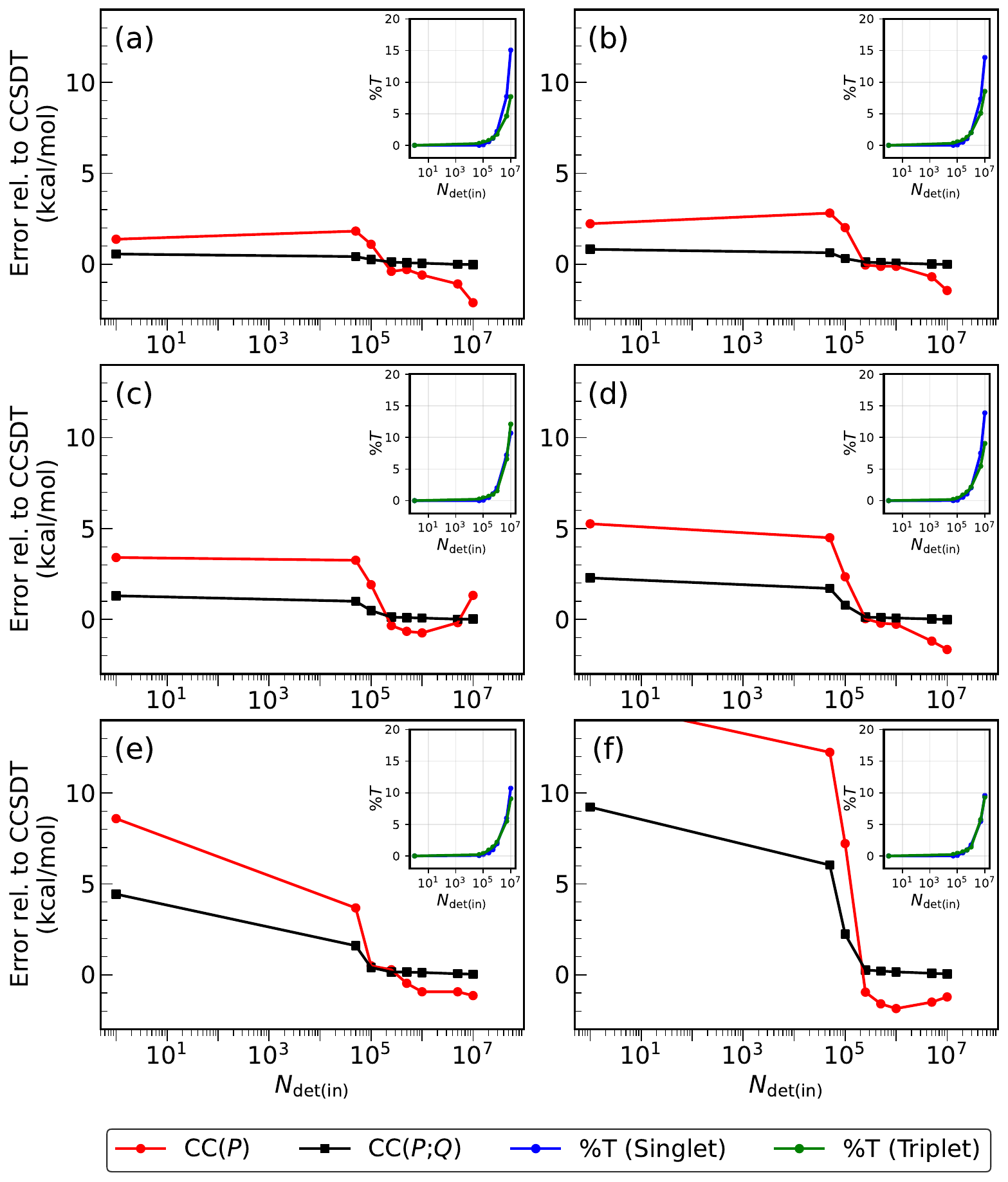}
\caption{
Graphical illustration of the convergence of the CC($P$) (red lines and circles) and CC($P$;$Q$)
(black lines and squares) singlet--triplet gaps $\Delta E_\text{S--T}= E_\text{S} - E_\text{T}$ of
cyclobutadiene, as described by the cc-pVDZ basis set, toward their CCSDT parents as functions of the
CIPSI input parameter $N_{\text{det(in)}}$ (common to the calculations for the lowest singlet and triplet
states) at (a) $\lambda = 0$, (b) $\lambda = 0.2$, (c) $\lambda = 0.4$, (d) $\lambda = 0.6$, (e)
$\lambda = 0.8$, and (f) $\lambda = 1$. The insets show the percentages of the triply excited
determinants of the $S_z=0$ $A_{g}(D_{2\text{h}})$ (blue lines and circles) and $S_z=1$
$B_{1g}(D_{2\text{h}})$ (green lines and circles) symmetries captured by the underlying CIPSI
runs as functions of $N_{\text{det(in)}}$.
}
\label{fig:figure4}
\end{figure}

\begin{figure}[!ht]
\centering
\includegraphics[scale=0.5]{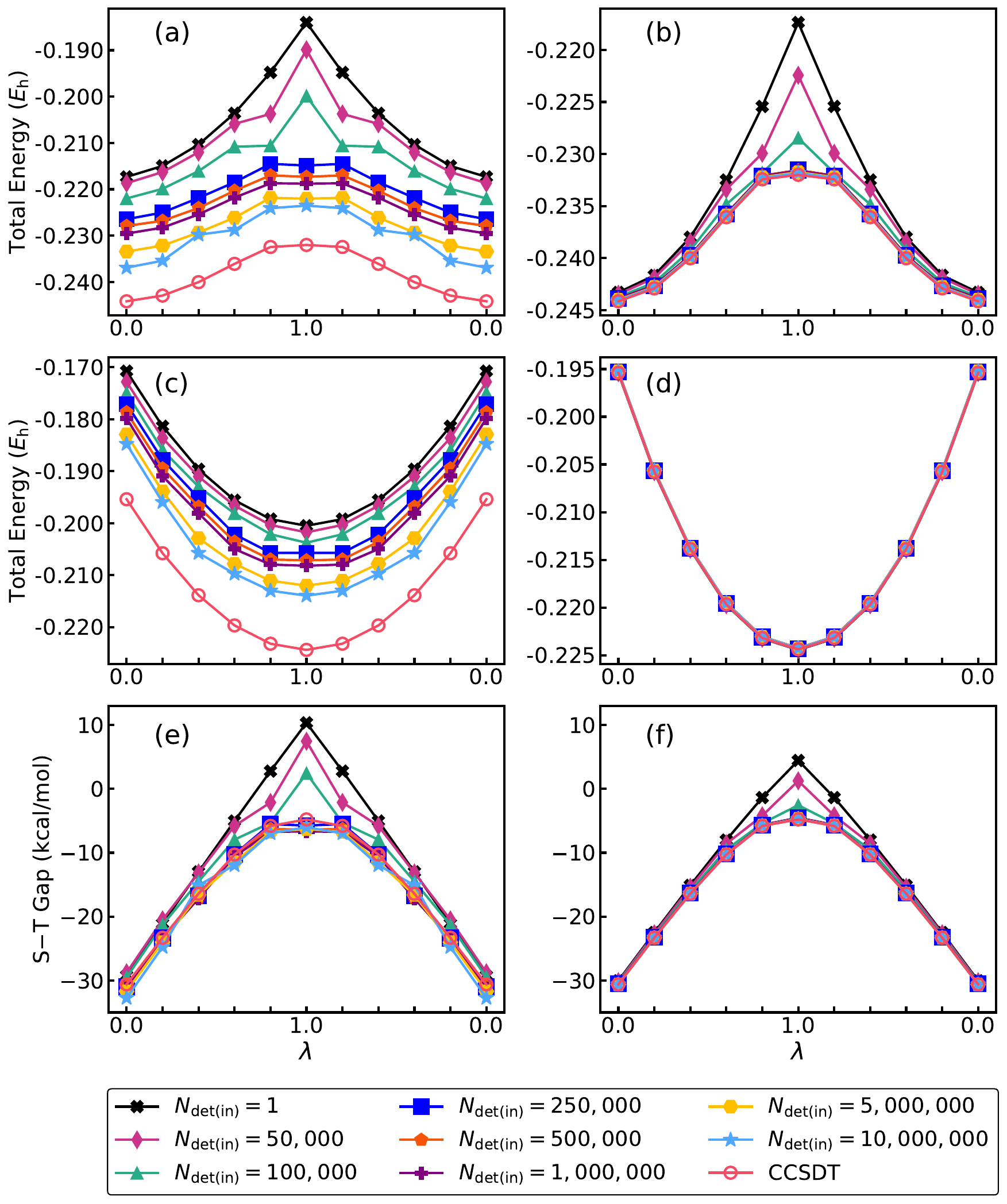}
\caption{Convergence of the CC($P$) and CC($P$;$Q$) energies $E$, reported as ($E + 154.0$) hartree,
of the lowest singlet [panels (a) and (b)] and triplet [panels (c) and (d)] states of cyclobutadiene, as
described by the cc-pVDZ basis set, and the $\Delta E_\text{S--T}$ gaps between them [panels (e) and (f)]
toward their CCSDT counterparts with the CIPSI wave function termination parameter $N_\text{det(in)}$
at selected values of the dimensionless variable $\lambda$ defining the automerization coordinate via
the interpolation formula given by Eq. (\ref{eq:ell}). The CC($P$) results are reported in panels
(a), (c), and (e). Panels (b), (d), and (f) show the corresponding CC($P$;$Q$) data.}
\label{fig:figure5}
\end{figure}

We begin by analyzing our CIPSI-driven CC($P$) and CC($P$;$Q$) computations employing
the cc-pVDZ basis set, used in most of the calculations reported in this work.
The results of our CIPSI-driven CC($P$)/cc-pVDZ and CC($P$;$Q$)/cc-pVDZ calculations for the
lowest-energy $^{1}A_{g}(D_{\rm 2h})$ and $^{3}B_{1g}(D_{\rm 2h})$ potentials of cyclobutadiene
along its automerization coordinate, the separation between them, and the associated $E_{\text{var}}$,
$E_{\text{var}}+\Delta E^{(2)}$, and $E_{\text{var}} + \Delta E^{(2)}_r$ data can be found in
Tables \ref{tab:table1}--\ref{tab:table3}. The convergence of the CC($P$)/cc-pVDZ
and CC($P$;$Q$)/cc-pVDZ energies of the lowest singlet and triplet states of cyclobutadiene and the
gaps between them, determined at $\lambda = 0$, 0.2, 0.4, 0.6, 0.8, and 1, toward their CCSDT/cc-pVDZ
parents as functions of the parameters $N_{\text{det(in)}}$ and $N_{\text{det(out)}}$ that control
[$N_{\text{det(in)}}$] and define [$N_{\text{det(out)}}$] the sizes of the terminal wave functions
$|\Psi^{(\text{CIPSI})}\rangle$ produced by the underlying CIPSI runs is also visualized in Figures
\ref{fig:figure2}--\ref{fig:figure4} and, in a summary form, in Figure \ref{fig:figure5}.

Let us first comment
on the calculations using $N_{\text{det(in)}} = 1$, which help us
appreciate the need for including the leading triply excited determinants in the $P$ spaces
employed in the CC($P$) and CC($P$;$Q$) computations, especially when $T_{3}$ effects and the
coupling between $T_{1}$ and $T_{2}$ clusters and their higher-rank $T_{3}$ counterpart
become significant. As explained in
Section \ref{sec2}, when $N_{\text{det(in)}} = 1$, the CIPSI-driven CC($P$) and CC($P$;$Q$)
approaches become equivalent to CCSD and CR-CC(2,3), respectively, {\it i.e.}, one solves the
CCSD equations for $T_{1}$ and $T_{2}$ clusters, as if they were decoupled from their higher-rank
$T_{3}$ counterpart, and corrects the resulting CCSD energies for the effects of connected triples
using the CR-CC(2,3) method. Upon examining the $N_{\text{det(in)}} = 1$ CC($P$) values in
Table \ref{tab:table1}
({\it cf.}, also, Figure \ref{fig:figure2}),
we observe that the $T_{3}$ correlation effects characterizing the lowest-energy ${}^1A_{g}(D_{2\text{h}})$
potential, estimated by forming differences between the CCSDT and CCSD energies, are not only
large, but also dramatically changing along the automerization coordinate, from $-26.827$
millihartree for the $\lambda = 0$ R species to $-47.979$ millihartree
for the $\lambda = 1$ TS structure when the cc-pVDZ basis set is employed.
As indicated by the $N_{\text{det(in)}} = 1$
CC($P$;$Q$) results shown in Table \ref{tab:table1}
and Figure \ref{fig:figure2},
the incorporation of $T_{3}$ correlations
using the noniterative CR-CC(2,3) corrections to the CCSD energies helps, reducing the 26.827,
27.964, 29.667, 32.473, 37.662, and 47.979 millihartree errors relative to CCSDT obtained at
$\lambda = 0$, 0.2, 0.4, 0.6, 0.8, and 1 with CCSD to 0.848, 1.253, 2.021, 3.582, 7.008, and
14.636 millihartree, respectively,
but substantial discrepancies between the CR-CC(2,3) and CCSDT
data, especially in the neighborhood of the barrier region, where they are as large as 7--15
millihartree when $\lambda \in [0.8,1]$, remain. As a result, the quality of the
${}^1A_{g}(D_{2\text{h}})$ potential obtained in the CR-CC(2,3) calculations, which is
characterized by a large, 13.788 millihartree, nonparallelity error (NPE) relative to its CCSDT
parent and which is shown in Figure \ref{fig:figure5} (b), is poor. Failure of CR-CC(2,3) and,
as demonstrated, for example, in Refs.\ \onlinecite{jspp-jcp2012,tailored3}, of other
noniterative triples corrections to CCSD, including CCSD(T)\cite{jspp-jcp2012,tailored3}
and CCSD(2)$_{\rm T}$,\cite{jspp-jcp2012} to accurately describe the lowest ${}^1A_{g}(D_{2\text{h}})$
state of cyclobutadiene in the automerization barrier region is,
in significant part, a consequence of the inability of all such methods to
capture the coupling of the lower-rank $T_{1}$ and $T_{2}$ clusters with $T_{3}$, which
in the vicinity of the TS geometry, where $T_{3}$ effects are large and nonperturbative,
is big enough to substantially alter $T_{1}$ and $T_{2}$ amplitudes compared to their CCSD values.
This means that to improve the quality of the ${}^1A_{g}(D_{2\text{h}})$ potential obtained with
CR-CC(2,3), one should relax $T_{1}$ and $T_{2}$ clusters, adjusting them to the presence of
$T_{3}$ correlations, prior to determining noniterative triples
corrections. The CIPSI-driven CC($P$;$Q$) methodology
allows us to do it in a computationally efficient manner, avoiding expensive CCSDT
iterations, by incorporating the leading triply excited determinants identified with the help of
the CIPSI runs using sufficiently large $N_{\text{det(in)}} > 1$ values into the underlying $P$ spaces
and correcting the resulting CC($P$) energies for the remaining $T_{3}$ effects using the
$\delta(\mbox{$P$;$Q$})$ corrections.

The situation with the lowest-energy $^{3}B_{1g}(D_{\rm 2h})$ potential is different. In this case,
as shown in Table \ref{tab:table2}
and Figure \ref{fig:figure3},
the $T_{3}$ correlation effects, estimated, after subtracting the CCSD
energies from their CCSDT counterparts
obtained with cc-pVDZ,
at about $(-25) - (-24)$ millihartree, barely vary with the
automerization coordinate $\lambda$ and are accurately described by the CR-CC(2,3) approach, which
reproduces the parent CCSDT energetics to within 60 microhartree across the entire $^{3}B_{1g}(D_{\rm 2h})$
PEC. Because of this very different behavior of CR-CC(2,3) compared to the lowest-energy singlet potential,
which can be seen by comparing the $N_{\text{det(in)}} = 1$ CC($P$;$Q$) PECs in the (b) and (d) panels of
Figure \ref{fig:figure5}, one ends up with a highly unbalanced description of the lowest singlet and triplet
states of cyclobutadiene by the CR-CC(2,3) method, especially in vicinity of the barrier on the
${}^1A_{g}(D_{2\text{h}})$ potential. This is reflected in the errors characterizing the singlet--triplet
gap values obtained with CR-CC(2,3), relative to CCSDT, which
in the calculations using the cc-pVDZ basis set
grow from
$0.553$ kcal/mol at $\lambda = 0$ to $9.222$ kcal/mol when the $\lambda = 1$ square TS structure
is considered [see the $N_{\text{det(in)}} = 1$ CC($P$;$Q$) values of $\Delta E_{\text{S--T}}$ in
Table \ref{tab:table3}; {\it cf.}, also,
Figures \ref{fig:figure4} and
\ref{fig:figure5} (f)]. Once again, the main problem
resides in the neglect of the coupling between $T_{1}$ and $T_{2}$ clusters and their higher-rank
$T_{3}$ counterpart in the CR-CC(2,3)
approach, which results in a poor description of the ${}^1A_{g}(D_{2\text{h}})$ potential in the
vicinity of the TS geometry that propagates into the similarly poor $\Delta E_{\text{S--T}}$ gap values.
To bring the results closer to those obtained with CCSDT, the input variable $N_{\text{det(in)}}$ that
controls the CIPSI runs preceding the CC($P$) and CC($P$;$Q$) steps must be increased. The results of
the CIPSI-driven
CC($P$)/cc-pVDZ and CC($P$;$Q$)/cc-pVDZ
computations for the lowest ${}^1A_{g}(D_{2\text{h}})$ and
$^{3}B_{1g}(D_{\rm 2h})$ potentials of cyclobutadiene and the gap between them using representative
$N_{\text{det(in)}} > 1$ values are discussed next.

As shown in Table \ref{tab:table1} and
Figures \ref{fig:figure2} and
\ref{fig:figure5} (b), the CC($P$;$Q$) computations for the
lowest ${}^1A_{g}(D_{2\text{h}})$ state, using CIPSI Hamiltonian diagonalizations to identify the leading
triply excited determinants for inclusion in the underlying $P$ spaces, display fast convergence toward
CCSDT with $N_{\text{det(in)}}$, independent of the value of $\lambda$.
In the case of the calculations performed using the cc-pVDZ basis set discussed here, with
as little as $101,361$--$196,965$
$S_z = 0$ determinants of the $A_{g}(D_{\rm 2h})$ symmetry in the terminal $|\Psi^{(\text{CIPSI})}\rangle$
wave functions generated by the inexpensive CIPSI runs using $N_{\text{det(in)}} = 100,000$, which capture
tiny fractions, on the order of 0.1--0.2\%, of the $14,483,876$ $A_{g}(D_{\rm 2h})$-symmetric $S_z=0$ triples,
the CC($P$;$Q$) method reduces the 0.848, 1.253, 2.021, 3.582, 7.008, and 14.636 millihartree errors relative
to CCSDT obtained at $\lambda = 0$, 0.2, 0.4, 0.6, 0.8, and 1 with CR-CC(2,3) to 0.431, 0.552, 0.776, 1.235,
0.628, and 3.539 millihartree, respectively. With the relatively small additional effort corresponding to
$N_{\text{det(in)}} = 250,000$, which results in $394,080$--$449,753$ $S_z=0$ determinants of the
$A_{g}(D_{\rm 2h})$ symmetry in the final CIPSI diagonalization spaces and only 0.5--0.6\% of all triples
in the underlying $P$ spaces, the differences between the CC($P$;$Q$) and CCSDT energies of the lowest
singlet state of cyclobutadiene at $\lambda = 0$, 0.2, 0.4, 0.6, 0.8, and 1 decrease to 0.278, 0.272,
0.271, 0.300, 0.336, and 0.458 millihartree, respectively.
Clearly, these are massive error reductions compared to the
CR-CC(2,3) computations, especially in the barrier region, which highlight the effectiveness of our
CIPSI-driven CC($P$;$Q$) strategy and the importance of relaxing $T_{1}$ and $T_{2}$ amplitudes in the
presence of the leading $T_{3}$ contributions compared to their CCSD values prior to determining the
noniterative corrections for the remaining $T_{3}$ effects. Similar comments
apply to the improvements in the troublesome 13.788 millihartree NPE relative to CCSDT characterizing the
${}^1A_{g}(D_{2\text{h}})$ potential obtained in the
CR-CC(2,3)/cc-pVDZ
calculations offered by the CIPSI-driven CC($P$;$Q$) runs. When the
CC($P$;$Q$)/cc-pVDZ
approach using $N_{\text{det(in)}} = 100,000$ is employed, the NPE
relative to CCSDT characterizing the resulting ${}^1A_{g}(D_{2\text{h}})$ potential becomes 3.108 millihartree,
which is a reduction of the
corresponding
CR-CC(2,3) NPE value by a factor of 4.4. The $N_{\text{det(in)}} = 250,000$
CC($P$;$Q$) computations, which extract the lists of triples from the relatively inexpensive Hamiltonian
diagonalizations in spaces that
in the case of the cc-pVDZ basis set
are 32--37 times smaller than the number of $T_{3}$ amplitudes used by
CCSDT, reduce the 13.788 millihartree NPE characterizing the lowest-energy ${}^1A_{g}(D_{2\text{h}})$
potential obtained with
CR-CC(2,3)/cc-pVDZ,
relative to its CCSDT counterpart, to an impressively small value of
0.187 millihartree. This is a 74-fold reduction in NPE compared to CR-CC(2,3).

It is clear from Table \ref{tab:table1} and
Figures \ref{fig:figure2} and
\ref{fig:figure5} (b) that the convergence of the
lowest-energy ${}^1A_{g}(D_{2\text{h}})$ potentials resulting from the CIPSI-driven CC($P$;$Q$)
calculations toward their CCSDT parent, including the challenging barrier region, with the CIPSI wave
function termination parameter $N_{\text{det(in)}}$, with the number of determinants in the final
Hamiltonian diagonalization space used to determine $|\Psi^{(\text{CIPSI})}\rangle$
[$N_{\text{det(out)}}$], and with the fraction of triply excited determinants in the $P$ space
captured by CIPSI is very fast, but one cannot say the same about the uncorrected CC($P$) energies.
As shown in Table \ref{tab:table1} and
Figures \ref{fig:figure2} and
\ref{fig:figure5} (a), and in line with the formal analysis
in Section \ref{sec2}, the CC($P$) energies improve the CCSD results and converge toward CCSDT, but they
do it at a much slower rate than their $\delta(\mbox{$P$;$Q$})$-corrected CC($P$;$Q$) counterparts.
For instance, the CIPSI-driven
CC($P$)/cc-pVDZ
computations for the lowest singlet state of cyclobutadiene
using $N_{\text{det(in)}}=100,000$
reduce the 26.827, 27.964, 29.667, 32.473, 37.662, and 47.979 millihartree
errors relative to CCSDT obtained at $\lambda = 0$, 0.2, 0.4, 0.6, 0.8, and 1 with
CCSD/cc-pVDZ
and the associated NPE value of 21.152 millihartree to 22.183, 23.029, 23.947,
25.279,
21.842, 32.125, and
10.283 millihartree, respectively. This should be compared to the much smaller error and NPE values
characterizing the corresponding
CC($P$;$Q$)/cc-pVDZ
calculations, which are 0.431--3.539
and 3.108 millihartree, respectively. The analogous CC($P$) calculations using $N_{\text{det(in)}} = 250,000$,
where the errors characterizing the $\delta(\mbox{$P$;$Q$})$-corrected CC($P$;$Q$)
energies in the entire $\lambda = 0$--1 region and the overall NPE relative to CCSDT are already at
the level of 0.2--0.5 millihartree, produce the
17.137--18.165
millihartree errors and the NPE of
1.028
millihartree. Even with the largest $N_{\text{det(in)}}$ value considered in this study, of
$10,000,000$, the 7.252--10.238 millihartree differences between the
CC($P$)/cc-pVDZ and CCSDT/cc-pVDZ
energies of the
lowest ${}^1A_{g}(D_{2\text{h}})$ state of cyclobutadiene in the $\lambda = 0$--1 region and the NPE of
2.986 millihartree that characterizes the resulting CC($P$) potential relative to its CCSDT parent remain. All
of this implies that while relaxing $T_{1}$ and $T_{2}$ amplitudes in the presence of the leading
triples is important, correcting the CC($P$) energies for the remaining $T_{3}$ effects, which the CC($P$)
computations using the $P$ spaces generated with the help of CIPSI do not describe, is critical to reach
submillihartree accuracy levels relative to CCSDT with small fractions of triples in these spaces.
We observed a similar behavior in the
semi-stochastic, CIQMC- and CCMC-driven,\cite{stochastic-ccpq-prl-2017,stochastic-ccpq-molphys-2020,%
stochastic-ccpq-jcp-2021,arnab-stgap-2022} and adaptive\cite{adaptiveccpq2023} CC($P$) and CC($P$;$Q$) calculations,
although based on the numerical evidence that we have generated to date, the CIPSI-driven CC($P$;$Q$)
methodology investigated in this study and its recently formulated adaptive analog seem to be more
effective in converging the target CC (in most of our work to date, CCSDT) energetics than
their semi-stochastic counterparts. This suggests that the sequences of Hamiltonian diagonalizations
utilized in the CIPSI-driven CC($P$) and CC($P$;$Q$) computations and the moment expansions
defining the $\delta(\mbox{$P$;$Q$})$ corrections that are used to construct excitation spaces
in the adaptive CC($P$) and CC($P$;$Q$) runs
are more efficient in identifying the leading higher--than--doubly excited determinants for
inclusion of the underlying $P$ spaces than the CIQMC/CCMC wave function propagations, although
this topic needs to be explored further and we will return to it in the future.

As shown in Table \ref{tab:table2} and
Figures \ref{fig:figure3} and
\ref{fig:figure5} (c) and (d), many of the above
observations apply to the CIPSI-driven CC($P$) and CC($P$;$Q$) calculations for the lowest-energy
$^{3}B_{1g}(D_{\rm 2h})$ potential, but, given the fact that the CR-CC(2,3) approach is already very
accurate in this case, producing errors relative to CCSDT that in absolute value do not exceed 60
microhartree
when the cc-pVDZ basis set is employed,
the CC($P$;$Q$) computations using $N_{\text{det(in)}} > 1$ offer no obvious advantages over
CR-CC(2,3). Nonetheless, it is reassuring that, in analogy to the $A_{g}(D_{\rm 2h})$-symmetric singlet
ground state, the differences between the energies of the $^{3}B_{1g}(D_{\rm 2h})$ state
obtained in the CC($P$) computations using
CIPSI Hamiltonian diagonalizations to create lists of the leading triply excited determinants
for inclusion in the underlying $P$ spaces and their CCSDT counterparts decrease as $N_{\text{det(in)}}$
increases, independent of $\lambda$. It is also encouraging that the $\delta(\mbox{$P$;$Q$})$ corrections
are as effective in bringing the CC($P$) energies to a virtually perfect agreement with the parent
CCSDT data as in the case of the CIPSI-driven
CC($P$;$Q$)/cc-pVDZ
calculations for the lowest
${}^1A_{g}(D_{2\text{h}})$ potential using $N_{\text{det(in)}} \geq 250,000$.
By inspecting the CC($P$;$Q$) column in Table \ref{tab:table2},
one may get the impression that the incorporation of the triply excited determinants identified by the
CIPSI runs using increasingly large $N_{\text{det(in)}}$ values in the preceding CC($P$) steps
worsens the CR-CC(2,3) results for the lowest $^{3}B_{1g}(D_{\rm 2h})$ state, which correspond to
$N_{\text{det(in)}} = 1$, but reading Table \ref{tab:table2} in this way would be misleading. Indeed,
in single-reference situations, such as that created by the lowest $^{3}B_{1g}(D_{\rm 2h})$ state
of cyclobutadiene, where many-electron correlation effects are essentially only dynamical, the triples
correction of CR-CC(2,3) [similarly to CCSD(T)] often overshoots the parent CCSDT energies (slightly).
Once one starts adding triply excited determinants to the $P$ space, the CC($P$;$Q$) energies
initially go up, becoming upper bounds to their CCSDT counterparts, but when the fraction of
triples in the $P$ space is large enough, the differences between the CC($P$;$Q$) and CCSDT energies 
decrease, steadily approaching 0. We see some of this behavior in Table \ref{tab:table2}, but
we have to keep in mind that the $P$ spaces used in our CIPSI-driven CC($P$;$Q$) calculations use
small fractions of triples, so that the residual, $\sim$0.1 millihartree, errors relative to CCSDT
remain. What is most important here is that the CC($P$;$Q$) computations for the
lowest triplet potential of cyclobutadiene using $N_{\text{det(in)}} > 1$ reported
in Table \ref{tab:table2} and 
Figures \ref{fig:figure3} and
\ref{fig:figure5} (d) do not substantially alter the
already excellent CR-CC(2,3) energetics. This allows us to conclude that the coupling
of the lower-rank $T_{1}$ and $T_{2}$ clusters with their higher-rank $T_{3}$ counterpart
is negligible in this case and the relaxation of the CCSD values of $T_{1}$ and $T_{2}$
amplitudes by including some triples in the iterative CC($P$) steps is not necessary for
obtaining high-accuracy CC($P$;$Q$) results.

Having demonstrated the excellent performance of the CIPSI-driven CC($P$;$Q$) approach in accurately
approximating the lowest-energy $^{1}A_{g}(D_{\rm 2h})$ and $^{3}B_{1g}(D_{\rm 2h})$ potentials of cyclobutadiene
along its automerization coordinate obtained with
CCSDT, we now turn
to the CC($P$) and CC($P$;$Q$) singlet--triplet gaps and their dependence on
$\lambda$ and $N_{\text{det(in)}}$ examined,
using the cc-pVDZ basis,
in Table \ref{tab:table3} and
Figures \ref{fig:figure4},
\ref{fig:figure5} (e)
[the uncorrected CC($P$) energetics],
and \ref{fig:figure5} (f)
[the CC($P$;$Q$) results].
As pointed out above, to obtain accurate $\Delta E_{\text{S--T}}$ values
for cyclobutadiene in the vicinity of the barrier region, one has to balance significant
nondynamical correlations associated with the multiconfigurational singlet state, which manifest themselves
in massive $T_{3}$ clusters that are strongly coupled to the one- and two-body
components of $T$, with predominantly dynamical correlations characterizing the lowest triplet state that
result in the generally smaller $T_{3}$ contributions having minimal effect on $T_{1}$ and $T_{2}$ amplitudes.
The singlet--triplet gap values reported in Table \ref{tab:table3},
Figure \ref{fig:figure4},
and Figure \ref{fig:figure5} (e) and (f) clearly
show that neither the CCSD approach nor the CR-CC(2,3) triples correction to CCSD, which are equivalent to
the CIPSI-driven CC($P$) and CC($P$;$Q$) calculations using $N_{\text{det(in)}} = 1$, can
do this. Both of these methods struggle with achieving a balanced description of the $^{1}A_{g}(D_{\rm 2h})$ and
$^{3}B_{1g}(D_{\rm 2h})$ states of cyclobutadiene as $\lambda$ approaches 1, producing errors relative to CCSDT
that
in the calculations using the cc-pVDZ basis set
are as large as 5.261 and 2.282 kcal/mol, respectively, at $\lambda = 0.6$, where the CCSDT value of
$\Delta E_{\text{S--T}}$ is $-10.295$ kcal/mol, 8.589 and 4.434 kcal/mol, respectively, at $\lambda = 0.8$,
where $\Delta E_{\text{S--T}}$ obtained with CCSDT is $-5.804$ kcal/mol, and 15.120 and 9.222 kcal/mol
at $\lambda = 1$, where the CCSDT result for $\Delta E_{\text{S--T}}$ is $-4.783$ kcal/mol. These large
error values in the singlet--triplet gaps resulting from the CCSD and CR-CC(2,3) computations in the barrier
region are a consequence of the dramatic increase in the magnitude of $T_{3}$ effects characterizing the
$^{1}A_{g}(D_{\rm 2h})$ state and a rapidly deteriorating description of this state by both CCSD and CR-CC(2,3)
as $\lambda \rightarrow 1$, seen in Table \ref{tab:table1},
Figure \ref{fig:figure2},
and Figure \ref{fig:figure5} (a) and (b), as opposed to
the nearly constant $T_{3}$ contributions and a virtually perfect agreement between the CR-CC(2,3) and CCSDT
$^{3}B_{1g}(D_{\rm 2h})$ potentials in the entire $\lambda = 0$--1 region shown in Table \ref{tab:table2},
Figure \ref{fig:figure3},
and Figure \ref{fig:figure5} (d) [while quantitatively inaccurate, the shape of the $^{3}B_{1g}(D_{\rm 2h})$
potential obtained with CCSD, shown in Figure \ref{fig:figure5} (c), is qualitatively correct too].
In analogy to the previously discussed calculations for the lowest-energy $^{1}A_{g}(D_{\rm 2h})$ potential,
in order to bring the above errors down, the input parameter $N_{\text{det(in)}}$, which controls the CIPSI
diagonalization sequences preceding the CC($P$) and CC($P$;$Q$) computations, must be increased, so that
the $P$ spaces used in these computations are augmented with the leading triply excited determinants,
the CCSD values of $T_{1}$ and $T_{2}$ amplitudes, used in CR-CC(2,3), are properly relaxed, and the quality
of the $\delta(\mbox{$P$;$Q$})$ corrections that capture the remaining $T_{3}$ effects improves.

This is precisely what we observe in the CIPSI-driven CC($P$) and CC($P$;$Q$) calculations of the
singlet--triplet gaps, especially the latter ones, reported in Table \ref{tab:table3},
Figure \ref{fig:figure4},
and Figure
\ref{fig:figure5} (e) and (f). Indeed, the CC($P$;$Q$) approach using
the cc-pVDZ basis and
$N_{\text{det(in)}}=100,000$,
which relies on small CIPSI diagonalization spaces [small $N_{\text{det(out)}}$ values], whose
dimensionalities are about 1\% of the $14,483,876$ $A_{g}(D_{\rm 2h})$-symmetric $S_z=0$ and $14,339,992$
$B_{1g}(D_{\rm 2h})$-symmetric $S_z=1$ triply excited amplitudes involved in the parent full CCSDT
computations, and which employs even smaller $P$ spaces having only 0.1--0.6\% of all triples, reduces
the 0.553, 0.813, 1.300, 2.282, 4.434, and 9.222 kcal/mol differences between the
CR-CC(2,3)/cc-pVDZ and CCSDT/cc-pVDZ
$\Delta E_{\text{S--T}}$ values at $\lambda = 0$, 0.2, 0.4, 0.6, 0.8, and 1 to 0.253, 0.301, 0.485, 0.784,
0.398, and 2.227 kcal/mol, respectively. When $N_{\text{det(in)}}$ is increased to $250,000$,
where the numbers of determinants included in the final CIPSI diagonalizations preceding
the CC($P$) and CC($P$;$Q$) steps are still only $\sim$2--3\% of all $T_{3}$ amplitudes used in the
target CCSDT/cc-pVDZ
calculations for the $^{1}A_{g}(D_{\rm 2h})$ and $^{3}B_{1g}(D_{\rm 2h})$ states, and where the
resulting $P$ spaces contain only 0.5--0.9\% of all triples, the errors in the CC($P$;$Q$)
$\Delta E_{\text{S--T}}$ gaps relative to their CCSDT counterparts
obtained
at the above values of $\lambda$
with the cc-pVDZ basis set
decrease even more, to 0.112, 0.108, 0.128, 0.128, 0.152, and 0.263 kcal/mol, respectively, bringing
the CC($P$;$Q$) and CCSDT results to a virtually perfect agreement, while reducing a computational effort
compared to the CCSDT runs by orders of magnitude. As in the previously discussed results for the lowest
singlet and triplet states of cyclobutadiene, especially for the challenging, $A_{g}(D_{\rm 2h})$-symmetric,
singlet ground state, the $\delta(\mbox{$P$;$Q$})$ corrections play a major role in the observed error
reductions, substantially improving the CC($P$) $\Delta E_{\text{S--T}}$ values, but, unlike in the CC($P$)
calculations of the $^{1}A_{g}(D_{\rm 2h})$ and $^{3}B_{1g}(D_{\rm 2h})$ potentials, the singlet--triplet
gaps obtained with the uncorrected CC($P$) approach using relatively small CIPSI diagonalization spaces
and tiny fractions of all triples in the associated $P$ spaces can be quite accurate in their own right,
reproducing the CCSDT values of $\Delta E_{\text{S--T}}$ across the entire $\lambda = 0$--1 region
to within $\sim$1--2 kcal/mol when $N_{\text{det(in)}} \gtrsim 250,000$
and the cc-pVDZ basis set is employed.
Clearly, this is a lot better than
the 5.261, 8.589, and 15.120 kcal/mol errors relative to CCSDT obtained
at $\lambda = 0.6$, 0.8, and 1, respectively, with CCSD
using the same basis,
demonstrating that the relaxation of $T_{1}$ and $T_{2}$ amplitudes in the presence
of the leading $T_{3}$ contributions in the CIPSI-driven CC($P$) computations using sufficiently large
$N_{\text{det(in)}}$ values results in a more balanced description of the many-electron correlation effects
in the $^{1}A_{g}(D_{\rm 2h})$ and $^{3}B_{1g}(D_{\rm 2h})$ states, especially when $\lambda$ approaches 1,
although, by inspecting Tables \ref{tab:table1}--\ref{tab:table3}
and Figures \ref{fig:figure2}--\ref{fig:figure4},
we can also see that much of
the improvement in the CCSD $\Delta E_{\text{S--T}}$ data offered by the CC($P$) approach
originates from error cancellations between the $^{1}A_{g}(D_{\rm 2h})$ and $^{3}B_{1g}(D_{\rm 2h})$
CC($P$) energies. In the case of the CIPSI-driven CC($P$;$Q$) computations, we do
not have to count on error cancellations, since both the total electronic energies of the lowest singlet
and triplet states of cyclobutadiene at various values of $\lambda$ and the gaps between them rapidly
converge toward their CCSDT parents with $N_{\text{det(in)}}$. The $\delta(\mbox{$P$;$Q$})$ corrections,
in addition to being highly effective in improving the CC($P$) energies of the $^{1}A_{g}(D_{\rm 2h})$ and
$^{3}B_{1g}(D_{\rm 2h})$ states and the associated $\Delta E_{\text{S--T}}$ values, are also very helpful in
curing the nonsystematic error patterns in the singlet--triplet gaps observed in the CC($P$) calculations
in Table \ref{tab:table3}
and Figure \ref{fig:figure4}
as $N_{\text{det(in)}}$ increases, where the differences
between the CC($P$) and CCSDT values of $\Delta E_{\text{S--T}}$ go up and down or oscillate. This does
not happen in the CC($P$;$Q$) calculations, where the resulting singlet--triplet gaps approach their
CCSDT parents systematically and very fast as $N_{\text{det(in)}}$ is made larger, independent of
$\lambda$. This is yet another demonstration of the ability of the CIPSI-driven CC($P$;$Q$) approach
to provide a highly accurate and well-balanced description of the
many-electron correlation effects characterizing the lowest-energy singlet and triplet states of
cyclobutadiene along its automerization coordinate, which the conventional CC methods, such as CCSD,
CR-CC(2,3), and other noniterative triples corrections to CCSD, cannot provide as one approaches the
barrier region.

\begin{figure}[!ht]
\centering
\includegraphics[scale=0.5]{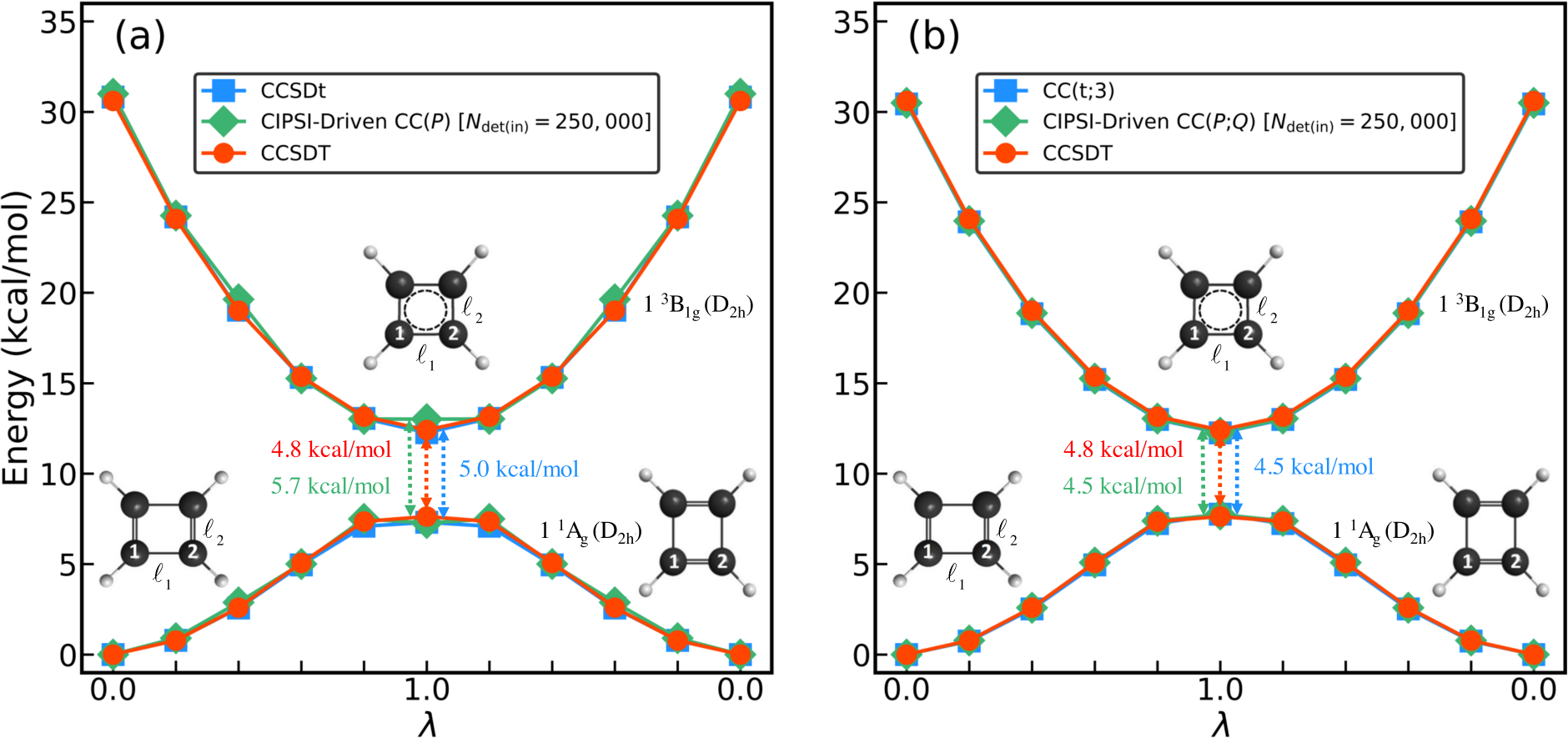}
\caption{
The PECs (in kcal/mol) characterizing the lowest-energy singlet and triplet states of cyclobutadiene,
as described by the cc-pVDZ basis set, along the $D_{\rm 2h}$-symmetric automerization pathway, defined
using the interpolation formula given by Eq. (\ref{eq:ell}) and parameterized by dimensionless variable $\lambda$. 
Panel (a) presents the active-orbital-based CCSDt (blue solid squares and lines), CIPSI-driven CC($P$)
(green solid diamonds and lines), and parent CCSDT data (red solid circles and lines) and panel (b)
shows the corresponding CC(t;3) (blue solid square and lines) and CIPSI-driven CC($P$;$Q$) (green solid
diamonds and lines) results, in addition to the full CCSDT energetics (red solid circles and lines).
The active space defining the subsets of triply excited determinants included in the CCSDt and CC(t;3)
calculations consisted of two orbitals of cyclobutadiene that correlate with the valence $e_g$
shell of the $D_{4\text{h}}$-symmetric TS ($\lambda= 1$) structure, whereas the lists of triples entering
the $P$ spaces employed in the CIPSI-based CC($P$) and CC($P$;$Q$) computations were extracted from the terminal
wave functions $|\Psi^{(\text{CIPSI})}\rangle$ obtained with $N_{\text{det(in)}}=250,000$. For each of
the methods in panels (a) and (b), the energy of the singlet ground state at the reactant (R, $\lambda = 0$)
geometry is set to 0. The numbers in the middle of each panel, colored in the same way as the corresponding PECs,
are the unsigned values of the singlet--triplet gaps determined at the $\lambda= 1$ TS structure.
}
\label{fig:figure6}
\end{figure}

Given our generally positive experiences with the active-orbital-based variant of the
CC($P$;$Q$) methodology abbreviated as CC(t;3),\cite{jspp-chemphys2012,jspp-jcp2012,jspp-jctc2012,%
nbjspp-molphys2017,ccpq-be2-jpca-2018,ccpq-mg2-mp-2019,adaptive-active-ccpq-2025} which corrects the CCSDt
energetics for those $T_{3}$ correlations that are not captured by the active-space CCSDt approach, it is
interesting to compare the lowest $^{1}A_{g}(D_{\rm 2h})$ and $^{3}B_{1g}(D_{\rm 2h})$ potentials of cyclobutadiene
and the gap between them obtained in the CIPSI-driven CC($P$) and CC($P$;$Q$) calculations with their CCSDt
and CC(t;3) counterparts. This is done in Figure \ref{fig:figure6}, which compares the PECs corresponding to
the lowest singlet and triplet states of cyclobutadiene, as described by the cc-pVDZ basis set, along the
$D_{\rm 2h}$-symmetric automerization pathway defined by Eq. (\ref{eq:ell}) resulting from the CIPSI-based
CC($P$) and CC($P$;$Q$) calculations employing $N_{\text{det(in)}} = 250,000$ with the analogous potentials
generated with the active-orbital-based CCSDt and CC(t;3) methods and full CCSDT. The numerical data used to
construct the CCSDt and CC(t;3) PECs shown in Figure \ref{fig:figure6} and to determine the gap between them
can be found in Tables S1--S3 of the Supporting Information.
We recall that in the language of the CC($P$) and CC($P$;$Q$) formalisms, the $P$ space
adopted in the CCSDt iterations consists of all singly and doubly excited determinants and the subset of triply
excited determinants that fit the formula $|\Phi_{ij{\bf K}}^{{\bf A}bc}\rangle$, where $i,j$ ($b,c$) designate
the spinorbitals occupied (unoccupied) in the reference determinant $|\Phi\rangle$ and ${\bf K}$ (${\bf A}$)
are the occupied (unoccupied) spinorbitals around the Fermi level belonging to the user-specified active set,
\cite{semi0b,semi1,semi2,ccsdtq3,semih2o,ghose,semi3c,semi4,semi4new,piecuch-qtp}
and the complementary $Q$ space needed to determine the CC(t;3) correction to CCSDt using Eq. (\ref{ccpq-6}) is
spanned by the remaining triply excited determinants $|\Phi_{ijk}^{abc}\rangle$ outside the
$|\Phi_{ij{\bf K}}^{{\bf A}bc}\rangle$ set. In the specific case of the CCSDt and CC(t;3) computations reported
in Figure \ref{fig:figure6} and Tables S1--S3 of the Supporting Information, the active space defining the subsets
of triply excited determinants included in the CCSDt calculations preceding the determination of the noniterative
CC(t;3) corrections consisted of two orbitals of cyclobutadiene that correlate with the valence $e_g$ shell of the
$D_{4\text{h}}$-symmetric TS ($\lambda= 1$) structure, meaning the highest occupied and lowest unoccupied RHF orbitals
for the lowest $^{1}A_{g}(D_{\rm 2h})$ state and the two singly occupied ROHF orbitals in the case of the lowest
$^{3}B_{1g}(D_{\rm 2h})$ state. As explained in the Supporting Information (see footnotes `b' in Tables S1 and S2),
with these choices of active orbitals, the $P$ space used in the CCSDt computations for the lowest singlet state
contained 1.5\% of the $S_z=0$ triples of the $A_{g}(D_{\rm 2h})$ symmetry involved in the parent full CCSDT work,
whereas that for the lowest triplet state contained 1.1\% of all $S_z=1$ $B_{1g}(D_{\rm 2h})$-symmetric triples. These
percentages should be compared to 0.5--0.6\% and 0.6--0.9\% of triples of the $S_z=0$ $A_{g}(D_{\rm 2h})$ and
$S_z=1$ $B_{1g}(D_{\rm 2h})$ symmetries, respectively, captured by the CIPSI runs preceding the CC($P$)
and CC($P$;$Q$) calculations with $N_{\text{det(in)}} = 250,000$, which, as shown in Figure \ref{fig:figure6}
({\it cf.}, also, Tables S1--S3 of the Supporting Information and Tables \ref{tab:table1}--\ref{tab:table3}), produce the
results that are in generally very good agreement with their CCSDt and CC(t;3) counterparts. One might argue that in spite
of having a somewhat larger fraction of triply excited determinants in the $P$ spaces employed in the CCSDt calculations
compared to the numbers of triples identified by the $N_{\text{det(in)}} = 250,000$ CIPSI runs, the CC(t;3) results
for the lowest-energy $^{3}B_{1g}(D_{\rm 2h})$ state reported in Table S2 of the Supporting Information are less
accurate than those obtained with the CIPSI-based CC($P$;$Q$) approach using $N_{\text{det(in)}} = 250,000$ (or
even smaller $N_{\text{det(in)}}$ values) shown in Table \ref{tab:table2}. One might also argue that with the
exception of the TS region, the usage of a larger fraction of triples in the $P$ spaces employed in the CCSDt
computations for the lowest $^{1}A_{g}(D_{\rm 2h})$ state compared to the numbers of the triples identified by
CIPSI using $N_{\text{det(in)}} = 250,000$ does not translate into substantial improvements in the CIPSI-driven
CC($P$;$Q$) results based on this $N_{\text{det(in)}}$ value by CC(t;3) ({\it cf.} Table S1 in the Supporting Information
and Table \ref{tab:table1}). None of this, however, alters our conclusion regarding the generally good agreement between
the CIPSI-driven CC($P$) and CC($P$;$Q$) calculations for the lowest singlet and triplet potentials of cyclobutadiene,
as described by the cc-pVDZ basis, and the gap between them obtained with $N_{\text{det(in)}} = 250,000$ and their
CCSDt and CC(t;3) counterparts, and none of this is surprising. Indeed, our choice of active orbitals in the CCSDt and
CC(t;3) calculations discussed here reflects on the multi-configurational character of the singlet ground state in the
vicinity of the $\lambda = 1$ TS region, but is not necessarily best for the lowest triplet state, which is
dominated by dynamical correlations at all values of $\lambda$, or the singlet
ground state as $\lambda \rightarrow 0$, where dynamical correlations dominate as well. Furthermore, unlike in
the CCSDt case, the lists of triply excited determinants included in the $P$ spaces adopted in the
CIPSI-driven CC($P$) calculations preceding the determination of the CC($P$;$Q$) corrections
may vary with the nuclear geometry (adjusting to the wave function content as the nuclear geometry changes), so
they tend to be more compact than those employed by CCSDt if the
active-orbital-based CC(t;3) and CIPSI-based CC($P$;$Q$) computations become similarly accurate [{\it cf.}
Ref.\ \onlinecite{adaptive-active-ccpq-2025} for the analogous
comments regarding the adaptive CC($P$;$Q$) framework {\it vs.} CC(t;3)].
Having said all this, the similarity between the PECs characterizing the lowest singlet and triplet states of
cyclobutadiene, as described by the cc-pVDZ basis, along its automerization pathway resulting from the
CIPSI-driven CC($P$;$Q$) calculations using $N_{\text{det(in)}} = 250,000$ and their counterparts obtained with
the CC(t;3) approach using a chemically motivated active space consisting of two valence orbitals, combined with the
observation that these two independent computations accurately approximate the parent CCSDT data, is reassuring.
It demonstrates that while there may be some differences between the subsets of triply excited determinants
identified with the help of active orbitals following the CCSDt recipe and those extracted from
the CIPSI runs, these differences are relatively small if the terminal diagonalization space used by the
CIPSI approach that drives the CC($P$) and CC($P$;$Q$) computations is sufficiently
large and the active space used to set up the CCSDt calculations preceding the determination of
the CC(t;3) corrections is reasonable.
In other words, both CCSDt and CIPSI are capable of capturing the leading triples for inclusion in the
$P$ spaces used by the CC($P$) and CC($P$;$Q$) calculations, but CIPSI allows us to do it in a more
black-box fashion, without having to resort to user-defined active orbitals, which is certainly appealing.
This remark is in line with one of
our earlier studies, reported in Ref.\ \onlinecite{stochastic-ccpq-molphys-2020}, where we compared the
manifolds of triply excited determinants captured in the context of the
semi-stochastic CC($P$;$Q$) considerations by CIQMC with those defined by the
$|\Phi_{ij{\bf K}}^{{\bf A}bc}\rangle$ formula of CCSDt.

With the exception of specific errors relative to CCSDT at various values of $N_{\text{det(in)}}$,
much of the above discussion applies to basis sets larger than cc-pVDZ. In fact, for larger basis sets,
the benefits of using the CIPSI-driven CC($P$;$Q$) approach to obtain the near-CCSDT energetics at small
fractions of the computational costs are expected to be even greater than those observed for cc-pVDZ since
one can continue using relatively small CIPSI diagonalization spaces to determine the subsets of triples
entering the CC($P$) computations, whereas the manifolds of all triply excited determinants and
amplitudes used by full CCSDT, which for a given number of electrons scale as cube of the number of
unoccupied orbitals, grow with the size of the one-electron basis very fast. This is illustrated in Tables
\ref{tab:table4}--\ref{tab:table6}, where we report the CIPSI-driven CC($P$) and CC($P$;$Q$)
calculations using the cc-pVTZ basis set, along with the associated variational ($E_{\text{var}}$) and
perturbatively corrected ($E_{\text{var}}+\Delta E^{(2)}$ and $E_{\text{var}} + \Delta E^{(2)}_r$)
CIPSI energies, for the lowest $^{1}A_{g}(D_{\rm 2h})$ and $^{3}B_{1g}(D_{\rm 2h})$ states of
cyclobutadiene and the gaps between them at the key R ($\lambda = 0$) and TS ($\lambda = 1$) geometries.
In this case, the CC($P$;$Q$) computations result in small $\sim$0.4--0.6 millihartree, $\sim$0.1 millihartree,
and $\sim$0.1--0.3 kcal/mol errors relative to CCSDT for the lowest singlet state, lowest triplet state,
and $\Delta E_{\text{S--T}}$, respectively, when $N_{\text{det(in)}} = 1,000,000$, {\it i.e.}, when the
$N_{\text{det(in)}}$ value is only 4 times larger than that leading to similar accuracies in the CC($P$;$Q$)/cc-pVDZ
calculations, but the total numbers of the $A_{g}(D_{\rm 2h})$-symmetric $S_z=0$ and $B_{1g}(D_{\rm 2h})$-symmetric
$S_z=1$ triply excited amplitudes used by the parent CCSDT/cc-pVTZ approach, which are 260,030,720
and 258,073,116, respectively, exceed those employed by its CCSDT/cc-pVDZ counterpart by a factor of 18.
As a result, the convergence of the CIPSI-driven CC($P$;$Q$) energetics toward CCSDT resulting from
the calculations employing the cc-pVTZ basis set reported in Tables \ref{tab:table4}--\ref{tab:table6} is impressive.
For example, as shown in Table \ref{tab:table4}, the CC($P$;$Q$)/cc-pVTZ calculation for the lowest-energy
singlet state at the challenging TS structure using the relatively small $N_{\text{det(in)}}$
value of 1,000,000, which relies on the CIPSI diagonalization space whose dimensionality [$N_{\text{det(out)}}$]
is a tiny 0.5\% of the $S_z=0$ triples of the $A_{g}(D_{\rm 2h})$ symmetry involved in the parent full
CCSDT/cc-pVTZ work, and which employs an even tinier $P$ space having only 0.1\% of all triples in it, reduces the
13.793 millihartree error relative to CCSDT obtained with CR-CC(2,3) to 0.591 millihartree.
Given that, analogous to the cc-pVDZ case, the CC($P$;$Q$) approach using the cc-pVTZ basis offers
a virtually perfect description of the lowest-energy triplet state, with the error relative to CCSDT
obtained with $N_{\text{det(in)}} = 1,000,000$ at $\lambda = 1$ of only 0.138 millihartree
(even though the dimensionality
of the associated CIPSI space is a tiny 0.6\% of the $B_{1g}(D_{\rm 2h})$-symmetric $S_z=1$ triply excited amplitudes
used by CCSDT and the underlying $P$ space contains only 0.2\% of all triples),
the observed excellent performance of the CIPSI-driven CC($P$;$Q$) methodology
for the singlet state translates into a highly accurate description of the singlet--triplet gap by
CC($P$;$Q$)/cc-pVTZ, which reduces the 8.685 kcal/mol error obtained at $\lambda = 1$ with CR-CC(2,3) to
0.285 kcal/mol when $N_{\text{det(in)}}$ is set to 1,000,000.
As one might anticipate in light of the previously discussed CC($P$;$Q$) calculations
using a smaller cc-pVDZ basis, the errors for the ``easier'' $\lambda = 0$ geometry resulting from the CIPSI-driven
CC($P$;$Q$)/cc-pVTZ calculations are generally smaller than their $\lambda = 1$ counterparts and
the results improve with increasing $N_{\text{det(in)}}$, but it is most encouraging to observe that
the CC($P$;$Q$)/cc-pVTZ approach is capable of providing an excellent description of the lowest singlet and
triplet states of cyclobutadiene and the gap between them at both the R and TS geometries, with errors
relative to CCSDT/cc-pVTZ on the order of small fractions of a millihartree or kilocalorie per mole, based on
the relatively small CIPSI diagonalization spaces, such as those corresponding to $N_{\text{det(in)}} = 1,000,000$,
which are only a few times larger than those employed in the similarly well converged CC($P$;$Q$) calculations
using cc-pVDZ. In analogy to the cc-pVDZ basis set,
it is reassuring to observe the remarkable efficiency of the $\delta(\mbox{$P$;$Q$})$ correction defined by
Eq. (\ref{ccpq-6}) in reducing errors obtained with the uncorrected CC($P$) approach at all values
of $N_{\text{det(in)}}$ included in Tables \ref{tab:table4}--\ref{tab:table6} when the larger cc-pVTZ basis
set is employed. Consistency of our observations regarding performance of the CIPSI-driven CC($P$) and CC($P$;$Q$)
methods for the lowest singlet and triplet states of cyclobutadiene and the gap between them in calculations
using a smaller cc-pVDZ and larger cc-pVTZ basis sets is reassuring too.

We conclude this section by noticing that the results reported in Tables \ref{tab:table1}--\ref{tab:table3}
for the cc-pVDZ basis and \ref{tab:table4}--\ref{tab:table6} for cc-pVTZ
also show that the convergence of the CIPSI-driven CC($P$) and CC($P$;$Q$) energies of the lowest
$^{1}A_{g}(D_{\rm 2h})$ and $^{3}B_{1g}(D_{\rm 2h})$ states of cyclobutadiene and the gap between them
toward their respective CCSDT parents with $N_{\text{det(in)}}$ or $N_{\text{det(out)}}$ is faster
than that characterizing the associated variational and perturbatively corrected CIPSI energetics
toward the extrapolated $E_{\text{var}} + \Delta E_{\text{r}}^{(2)}$ values
(this is particularly true for the larger cc-pVTZ basis set, where we would have to consider
much larger $N_{\text{det(in)}}$ values and diagonalization spaces in CIPSI than those used in this study to
obtain more accurate estimates of the extrapolated $E_{\text{var}} + \Delta E_{\text{r}}^{(2)}$ energies).
This observation is
consistent with our initial study announcing the CIPSI-based CC($P$) and CC($P$;$Q$) methodologies
\cite{cipsi-ccpq-2021} and the fact that the CC($P$;$Q$) calculations are capable of accurately
approximating the parent CCSDT energetics out of the unconverged CIPSI runs using relatively small Hamiltonian
diagonalization spaces, even when $T_{3}$ correlations are large and nonperturbative and electronic
quasi-degeneracies become substantial, as is the case in the barrier region of the ground-state
${}^1A_{g}(D_{2\text{h}})$ potential. While this may not be a general remark, we also observe that
perturbatively corrected $E_{\text{var}} + \Delta E^{(2)}$ and $E_{\text{var}} + \Delta E^{(2)}_r$ energies
of the lowest singlet and triplet states of cyclobutadiene and the gap between them converge toward
their extrapolated limits at a rate similar to that characterizing our uncorrected CC($P$) calculations
toward CCSDT. The $\delta(\mbox{$P$;$Q$})$-corrected CC($P$;$Q$) energetics converge to CCSDT much
faster. This might be yet another way of looking at the effectiveness of the $\delta(\mbox{$P$;$Q$})$
moment corrections in improving the underlying CC($P$) results. That being said, we should keep in mind
that the algorithms used to obtain the CCSDT and the perturbatively corrected and extrapolated
CIPSI energies are fundamentally different procedures. Furthermore, and
more importantly given the objectives of this study, where
we are interested in exploring the CIPSI-driven CC($P$) and CC($P$;$Q$) methodologies, not the CIPSI
approach itself, the CIPSI wave function growth in the calculations reported in Tables
\ref{tab:table1}--\ref{tab:table6} (especially in Tables \ref{tab:table4}--\ref{tab:table6} for the cc-pVTZ basis)
was terminated long before our CIPSI runs were well converged, as we only
needed information about the leading triply excited determinants and were not interested in saturating
the triply excited manifolds of the relevant many-electron Hilbert spaces. Last but not least, to highlight
the robustness of our CC($P$;$Q$) framework, all of the calculations reported in this work relied on the
RHF and ROHF orbitals, {\it i.e.}, we made no attempt to further optimize orbitals to make them consistent with
correlated computations, which would improve
CIPSI's performance and which might also help our CC($P$) and CC($P$;$Q$)
results using smaller $N_{\text{det(in)}}$ values, especially in the vicinity of the square TS geometry.
We intend to look into potential benefits that
might be offered by orbital optimizations in the CIPSI-driven CC($P$) and CC($P$;$Q$) calculations in
a future study.

%
\section{Summary and Concluding Remarks}
\label{sec4}

An accurate determination of singlet--triplet gaps in biradicals represents a formidable test for
\emph{ab initio} electronic structure methodologies, as it requires balancing strong nondynamical
many-electron correlation effects, needed for a reliable description of the low-spin singlet states
that have a manifestly multiconfigurational nature, with the generally weaker, largely dynamical,
correlations characterizing the high-spin triplet states. Although high-level CC methods with a full
treatment of higher--than--two-body clusters, such as CCSDT or CCSDTQ, are often powerful enough to
capture the dynamical and nondynamical correlation effects relevant in such studies, their applications
are hindered by the demanding computational steps and memory requirements, which are prohibitively
expensive when larger many-electron systems are examined. One of the promising ideas aimed at
addressing this situation within the single-reference CC framework is the CC($P$;$Q$) formalism,
in which one solves the CC amplitude equations in a suitably defined subspace of the many-electron
Hilbert space, referred to as the $P$ space, and improves the resulting CC($P$) energies using the
{\it a posteriori} moment corrections, designated as $\delta(P;Q)$, calculated with the help of
the complementary $Q$ space.\cite{jspp-chemphys2012,jspp-jcp2012,jspp-jctc2012,nbjspp-molphys2017,%
ccpq-be2-jpca-2018,ccpq-mg2-mp-2019,stochastic-ccpq-prl-2017,stochastic-ccpq-molphys-2020,%
stochastic-ccpq-jcp-2021,arnab-stgap-2022,cipsi-ccpq-2021,adaptiveccpq2023,adaptive-active-ccpq-2025}
Among the most attractive
features of the CC($P$;$Q$) methodology is its flexibility, so that in addition to conventional choices
of the $P$ and $Q$ spaces using truncations based on excitation ranks, which in the past resulted
in the development of the biorthogonal CR-CC methods, such as the CR-CC(2,3) triples correction to CCSD,
\cite{crccl_jcp,crccl_cpl,crccl_molphys,crccl_jpc}
one can consider various unconventional ways of setting up these spaces that can improve the
CR-CC(2,3), CCSD(T), and similar energetics for
systems with substantial electronic quasi-degeneracies
by relaxing the $T_{1}$ and $T_{2}$ components of the cluster operator $T$
in the presence of their higher-rank $T_{n}$ counterparts with $n > 2$, such as $T_{3}$, which
become large, nonperturbative, and strongly coupled to $T_{1}$ and $T_{2}$ in such situations. This
can be done without major increases in the computational effort by incorporating the leading
higher--than--doubly excited determinants in the $P$ spaces
and using corrections $\delta(P;Q)$ to capture the remaining correlations of interest.

In this work, we have examined the hybrid variant of the CC($P$;$Q$) methodology introduced
in Ref.\ \onlinecite{cipsi-ccpq-2021}, in which the leading higher--than--doubly excited determinants
in the $P$ space are identified, in an automated fashion, using the sequences of Hamiltonian
diagonalizations generated with the CIPSI algorithm.\cite{sci_3,cipsi_1,cipsi_2} In order to thoroughly
test the CIPSI-driven CC($P$;$Q$) formalism and obtain useful insights into its performance, we have
focused on recovering the lowest-energy singlet and triplet potentials of cyclobutadiene along its
automerization coordinate and the gap between them resulting from the full CCSDT computations which,
based on comparisons with the perturbatively corrected and extrapolated CIPSI and DEA-EOMCC(4p-2h)-level
data, obtained in this study as well, provide reliable information. To do so, we have constructed an
approximate, $D_{\rm 2h}$-symmetric, one-dimensional automerization pathway connecting the rectangular
reactant and product species via the square TS structure on the ground-state singlet potential using
the information taken from Ref.\ \onlinecite{MR-AQCC} and performed a large number of CC($P$;$Q$)
calculations for the lowest $^{1}A_{g}(D_{\rm 2h})$ and $^{3}B_{1g}(D_{\rm 2h})$ states of cyclobutadiene
at the selected nuclear geometries along the resulting path, where for each state and for each geometry,
we have explored a wide range of values of the CIPSI wave function termination parameter
$N_{\text{det(in)}}$ that controls the Hamiltonian diagonalization sequences preceding the CC($P$) and
CC($P$;$Q$) runs. We have demonstrated that the CIPSI-driven CC($P$;$Q$) calculations are capable of
accurately approximating the high-level CCSDT energetics of the lowest singlet and triplet states of
cyclobutadiene across the entire automerization pathway, to within small fractions of a millihartree
for total energies and
0.1--0.3
kcal/mol for the singlet--triplet gaps, using tiny fractions
of the triply excited determinants, on the order of 1\% of all triples
for the cc-pVDZ basis set and 0.1--0.2\% for cc-pVTZ,
in the underlying $P$ spaces
extracted from the relatively inexpensive CIPSI diagonalizations in spaces that are orders of magnitude
smaller than the numbers of cluster amplitudes used by CCSDT. This extraordinary performance of
the CIPSI-driven CC($P$;$Q$) approach applies to both the less demanding reactant/product region,
where the lowest singlet and triplet states of cyclobutadiene are largely single-configurational
and the absolute values of the singlet--triplet gap exceed 30 kcal/mol, and the vicinity of the TS
structure on the ground-state singlet potential, where the high-spin triplet state retains its weakly
correlated, single-determinantal, nature, but the singlet state, separated from its triplet counterpart
by only
a few kcal/mol,
becomes multiconfigurational, strongly correlated, and characterized by large
and highly nonperturbative $T_{3}$ correlations, which are strongly coupled to the one- and two-body
components of $T$ and which result in failure of CR-CC(2,3) and other noniterative triples corrections
to CCSD. Interestingly, the uncorrected CC($P$) computations using similarly compact excitation spaces
can be accurate as well, reproducing the CCSDT values of the
singlet--triplet gap across the entire automerization pathway to within $\sim$1--2 kcal/mol, improving
the poor CCSD and CR-CC(2,3) results in the vicinity of the TS geometry, and reaffirming the usefulness of
incorporating the leading triply excited determinants into the underlying $P$ spaces, but the total energies
of the lowest  $^{1}A_{g}(D_{\rm 2h})$ and $^{3}B_{1g}(D_{\rm 2h})$ states of cyclobutadiene obtained
with the CC($P$) approach converge to their CCSDT parents with $N_{\text{det(in)}}$ very slowly, so one
has to rely on error cancellations to obtain accurate singlet--triplet gaps with CC($P$).
The $\delta(\mbox{$P$;$Q$})$ corrections are very helpful in this regard. They reduce errors
in the total CC($P$) energies of the $^{1}A_{g}(D_{\rm 2h})$ and $^{3}B_{1g}(D_{\rm 2h})$ states of
cyclobutadiene by orders of magnitude while substantially improving the resulting singlet--triplet
gaps, making their convergence toward CCSDT smoother and more systematic. Given the relatively
low costs of determining the $\delta(\mbox{$P$;$Q$})$ corrections compared to the preceding CIPSI and
CC($P$) steps and the enormous benefits resulting from their application in the CIPSI-driven CC($P$;$Q$)
calculations, we recommend using CC($P$;$Q$).

The excellent performance of the CIPSI-driven CC($P$;$Q$) approach in converging the lowest-energy singlet
and triplet potentials of cyclobutadiene obtained with CCSDT, observed in this study, along with the
promising initial results reported in Ref.\ \onlinecite{cipsi-ccpq-2021}, motivate us to pursue the hybrid
CC methodologies combining the CC($P$;$Q$) framework with selected CI even further. We will, for example,
investigate how much the CC($P$;$Q$) singlet and triplet potentials reported in this work, especially the
ground-state singlet potential in the vicinity of the TS geometry, can benefit from replacing the RHF and
ROHF orbitals exploited in the calculations reported in this work by the suitably optimized orbitals
consistent with the CC($P$) or CIPSI wave functions. We will also
examine how much each of the singlet and triplet potentials of cyclobutadiene obtained with the CIPSI-driven
CC($P$;$Q$) approach using a given value of $N_{\text{det(in)}}$, especially its smoothness, can
improve by consolidating the $P$ spaces corresponding to the different geometries along the automerization
path or, to be more precise in the context of the CC($P$;$Q$) calculations aimed at recovering the CCSDT
energetics performed in this study, by merging the triple excitation manifolds incorporated in those spaces.
Among other topics worth exploring, it will be interesting to examine if anything substantial can be gained
by replacing the CIPSI algorithm in our CC($P$;$Q$) considerations by other selected CI techniques, such as
heat-bath CI,\cite{shci_1,shci_2,shci_3} adaptive CI,\cite{adaptive_ci_1,adaptive_ci_2} or adaptive sampling
CI.\cite{asci_1,asci_2} Last but not least, following the strategy adopted in our previous work on the
semi-stochastic, CIQMC-driven, CC($P$;$Q$) approaches,
\cite{stochastic-ccpq-prl-2017,stochastic-ccpq-molphys-2020,stochastic-ccpq-jcp-2021} we are planning to
extend the CIPSI-driven CC($P$;$Q$) methodology investigated in this study and Ref.\ \onlinecite{cipsi-ccpq-2021},
to higher CC levels, especially CCSDTQ, and excited electronic states, with an initial focus on converging the
EOMCCSDT\cite{eomccsdt1,eomccsdt2,eomccsdt3} energetics, while seeking additional savings in the computational
effort by replacing the unconstrained CIPSI algorithm, which is allowed to explore the entire many-electron Hilbert
space, by its truncated analogs consistent with the determinantal spaces needed in the target CC calculations
({\it e.g.}, the CISDT or CISDTQ analogs of CIPSI when attempting to use the CIPSI-driven CC($P$;$Q$) framework
to converge the CCSDT or CCSDTQ energetics).

\section*{Appendix: Key Elements of the Algorithm Used to Implement the
CC($\bm{P}$) Amplitude Equations, Along with Illustrative Timings}

\renewcommand{\theequation}{A.\arabic{equation}}
\setcounter{equation}{0}
As pointed out in Section \ref{sec2}, the key to achieving computational efficiency in the
CC($P$;$Q$) calculations, including the CIPSI-driven CC($P$;$Q$) approach aimed at converging the
CCSDT energetics examined in this study, lies in the development of an algorithm capable of offering
substantial speedups compared to the parent CC method when the lists of higher--than--doubly excited
determinants included in the $P$ spaces used in the CC($P$) iterations do not necessarily form
continuous manifolds. The most essential ingredients of our strategy for implementing
the CC($P$) amplitude equations, Eq. (\ref{ccpq-2}), and the companion left-eigenstate system,
Eq. (\ref{ccpq-10}), in which the $P$ space used to define the cluster operator $T^{(P)}$ and its
deexcitation $\Lambda^{(P)}$ counterpart consists of all singly and doubly excited determinants,
$|\Phi_{i}^{a} \rangle$ and $|\Phi_{ij}^{ab} \rangle$, respectively, and a potentially spotty subset
of triply excited determinants $|\Phi_{ijk}^{abc} \rangle$, identified in this work by CIPSI, are
summarized in this appendix. In the interest of space, in the description below, we focus on the CC($P$)
amplitude equations that are used to determine the cluster operator
\begin{equation}
T^{(P)}=T_{1}+T_{2}+T_{3}^{(P)} ,
\label{A:TP}
\end{equation}
where, consistent with the above definition of the $P$ space, designated as $\mathscr{H}^{(P)}$,
the one- and two-body components of $T^{(P)}$,
\begin{equation}
T_{1} = \sum_{i,a} t_{a}^{i} E_{i}^{a}
\label{A:T1}
\end{equation}
and
\begin{equation}
T_{2} = \sum_{i<j,a<b} t_{ab}^{ij} E_{ij}^{ab} ,
\label{A:T2}
\end{equation}
respectively, are treated fully, but the three-body component
\begin{equation}
T_{3}^{(P)} = \sum_{|\Phi_{ijk}^{abc} \rangle \in \mathscr{H}^{(P)}} t_{abc}^{ijk} E_{ijk}^{abc}
\label{A:T3P}
\end{equation}
is defined on a subset of triply excited determinants that do not have to form a continuous manifold
(if the list of triples in $T_{3}^{(P)}$ is extracted from
the terminal CIPSI wave function $|\Psi^\text{(CIPSI)}\rangle$, $T_{3}^{(P)}$ becomes the
$T_3^\text{(CIPSI)}$ operator introduced in Section \ref{sec2}). Following the notation used
in Section \ref{sec2}, $E_{i}^{a}$, $E_{ij}^{ab}$, and $E_{ijk}^{abc}$ in Eqs.\
(\ref{A:T1})--(\ref{A:T3P}) are the elementary particle--hole excitation operators that generate the
$|\Phi_{i}^{a} \rangle$, $|\Phi_{ij}^{ab} \rangle$, and $|\Phi_{ijk}^{abc} \rangle$ determinants
when acting on the reference function $|\Phi\rangle$ and $t_{a}^{i}$, $t_{ab}^{ij}$, and
$t_{abc}^{ijk}$ are the cluster amplitudes defining $T_{1}$, $T_{2}$, and $T_{3}^{(P)}$,
respectively. Our implementation of the left-eigenstate
CC($P$) system, needed to obtain the deexcitation operator
$\Lambda^{(P)} = \Lambda_1 + \Lambda_2 + \Lambda_3^\text{(P)}$,
in which the triples entering $\Lambda_3^\text{(P)}$ are the same as those included in $T_{3}^{(P)}$,
uses the same philosophy as that adopted in handling the CC($P$) amplitude equations, so we are not
discussing it here. A more complete description of our algorithm used to efficiently handle the
right and left CC($P$) equations, Eqs. (\ref{ccpq-2}) and (\ref{ccpq-10}), assuming the above
definitions of $T^{(P)}$, $\Lambda^{(P)}$, and the underlying $P$ space, and of CCpy,
in which the CIPSI-driven, adaptive, and active-orbital-based CC($P$;$Q$)
approaches targeting CCSDT have been implemented, will be discussed in a separate publication.

To formulate our algorithm for the CC($P$) amplitude equations that leads to
the desired speedups compared to the parent CCSDT approach, we first isolate the contributions due to
the three-body component $T_{3}^{(P)}$ of the cluster operator $T^{(P)}$ by expanding
Eq.\ (\ref{ccpq-2}), in which $T^{(P)}$ is defined by Eq. (\ref{A:TP}), as follows:
\begin{equation}
\mathfrak{M}_{K}(2) + \langle \Phi_{K} | [\overline{H}^{(2)}, T_{3}^{(P)}]| \Phi \rangle = 0,\:\:
|\Phi_K\rangle \in \mathscr{H}^{(P)} .
\label{A:ampeqs2}
\end{equation}
The $|\Phi_K\rangle$s in Eq.\ (\ref{A:ampeqs2}) are the singly, doubly, and selected triply excited
determinants included in the $P$ space,
\begin{equation}
\overline{H}^{(2)} = e^{-T_{1}-T_{2}} H e^{T_{1}+T_{2}}
\label{A:Hbar2}
\end{equation}
is the Hamiltonian transformed with the $e^{T_{1}+T_{2}}$ part of $e^{T^{(P)}}$, and
\begin{equation}
\mathfrak{M}_K(2) = \langle \Phi_K | \overline{H}^{(2)} | \Phi \rangle
\label{A:mk2}
\end{equation}
are the quantities resembling the generalized moments of the CCSD equations, except that the
$T_{1}$ and $T_{2}$ amplitudes used to construct them originate from the CC($P$) iterations, {\it i.e.},
they are relaxed in the presence of the $T_{3}^{(P)}$ contribution to $T^{(P)}$.
It should be noted that with the definitions of
$\mathscr{H}^{(P)}$ and $T^{(P)}$ considered here, terms nonlinear in $T_{3}^{(P)}$ 
do not contribute to the CC($P$) amplitude equations.
Thus, after straightforward manipulations following the insertion of Eq.\ (\ref{A:T3P})
for $T_{3}^{(P)}$ into the commutator (or the equivalent connected product of $\overline{H}^{(2)}$
and $T_{3}^{(P)}$) appearing in Eq. (\ref{A:ampeqs2}), we obtain
\begin{widetext}
\begin{equation}
\underbrace{\mathfrak{M}_{K}(2)}_{\text{(I)}}
+ \underbrace{\sum_{|\Phi_{lmn}^{def} \rangle \in \mathscr{H}^{(P)}}
\langle \Phi_{K} | \overline{H}_{N}^{(2)} | \Phi_{lmn}^{def} \rangle t_{def}^{lmn}}_{\text{(II)}} = 0,
\:\: |\Phi_K\rangle \in \mathscr{H}^{(P)},
\label{A:ampeqs3}
\end{equation}
\end{widetext}
where $\overline{H}_{N}^{(2)} = \overline{H}^{(2)} - E^{(P)}\textbf{1}$ is the
$\overline{H}^{(2)}$ operator in the normal-product form with respect to the Fermi vacuum
$|\Phi\rangle$, with $E^{(P)}$ representing the ground-state CC($P$) energy, Eq. (\ref{ccpq-5}),
and \textbf{1} designating the unit operator (because of the absence of higher--than--two-body interactions
in the Hamiltonians used in quantum chemistry, the $T_{3}^{(P)}$ contribution to $T^{(P)}$ does not
enter the formula for $E^{(P)}$, {\it i.e.}, $E^{(P)} = \langle \Phi | \overline{H}^{(2)} | \Phi \rangle$).

Equation (\ref{A:ampeqs3}) represents the core equation underlying our CC($P$) algorithm. The
programmable expressions for the matrix elements defining the one- and two-body components of 
$\overline{H}_{N}^{(2)}$, denoted as $\bar{h}_{p}^{q}$ and $\bar{h}_{pq}^{rs}$, respectively, in
terms of the one- and two-body cluster amplitudes $t_{a}^{i}$ and $t_{ab}^{ij}$ and one- and two-electron
integrals in a molecular spin-orbital basis, designated as $f_{p}^{q} = \langle p | f | q \rangle$, where
$f$ is the Fock operator, and $v_{pq}^{rs} = \langle pq | v | rs \rangle - \langle pq | v | sr \rangle$,
where $v$ is the electron--electron interaction, which define the Hamiltonian in the
normal-ordered form, $H_{N} = H - \langle \Phi |H| \Phi \rangle$, in a usual manner,
are well known and can be found, for example, in Table I of Ref.\ \onlinecite{crccl_ijqc2}
(indices $p,q,r,s$ refer to generic -- meaning occupied as well as unoccupied -- molecular spin-orbitals).
In each iteration of the CC($P$) procedure, the $\bar{h}_{p}^{q}$ and $\bar{h}_{pq}^{rs}$
matrix elements are recalculated with the current values of $T_{1}$ and
$T_{2}$ and stored as computational intermediates, which allows us to evaluate
Eq. (\ref{A:ampeqs3}) in a factorized form. In our current implementation, the CC($P$) system 
is constructed as the sum of two distinct contributions, referred to in Eq. (\ref{A:ampeqs3})
as terms (I) and (II). Each of these two terms is computed in an efficient fashion using
a dedicated strategy tailored to its structure, which we discuss next.

\noindent
\emph{Evaluation of term (I) in Eq. (\ref{A:ampeqs3})}

As already alluded to above, the first term in Eq. (\ref{A:ampeqs3}) resembles the generalized moments of the CCSD
equations. In particular, when $|\Phi_K\rangle = |\Phi_{i}^{a}\rangle$ or $|\Phi_{ij}^{ab}\rangle$, the resulting
one- and two-body moments,
\begin{equation}
\mathfrak{M}_{a}^{i}(2) = \langle \Phi_{i}^{a} | \overline{H}^{(2)} | \Phi \rangle \equiv \bar{h}_{a}^{i}
\label{moment-ai}
\end{equation}
and
\begin{equation}
\mathfrak{M}_{ab}^{ij}(2) = \langle \Phi_{ij}^{ab} | \overline{H}^{(2)} | \Phi \rangle \equiv \bar{h}_{ab}^{ij} ,
\label{moment-abij}
\end{equation}
respectively, are equivalent to the left-hand sides of the standard CCSD amplitude equations, in which the singly and
doubly excited cluster amplitudes [that in CCSD originate from setting
$\mathfrak{M}_{a}^{i}(2)$ and $\mathfrak{M}_{ab}^{ij}(2)$
to zero] are obtained in the process of solving the CC($P$) equations, Eq. (\ref{A:ampeqs2}) or (\ref{A:ampeqs3}),
for the cluster operator $T^{(P)}$ defined by Eq. (\ref{A:TP}).
As is well established, the $\mathfrak{M}_{a}^{i}(2)$ and $\mathfrak{M}_{ab}^{ij}(2)$
expressions can be computed in a fully vectorized fashion ({\it i.e.}, avoiding the use of explicit loops) by taking advantage
of efficient matrix multiplication and transposition routines provided by BLAS. When the $P$ space $\mathscr{H}^{(P)}$ contains
all singly and doubly excited determinants, which is the case in the CIPSI-driven CC($P$;$Q$) calculations aimed at converging
the CCSDT energetics, such as those discussed in the present article, the evaluation of $\mathfrak{M}_{a}^{i}(2)$ and
$\mathfrak{M}_{ab}^{ij}(2)$ involves the usual $n_{o}^2 n_{u}^4$ and other, less expensive, computational steps
characterizing CCSD, where $n_{o}$ ($n_{u}$) is the number of correlated occupied (unoccupied) spin-orbitals in $|\Phi\rangle$.

With the exception of the source of $T_{1}$ and $T_{2}$,
which in the CC($P$) calculations originate from solving the system given by Eq. (\ref{A:ampeqs2}) or (\ref{A:ampeqs3}),
and besides the fact that the projections on the triply excited determinants $|\Phi_{ijk}^{abc}\rangle$ in
Eq. (\ref{A:mk2}) entering Eq. (\ref{A:ampeqs3}) are limited to the $|\Phi_{ijk}^{abc}\rangle$s included in
the $P$ space $\mathscr{H}^{(P)}$, the programmable expressions for the three-body moments
\begin{equation}
\mathfrak{M}_{abc}^{ijk}(2) = \langle \Phi_{ijk}^{abc} | \overline{H}^{(2)} | \Phi \rangle ,
\label{moment-abcijk}
\end{equation}
in terms of the singly and doubly cluster amplitudes and one- and two-body matrix elements of $\overline{H}_{N}^{(2)}$,
are identical to those exploited in methods such as CR-CC(2,3). Assuming the Einstein summation convention over
repeated upper and lower indices used in the remainder of this appendix, they are
[{\it cf.}, {\it e.g.}, Eqs. (59) and (62) in Ref.\ \onlinecite{crccl_ijqc2}]
\begin{equation}
\mathfrak{M}_{abc}^{ijk}(2) = \frac{1}{2} \mathscr{A}^{i/jk}\mathscr{A}_{abc} 
(\bar{h}_{ab}^{ie} t_{ec}^{jk} - I_{am}^{ij} t_{bc}^{mk}) ,
\label{A:moment3}
\end{equation}
where
$I_{am}^{ij} = \bar{h}_{am}^{ij} - \bar{h}_{m}^{e} t_{ae}^{ij}$ and
$\mathscr{A}_{p/qr} \equiv \mathscr{A}^{p/qr} = 1 - (pq) - (pr)$ and
$\mathscr{A}_{pqr} \equiv \mathscr{A}^{pqr} = 1 - (pq) - (pr) - (qr) + (pqr) + (prq)$
are index antisymmetrizers [with $(pq)$ representing a transposition of $p$ and $q$].
For each $|\Phi_{ijk}^{abc}\rangle \in \mathscr{H}^{(P)}$, Eq. (\ref{A:moment3}) is computed
by forming dot products of $\bar{h}_{ab}^{ie}$ with $t_{ec}^{jk}$ (summed over $e$) and
$I_{am}^{ij}$ with $t_{bc}^{mk}$ (summed over $m$), which can be efficiently done with fast matrix
multiplication routines from the BLAS library, and subtracting the latter from the former. Similar
remarks apply to the $\displaystyle (-\bar{h}_{m}^{e} t_{ae}^{ij})$ term in the definition of the $I_{am}^{ij}$
intermediate (which is a dot product involving summation over $e$) and the
fully vectorized expressions for the one- and two-body matrix elements of
$\overline{H}_{N}^{(2)}$ entering $I_{am}^{ij}$ and the right-hand side of Eq. (\ref{A:moment3}),
provided, for example, in Table I of Ref.\ \onlinecite{crccl_ijqc2}.
In the limit of the $P$ space including all triply excited determinants (as in, {\it e.g.}, full CCSDT),
evaluation of Eq. (\ref{A:moment3}) involves $n_{o}^3 n_{u}^4$ operations.
However, when the $P$ space contains only a small subset of triply excited determinants, as
is the case in our CIPSI-driven CC($P$;$Q$) calculations, the cost
of evaluating Eq. (\ref{A:moment3}) is reduced relative to CCSDT by a factor of $(D/d)$, where $D$
is the number of all triply excited determinants $|\Phi_{ijk}^{abc}\rangle$ and $d$ is the subset of those
$|\Phi_{ijk}^{abc}\rangle$s that are included in the $P$ space $\mathscr{H}^{(P)}$.

\noindent
\emph{Evaluation of term (II) in Eq. (\ref{A:ampeqs3})}

We now turn our attention to term (II) in Eq. (\ref{A:ampeqs3}), which captures the contributions to the CC($P$)
amplitude equations that are linear in $T_{3}^{(P)}$. A simple recipe for evaluating this term, which we have adopted
in our CC($P$) algorithm implemented in CCpy, consists of computing matrix elements of $\overline{H}_{N}^{(2)}$
in the singles--triples (ST) sector, $\langle \Phi_{i}^{a} | \overline{H}_{N}^{(2)} | \Phi_{lmn}^{def} \rangle$,
double--triples (DT) sector, $\langle \Phi_{ij}^{ab} | \overline{H}_{N}^{(2)} | \Phi_{lmn}^{def} \rangle$, and
triples--triples (TT) sector, $\langle \Phi_{ijk}^{abc} | \overline{H}_{N}^{(2)} | \Phi_{lmn}^{def} \rangle$, where
the triply excited determinants $|\Phi_{ijk}^{abc}\rangle$ and $|\Phi_{lmn}^{def}\rangle$ are limited to those
included in the $P$ space, and multiplying the resulting ST, DT, and TT blocks of the matrix representing
$\overline{H}_{N}^{(2)}$ with the vector of three-body cluster amplitudes corresponding to $|\Phi_{lmn}^{def}\rangle$s
belonging to $\mathscr{H}^{(P)}$ using a standard definition of the matrix--vector product. The programmable
expressions for the ST, DT, and TT blocks of the matrix representing $\overline{H}_{N}^{(2)}$ exploited in our
work, focusing on the contributions that can be written in terms of the one- and two-body
matrix elements $\bar{h}_{p}^{q}$ and $\bar{h}_{pq}^{rs}$ (which can be efficiently determined using the
vectorized expressions provided in Table I of Ref.\ \onlinecite{crccl_ijqc2}), {\it i.e.}, excluding
terms in the TT block that involve the three-body component of the $\overline{H}_{N}^{(2)}$ operator,
which, as further elaborated on below, may require a different treatment, are shown in Table \ref{A:table1}.

In our present implementation of Eq. (\ref{A:ampeqs3}), we evaluate term (II) using two nested loops, where the
outer loop runs over the bra states $\langle \Phi_{K}|$ that correspond to the projections
on all singly and doubly excited determinants, $|\Phi_{i}^{a}\rangle$ and $|\Phi_{ij}^{ab}\rangle$, respectively,
and the subset of triply excited determinants $|\Phi_{ijk}^{abc}\rangle$ included in the $P$ space, and
an inner loop enumerates the ket states associated with the triply excited determinants
$|\Phi_{lmn}^{def}\rangle \in \mathscr{H}^{(P)}$ matching the content of $T_{3}^{(P)}$. For all pairs of the
bra ($\langle \Phi_{K}|$) and ket ($|\Phi_{lmn}^{def}\rangle$) determinants specified within these two loops,
the corresponding matrix elements $\langle \Phi_{K} | \overline{H}_{N}^{(2)} | \Phi_{lmn}^{def}\rangle$ can be
evaluated using the formulas provided in Table \ref{A:table1}, which are subsequently multiplied by the triply
excited cluster amplitudes $t_{def}^{lmn}$ defining $T_{3}^{(P)}$ associated with $|\Phi_{lmn}^{def}\rangle \in
\mathscr{H}^{(P)}$. It should be noted though that in addition to the terms that originate from the one- and two-body
components of the $\overline{H}_{N}^{(2)}$ operator considered in Table \ref{A:table1}, the TT block of the
matrix representing $\overline{H}_{N}^{(2)}$ also contains contributions that engage its three-body component
arising from $\langle \Phi_{ijk}^{abc} | [[H_{N}, T_{2}], T_{3}^{(P)}] | \Phi \rangle$, which are not
accounted for in Table \ref{A:table1}. One could determine
these contributions within the two nested loops over $\langle \Phi_{K}|$ and $|\Phi_{lmn}^{def}\rangle$
discussed here (after augmenting Table \ref{A:table1} with the expressions due to the three-body component
of $\overline{H}_{N}^{(2)}$), but it is much more efficient, in terms of CPU time and storage costs,
to determine them during the previously described evaluation of moments $\mathfrak{M}_{abc}^{ijk}(2)$
that enter term (I) of Eq. (\ref{A:ampeqs3}). This can be accomplished by
rewriting the contributions to $\langle \Phi_{ijk}^{abc} | [[H_{N}, T_{2}], T_{3}^{(P)}] | \Phi \rangle$
that originate from the three-body component of $\overline{H}_{N}^{(2)}$, such that the Hamiltonian
is first contracted with $T_{3}^{(P)}$ rather than $T_{2}$, and combining the resulting expression
with the three-body moment $\mathfrak{M}_{abc}^{ijk}(2)$. In practice, all one has to do is to dress
$\bar{h}_{ab}^{ie}$ and $I_{am}^{ij}$ in Eq. (\ref{A:moment3}) with the suitably defined
$T_{3}^{(P)}$-dependent terms and compute
\begin{equation}
\tilde{{\mathfrak{M}}}_{abc}^{ijk}(2) = \frac{1}{2} \mathscr{A}^{i/jk}\mathscr{A}_{abc} 
(\tilde{h}_{ab}^{ie} t_{ec}^{jk} - \tilde{I}_{am}^{ij} t_{bc}^{mk}) ,
\label{A:moment3augmented}
\end{equation}
where $\tilde{h}_{ab}^{ie} = \bar{h}_{ab}^{ie} - \tfrac{1}{2}\bar{h}_{mn}^{ef} t_{abf}^{imn}
\equiv \bar{h}_{ab}^{ie} - \tfrac{1}{2 }v_{mn}^{ef} t_{abf}^{imn}$ and
$\tilde{I}_{am}^{ij} = I_{am}^{ij} + \tfrac{1}{2} \bar{h}_{mn}^{ef} t_{aef}^{ijn}
\equiv I_{am}^{ij} + \tfrac{1}{2} v_{mn}^{ef} t_{aef}^{ijn}$, instead of the original
moments $\mathfrak{M}_{abc}^{ijk}(2)$ given by Eq. (\ref{A:moment3}). In this way, we can generate the
contributions to the $\langle \Phi_{ijk}^{abc} | [[H_{N}, T_{2}], T_{3}^{(P)}] | \Phi \rangle$ term due to the
three-body component of $\overline{H}_{N}^{(2)}$, which would otherwise have to be incorporated in the TT block
of the matrix representing $\overline{H}_{N}^{(2)}$, outside the two loops over $\langle \Phi_{K}|$ and
$|\Phi_{lmn}^{def}\rangle$ from $\mathscr{H}^{(P)}$ used to treat the rest of term (II)
and without having to calculate and store the memory-demanding six-index
matrix elements associated with the second-quantized formula for the three-body component of the
$\overline{H}_{N}^{(2)}$ operator. To keep the costs of our computations as low as possible,
the matrix multiplications involving the summations over $m$, $n$, and $f$ in the
$(-\tfrac{1}{2}\bar{h}_{mn}^{ef} t_{abf}^{imn})$ contribution to $\tilde{h}_{ab}^{ie}$ and
$n$, $e$, and $f$ in the $\tfrac{1}{2} \bar{h}_{mn}^{ef} t_{aef}^{ijn}$ contribution to $\tilde{I}_{am}^{ij}$
are executed by looping over the triply excited amplitudes seen in these expressions, which span the
subset of triples included in $T_{3}^{(P)}$. The determination of the $T_{3}^{(P)}$ contributions to the
CC($P$) amplitude equations represented in Eq. (\ref{A:ampeqs3}) by term (II) via the matrix-vector products
involving the ST, DT, and TT blocks of the matrix representing $\overline{H}_{N}^{(2)}$
is reminiscent of the strategies adopted in selected CI codes. It
allows us to accommodate arbitrary or irregular lists of triply excited determinants included in the
$P$ space and defining the $T_{3}^{(P)}$ cluster operator, which may not form continuous excitation manifolds
labeled by occupied and unoccupied orbitals from the respective ranges of indices. As illustrated by the
computational timings provided at the end of this appendix, the benefits of implementing the CC($P$) approach
in this manner become enormous when the fraction of triples included in the $P$ space is very small, which
is the main point of all CC($P$;$Q$) methods that we have pursued so far, including the CIPSI-driven
variant examined in this study.

In order to reduce the computational effort involved in determining term (II) in Eq. (\ref{A:ampeqs3}) 
using the algorithm discussed in the preceding two paragraphs
even further, in our implementation
of the CC($P$) approach in CCpy, we also take advantage of the sparsity of the matrix representing
$\overline{H}_{N}^{(2)}$ in the subspace of the many-electron Hilbert space spanned by singly, doubly,
and triply excited determinants. In particular, close inspection of the Kronecker deltas entering the
formulas for the matrix elements of the $\overline{H}_{N}^{(2)}$ operator listed in Table \ref{A:table1}
shows that $\langle \Phi_{K} | \overline{H}_{N}^{(2)} | \Phi_{lmn}^{def} \rangle$ is zero unless the bra
and ket determinants in it share some or all of their hole and particle indices. In analogy to the well-known
0-, 1-, and 2-electron Slater rules used to evaluate the nonzero matrix elements of the bare electronic
Hamiltonian, the nonzero matrix elements of the similarity-transformed Hamiltonian $\overline{H}_{N}^{(2)}$
in the ST, DT, and TT sectors entering term (II) of Eq. (\ref{A:ampeqs3}) can be classified according to
the numbers of differences occurring in the hole and particle indices characterizing
$\langle \Phi_{K}|$ and $|\Phi_{lmn}^{def} \rangle$. Thus, in our CC($P$) algorithm, we adopt the language in which
if the bra and ket determinants entering $\langle \Phi_{K} | \overline{H}_{N}^{(2)} | \Phi_{lmn}^{def} \rangle$
differ in $\mu$ of their particle indices and $\nu$ of their hole indices, we call such
$\langle \Phi_{K} | \overline{H}_{N}^{(2)} | \Phi_{lmn}^{def} \rangle$ a $\mu p$-$\nu h$--difference matrix element.
As shown in Table \ref{A:table1}, each
$\mu p$-$\nu h$--difference matrix element $\langle \Phi_{K} | \overline{H}_{N}^{(2)} | \Phi_{lmn}^{def} \rangle$
engages a specific type of $\bar{h}_{p}^{q}$ or $\bar{h}_{pq}^{rs}$ and satisfies a distinct condition
on the hole and particle indices in its bra and ket determinants to produce a nonzero contribution.
To be more specific, if we define the set of hole (particle) indices characterizing the bra determinant
$\langle \Phi_{K}|$ as $I_h$ ($I_p$) and the corresponding set of hole (particle) indices in the
ket determinant $|\Phi_{lmn}^{def} \rangle$ as $J_h$ ($J_p$), then
the index constraint applicable to each nonzero $\mu p$-$\nu h$--difference matrix element
$\langle \Phi_{K} | \overline{H}_{N}^{(2)} | \Phi_{lmn}^{def} \rangle$ considered in Table
\ref{A:table1} can be mathematically expressed using the numbers of elements in the set intersections
$S_h = I_h \cap J_h$ and $S_p = I_p \cap J_p$, {\it i.e.}, via the cardinal numbers of $S_{h}$ and $S_{p}$,
designated in Table \ref{A:table1} as $|S_{h}|$ and $|S_{p}|$, respectively. The restrictions on
these cardinal numbers, which must be imposed in order to obtain nonzero values of the matrix elements
$\langle \Phi_{K} | \overline{H}_{N}^{(2)} | \Phi_{lmn}^{def} \rangle$ listed in Table \ref{A:table1},
allow us to efficiently organize our work, identify the nonzero matrix elements very fast, and
minimize the CPU operations involved in the evaluation and processing of the ST, DT, and TT blocks
of $\overline{H}_{N}^{(2)}$ that enter term (II) of Eq. (\ref{A:ampeqs3}). An example illustrating
this statement, especially how the restrictions on $|S_{h}|$ and $|S_{p}|$ help, is
discussed next.

Consider the evaluation of 2$p$-0$h$--difference matrix elements in the
$\langle \Phi_{ijk}^{abc} | \overline{H}_{N}^{(2)} | \Phi_{lmn}^{def} \rangle$ category (belonging to the TT block
of $\overline{H}_{N}^{(2)}$), which provide the contributions corresponding to the most expensive diagram in the
CCSDT amplitude equations that in full CCSDT scales as $n_{o}^{3} n_{u}^{5}$. As shown in Table \ref{A:table1},
matrix elements of this type are evaluated according to the formula
\begin{align}
& \mathscr{A}_{c/ab}\mathscr{A}^{f/de} \bar{h}_{ab}^{de} \delta^{k}_{n} \delta^{i}_{l} \delta^{j}_{m} \delta_{c}^{f}
\nonumber \\
& = \delta^{k}_{n} \delta^{i}_{l} \delta^{j}_{m} (\bar{h}_{ab}^{de} \delta_{c}^{f} - \bar{h}_{cb}^{de} \delta_{a}^{f}
- \bar{h}_{ac}^{de} \delta_{b}^{f}
\nonumber \\
& - \bar{h}_{ab}^{fe} \delta_{c}^{d} + \bar{h}_{cb}^{fe} \delta_{a}^{d} + \bar{h}_{ac}^{fe} \delta_{b}^{d}
\nonumber \\
& - \bar{h}_{ab}^{df} \delta_{c}^{e} + \bar{h}_{cb}^{df} \delta_{a}^{e} + \bar{h}_{ac}^{df} \delta_{b}^{e}),
\label{A:pppp}
\end{align}
where $\delta_{p}^{q}$ is the Kronecker delta [since all indices in Eq. (\ref{A:pppp}) and other similar expressions
in Table \ref{A:table1} are fixed, and to remain consistent with the Einstein summation convention adopted in this
work, upper and lower indices in Eq. (\ref{A:pppp}) and Table \ref{A:table1} have been arranged such that the indices
appearing on the same line are not summed over]. For a given pair of triply excited determinants
$|\Phi_{ijk}^{abc}\rangle$ and $|\Phi_{lmn}^{def}\rangle$ belonging to the $P$ space, Eq. (\ref{A:pppp})
will evaluate to zero unless $|S_h| = 3$ and $|S_p| = 1$, where $S_h = \{i,j,k\} \cap \{l,m,n\}$ and
$S_p = \{a,b,c\} \cap \{d,e,f\}$. Our algorithm takes advantage of these restrictions on the cardinal
numbers $|S_h|$ and $|S_p|$ required to obtain nonzero values of 2$p$-0$h$--difference matrix elements
$\langle \Phi_{ijk}^{abc} | \overline{H}_{N}^{(2)} | \Phi_{lmn}^{def} \rangle$ by partitioning the list of triply
excited determinants entering the $P$ space into nonoverlapping ``buckets'', where each bucket contains triply excited
determinants that share the same $|S_h|$ (in this case, 3) hole indices and the same $|S_p|$ (in this case, 1)
particle indices. For example, if we have 4 electrons in a system and all determinants $|\Phi_{ijk}^{abc}\rangle$
in which $i = 1$, $j =2$, and $k = 3$ and $c = 8$ are in the $P$ space (in listing $P$-space triples, we always
assume that $i < j < k$ and $a < b < c$), the
three determinants $|\Phi_{123}^{568}\rangle$, $|\Phi_{123}^{578}\rangle$, and $|\Phi_{123}^{678}\rangle$ that
share three hole indices $i$, $j$, and $k$ and one particle index $c$ form one of the buckets.
When the list of the $P$-space triples is organized in this fashion -- with each bucket having $|S_h|$ hole
indices and $|S_p|$ particle indices in common -- we can simply skip the evaluation of 2$p$-0$h$--difference
matrix elements of the $\langle \Phi_{ijk}^{abc} | \overline{H}_{N}^{(2)} | \Phi_{lmn}^{def} \rangle$ type in which
the bra and ket determinants belong to different buckets, as these will automatically evaluate to zero.
In other words, we only determine those 2$p$-0$h$--difference matrix elements
$\langle \Phi_{ijk}^{abc} | \overline{H}_{N}^{(2)} | \Phi_{lmn}^{def} \rangle$, in which the
$|\Phi_{ijk}^{abc} \rangle$ and $|\Phi_{lmn}^{def}\rangle$ determinants belong to the same bucket,
repeating this process for all the buckets into which the list of the $P$ space triples has been partitioned.
We can similarly exploit the sparsity patterns characterizing other $\mu p$-$\nu h$--difference matrix elements
of $\overline{H}_{N}^{(2)}$ in the ST, DT, and TT categories listed in Table \ref{A:table1} by judiciously
organizing the triply excited determinants included in the $P$ space into the appropriately defined buckets
based on common hole/particle indices, as indicated by the relevant values of $|S_h|$ and $|S_p|$. In this
way, we only have to execute the minimum number of CPU operations needed to evaluate nonzero contributions
to term (II) in Eq. (\ref{A:ampeqs3}), which is the key to realizing the immense
computational speedups offered by the
short (but not necessarily regular) lists of triply excited determinants included in the $P$ space.
A detailed description of the numerical procedures used to partition the $P$ space into the buckets
of triply excited determinants relevant to the various
types of $\langle \Phi_{K} | \overline{H}_{N}^{(2)} | \Phi_{lmn}^{def} \rangle$ matrix elements and
$\mu p$-$\nu h$--difference cases listed in Table \ref{A:table1}, along with the associated
spin-orbital index manipulations, will be presented in a future publication dedicated to the
CC($P$) and CC($P$;$Q$) algorithms, as implemented in CCpy.

The above approach to handling term II in Eq. (\ref{A:ampeqs3}), in which we efficiently identify, sort,
and compute the nonzero matrix elements of the ST, DT, and TT blocks of $\overline{H}_{N}^{(2)}$, results
in a possibility of speeding up CC($P$) calculations by factors of $(D/d)$ for the ST and DT blocks and
$(D/d)^2$ for the TT block relative to their full CCSDT counterparts,
where we use the same notation as in our discussion of term (I), in which $D$
defines the number of all triply excited determinants and $d$ is the number of triples
included in the $P$ space. Combined with the substantial savings offered
by our way of handling term (I) discussed above,
we obtain a highly efficient algorithm for constructing and solving the CC($P$) amplitude equations,
even when the triply excited determinants included in the $P$ space do not form a continuous manifold.
As already alluded to above, similar savings in the computational effort apply to the companion
left-eigenstate CC($P$) system based on Eq. (\ref{ccpq-10}), needed to obtain the deexcitation operator
$\Lambda^{(P)} = \Lambda_1 + \Lambda_2 + \Lambda_3^\text{(P)}$, in which the subset of triples entering
$\Lambda_3^\text{(P)}$ is the same as that defining $T_{3}^{(P)}$.

We end this appendix by illustrating the computational benefits offered by our CC($P$;$Q$) algorithm in CCpy,
especially its key CC($P$) part discussed above, by comparing the CPU times needed to solve the CC($P$)
amplitude and left-eigenstate equations and to form the noniterative $\delta(\mbox{$P$;$Q$})$ corrections
with those required by the parent full CCSDT approach for the singlet ground state of cyclobutadiene, as
described by the cc-pVDZ and cc-pVTZ basis sets, at the challenging TS structure along its automerization
coordinate corresponding in Eq. (\ref{eq:ell}) to $\lambda = 1$, where $T_{3}$ correlations become large,
nonperturbative, and difficult to capture. All the CPU times reported below
correspond to single-core runs on a PowerEdge R940 server from Dell equipped with Intel Xeon Gold 6252 2.1 GHz
processor boards. The CC($P$) and CC($P$;$Q$) calculations using the lists of triply excited determinants included
in the underlying $P$ spaces generated by CIPSI were performed with CCpy, whereas the parent CCSDT computations
were carried out using our highly efficient, fully vectorized, CCSDT codes available in GAMESS.
As in all of the CIPSI-driven CC($P$)
CC($P$;$Q$) computations discussed in the main text, no advantage of the $D_{\rm 4h}$ symmetry of the TS structure
of cyclobutadiene or its $D_{\rm 2h}$ Abelian subgroup was taken in any of the post-RHF calculations. In presenting
the timings characterizing the CC($P$) and CC($P$;$Q$) computations, the computational times associated with the
execution of the integral, RHF, and integral transformation and sorting routines are ignored.

Our first set of timings involves the calculations using the cc-pVDZ basis set. In this case, we needed 176.5 CPU
minutes and 30 iterations to converge the CCSDT energy of the cyclobutadiene TS species to $10^{-7}$ hartree. The
corresponding CC($P$) amplitude equations using 0.5\% of triply excited determinants in the $P$ space identified
by the CIPSI run with $N_{\rm det(in)} = 250,000$, which, as shown in Table \ref{tab:table1},
after correcting for the remaining triples not included in the $P$ space
via the CC($P$;$Q$) correction $\delta(\mbox{$P$;$Q$})$, reproduce the CCSDT energy to within 0.458 millihartree,
needed only 2.3 CPU minutes and 19 iterations to converge. This is a speedup by a factor of 77 compared to CCSDT.
The combined time spent on solving the CC($P$) amplitude equations and the associated left-eigenstate problem that
provides the information used to determine the $\ell_K(P)$ coefficients multiplying moments
$\mathfrak{M}_K(P)$ in Eq. (\ref{ccpq-6}) for $\delta(\mbox{$P$;$Q$})$ was 4.1 CPU minutes and the time needed to
determine the noniterative $\delta(\mbox{$P$;$Q$})$ correction was 0.9 CPU minutes, which are again considerable
savings in the computational effort compared to CCSDT. The latter time is somewhat longer
than the 0.5 CPU minutes needed to calculate the noniterative triples correction of CR-CC(2,3), since moments
\begin{equation}
\mathfrak{M}_{abc}^{ijk}(P) = \langle \Phi_{ijk}^{abc} | \overline{H}^{(P)} | \Phi \rangle , \:\:
|\Phi_{ijk}^{abc} \rangle \in \mathscr{H}^{(Q)}
\label{moment-abcijk-P}
\end{equation}
and the associated coefficients
\begin{equation}
\ell_{ijk}^{abc}(P) = \langle\Phi|(1+\Lambda^{(P)})\overline{H}^{(P)}|\Phi_{ijk}^{abc}\rangle /
D_{abc}^{ijk}{(P)} ,
\label{ell-abcijk-P}
\end{equation}
where $D_{abc}^{ijk}{(P)} = E^{(P)} - \langle \Phi_{ijk}^{abc} | \overline{H}^{(P)} | \Phi_{ijk}^{abc}\rangle$,
used to compute the $\delta(\mbox{$P$;$Q$})$ correction when there are some triply excited determinants in the
$P$ space, engage the $T_{1}$, $T_{2}$, and $T_{3}^{(P)}$ components of the cluster operator $T^{(P)}$
in constructing the similarity-transformed Hamiltonian $\overline{H}^{(P)}$, as opposed to only $T_{1}$ and
$T_{2}$ obtained with CCSD used to construct $\mathfrak{M}_{abc}^{ijk}(2)$, Eq. (\ref{moment-abcijk}),
in CR-CC(2,3), and the three-body component of the deexcitation operator $\Lambda^{(P)}$, not used by CR-CC(2,3)
either, but it is still almost 200 times shorter than the time
needed to converge CCSDT and 6 times shorter than the timing characterizing a single CCSDT
iteration. In fact, the combined time required to solve the CC($P$) amplitude equations and the corresponding
left-eigenstate problem and to construct the CC($P$;$Q$) correction $\delta(\mbox{$P$;$Q$})$ turned out to be shorter
than that associated with a single CCSDT iteration too.

The timings characterizing the CC($P$) and CC($P$;$Q$) computations for the cyclobutadiene TS species using
the cc-pVTZ basis are similarly encouraging. The CCSDT amplitude equations converged to $10^{-7}$ hartree in
7,356.9 CPU minutes, requiring 30 iterations. The analogous CC($P$) calculations using 0.1\% of triply
excited determinants in the $P$ space identified by the CIPSI run with $N_{\rm det(in)} = 1,000,000$, which,
after accounting for the remaining triples outside the $P$ space using the CC($P$;$Q$) correction
$\delta(\mbox{$P$;$Q$})$, reproduce the CCSDT energy to within 0.591 millihartree ({\it cf.} Table \ref{tab:table4}),
needed 19 iterations and only 122.5 CPU minutes to converge, speeding up the CCSDT computations by a factor of 60.
The cumulative time required to solve the CC($P$) amplitude equations and the associated left-eigenstate problem,
of 232.4 CPU minutes, and the 8.3 CPU minutes used to calculate the $\delta(\mbox{$P$;$Q$})$ correction represent
substantial savings in the computational effort compared to CCSDT as well. Once again, the 8.3 CPU minutes spent
on constructing the noniterative correction $\delta(\mbox{$P$;$Q$})$ is more than the 5.2 CPU minutes needed to
determine the triples correction of CR-CC(2,3), which is a consequence of using moments $\mathfrak{M}_{abc}^{ijk}(P)$,
Eq. (\ref{moment-abcijk-P}), involving $T_{3}^{(P)}$, in addition to $T_{1}$ and $T_{2}$, instead of the less
demanding $\mathfrak{M}_{abc}^{ijk}(2)$ using only $T_{1}$ and $T_{2}$ obtained with CCSD exploited
in CR-CC(2,3), and coefficients $\ell_{ijk}^{abc}(P)$, Eq. (\ref{ell-abcijk-P}), engaging $\overline{H}^{(P)}$ and
$\Lambda_{3}^{(P)}$, but the overall message that both the CC($P$) computations and the noniterative steps of CC($P$;$Q$)
are orders of magnitude less expensive than the parent CCSDT runs remains. Again, the cumulative time associated
with solving the CC($P$) amplitude and left-eigenstate equations and to determine the CC($P$;$Q$) correction
$\delta(\mbox{$P$;$Q$})$ turned out to be somewhat shorter than the time spent on a single CCSDT iteration.

It should be clear from the above analysis that our CC($P$;$Q$) codes in CCpy aimed at accurately approximating
the CCSDT energetics using small fractions of triples in the underlying $P$ spaces, which were identified in this
work with CIPSI, can offer significant savings in the computational effort compared to full CCSDT, even when $T_{3}$
effects and electronic quasidegeneracies become substantial and noniterative corrections to CCSD struggle.
We have also demonstrated
that a significant part of this success is due to our novel approach to implementing the CC($P$) equations, which
can efficiently handle small but generally spotty subsets of triply excited determinants in the underlying $P$
spaces. We will continue examining if our CC($P$;$Q$) codes in CCpy, especially the routines that construct
and solve the CC($P$) equations, can achieve even greater speedups compared to CCSDT.
%

\vspace*{2em}
\section*{Supporting Information}
The Supporting Information, which contains
the data used to construct the CCSDt and CC(t;3) potentials presented in Figure \ref{fig:figure6},
is available free of charge at ... .

\section*{Notes}
The authors declare no competing financial interest.

\begin{acknowledgments}
We dedicate this article to the memory of the late Professor John F. Stanton,
whose seminal contributions to quantum chemistry, especially coupled-cluster theory,
have inspired us over the years.
This work has been supported by the Chemical Sciences, Geosciences
and Biosciences Division, Office of Basic Energy Sciences, Office
of Science, U.S. Department of Energy (Grant No. DE-FG02-01ER15228 to P.P.).
\end{acknowledgments}

\section*{References}
\bibliographystyle{achemso-100etal}
\bibliography{refs}

\providecommand{\latin}[1]{#1}
\makeatletter
\providecommand{\doi}
  {\begingroup\let\do\@makeother\dospecials
  \catcode`\{=1 \catcode`\}=2 \doi@aux}
\providecommand{\doi@aux}[1]{\endgroup\texttt{#1}}
\makeatother
\providecommand*\mcitethebibliography{\thebibliography}
\csname @ifundefined\endcsname{endmcitethebibliography}
  {\let\endmcitethebibliography\endthebibliography}{}
\begin{mcitethebibliography}{167}
\providecommand*\natexlab[1]{#1}
\providecommand*\mciteSetBstSublistMode[1]{}
\providecommand*\mciteSetBstMaxWidthForm[2]{}
\providecommand*\mciteBstWouldAddEndPuncttrue
  {\def\EndOfBibitem{\unskip.}}
\providecommand*\mciteBstWouldAddEndPunctfalse
  {\let\EndOfBibitem\relax}
\providecommand*\mciteSetBstMidEndSepPunct[3]{}
\providecommand*\mciteSetBstSublistLabelBeginEnd[3]{}
\providecommand*\EndOfBibitem{}
\mciteSetBstSublistMode{f}
\mciteSetBstMaxWidthForm{subitem}{(\alph{mcitesubitemcount})}
\mciteSetBstSublistLabelBeginEnd
  {\mcitemaxwidthsubitemform\space}
  {\relax}
  {\relax}

\bibitem[Myers \latin{et~al.}(1992)Myers, Dragovich, and Kuo]{thermochem}
Myers,~A.~G.; Dragovich,~P.~S.; Kuo,~E.~Y. Studies on the Thermal Generation
  and Reactivity of a Class of ($\sigma,\pi$)-1,4-Biradicals. \emph{J. Am.
  Chem. Soc.} \textbf{1992}, \emph{114}, 9369--9386\relax
\mciteBstWouldAddEndPuncttrue
\mciteSetBstMidEndSepPunct{\mcitedefaultmidpunct}
{\mcitedefaultendpunct}{\mcitedefaultseppunct}\relax
\EndOfBibitem
\bibitem[Pedersen \latin{et~al.}(1994)Pedersen, Herek, and Zewail]{sjs-ref1}
Pedersen,~S.; Herek,~J.~L.; Zewail,~A.~H. The Validity of the ``Diradical"
  Hypothesis: Direct Femtosecond Studies of the Transition-State Structures.
  \emph{Science} \textbf{1994}, \emph{266}, 1359--1364\relax
\mciteBstWouldAddEndPuncttrue
\mciteSetBstMidEndSepPunct{\mcitedefaultmidpunct}
{\mcitedefaultendpunct}{\mcitedefaultseppunct}\relax
\EndOfBibitem
\bibitem[Berson(1978)]{photochem1}
Berson,~J.~A. The Chemistry of Trimethylenemethanes, A New Class of Biradical
  Reactive Intermediates. \emph{Acc. Chem. Res.} \textbf{1978}, \emph{11},
  446--453\relax
\mciteBstWouldAddEndPuncttrue
\mciteSetBstMidEndSepPunct{\mcitedefaultmidpunct}
{\mcitedefaultendpunct}{\mcitedefaultseppunct}\relax
\EndOfBibitem
\bibitem[Zgierski \latin{et~al.}(2005)Zgierski, Patchkovskii, and
  Lim]{photochem2}
Zgierski,~M.~Z.; Patchkovskii,~S.; Lim,~E.~C. \textit{Ab Initio} Study of a
  Biradical Radiationless Decay Channel of the Lowest Excited Electronic State
  of Cytosine and Its Derivatives. \emph{J. Chem. Phys.} \textbf{2005},
  \emph{123}, 081101\relax
\mciteBstWouldAddEndPuncttrue
\mciteSetBstMidEndSepPunct{\mcitedefaultmidpunct}
{\mcitedefaultendpunct}{\mcitedefaultseppunct}\relax
\EndOfBibitem
\bibitem[Zgierski \latin{et~al.}(2005)Zgierski, Patchkovskii, Fujiwara, and
  Lim]{photochem3}
Zgierski,~M.~Z.; Patchkovskii,~S.; Fujiwara,~T.; Lim,~E.~C. On the Origin of
  the Ultrafast Internal Conversion of Electronically Excited Pyrimidine Bases.
  \emph{J. Phys. Chem. A} \textbf{2005}, \emph{109}, 9384--9387\relax
\mciteBstWouldAddEndPuncttrue
\mciteSetBstMidEndSepPunct{\mcitedefaultmidpunct}
{\mcitedefaultendpunct}{\mcitedefaultseppunct}\relax
\EndOfBibitem
\bibitem[Park \latin{et~al.}(2021)Park, Shen, Lee, Piecuch, Filatov, and
  Choi]{photochem4}
Park,~W.; Shen,~J.; Lee,~S.; Piecuch,~P.; Filatov,~M.; Choi,~C.~H. Internal
  Conversion Between Bright ($1^1{B}_u^+$) and Dark ($2^1{A}_g^-$) States in
  s-\textit{trans}-Butadiene and s-\textit{trans}-Hexatriene. \emph{J. Phys.
  Chem. Lett.} \textbf{2021}, \emph{12}, 9720--9729\relax
\mciteBstWouldAddEndPuncttrue
\mciteSetBstMidEndSepPunct{\mcitedefaultmidpunct}
{\mcitedefaultendpunct}{\mcitedefaultseppunct}\relax
\EndOfBibitem
\bibitem[Abe \latin{et~al.}(2012)Abe, Ye, and Mishima]{abe2012}
Abe,~M.; Ye,~J.; Mishima,~M. The Chemistry of Localized Singlet 1,3-Diradicals
  (Biradicals): From Putative Intermediates to Persistent Species and Unusual
  Molecules with a $\pi$-Single Bonded Character. \emph{Chem. Soc. Rev.}
  \textbf{2012}, \emph{41}, 3808--3820\relax
\mciteBstWouldAddEndPuncttrue
\mciteSetBstMidEndSepPunct{\mcitedefaultmidpunct}
{\mcitedefaultendpunct}{\mcitedefaultseppunct}\relax
\EndOfBibitem
\bibitem[Dougherty(1991)]{molmagnet1}
Dougherty,~D.~A. Spin Control in Organic Molecules. \emph{Acc. Chem. Res.}
  \textbf{1991}, \emph{24}, 88--94\relax
\mciteBstWouldAddEndPuncttrue
\mciteSetBstMidEndSepPunct{\mcitedefaultmidpunct}
{\mcitedefaultendpunct}{\mcitedefaultseppunct}\relax
\EndOfBibitem
\bibitem[Malrieu \latin{et~al.}(2014)Malrieu, Caballol, Calzado, de~Graaf, and
  Guih\'ery]{molmagnet2}
Malrieu,~J.~P.; Caballol,~R.; Calzado,~C.~J.; de~Graaf,~C.; Guih\'ery,~N.
  Magnetic Interactions in Molecules and Highly Correlated Materials: Physical
  Content, Analytical Derivation, and Rigorous Extraction of Magnetic
  Hamiltonians. \emph{Chem. Rev.} \textbf{2014}, \emph{114}, 429--492\relax
\mciteBstWouldAddEndPuncttrue
\mciteSetBstMidEndSepPunct{\mcitedefaultmidpunct}
{\mcitedefaultendpunct}{\mcitedefaultseppunct}\relax
\EndOfBibitem
\bibitem[Cho \latin{et~al.}(2016)Cho, Ko, and Lee]{molmagnet3}
Cho,~D.; Ko,~K.~C.; Lee,~J.~Y. Quantum Chemical Approaches for Controlling and
  Evaluating Intramolecular Magnetic Interactions in Organic Diradicals.
  \emph{Int. J. Quantum Chem.} \textbf{2016}, \emph{116}, 578--597\relax
\mciteBstWouldAddEndPuncttrue
\mciteSetBstMidEndSepPunct{\mcitedefaultmidpunct}
{\mcitedefaultendpunct}{\mcitedefaultseppunct}\relax
\EndOfBibitem
\bibitem[Sun \latin{et~al.}(2014)Sun, Zeng, and Wu]{battery2014}
Sun,~Z.; Zeng,~Z.; Wu,~J. Zethrenes, Extended \textit{p}-Quinodimethanes, and
  Periacenes with a Singlet Biradical Ground State. \emph{Acc. Chem. Res.}
  \textbf{2014}, \emph{47}, 2582--2591\relax
\mciteBstWouldAddEndPuncttrue
\mciteSetBstMidEndSepPunct{\mcitedefaultmidpunct}
{\mcitedefaultendpunct}{\mcitedefaultseppunct}\relax
\EndOfBibitem
\bibitem[Minami and Nakano(2012)Minami, and Nakano]{photovol1}
Minami,~T.; Nakano,~M. Diradical Character View of Singlet Fission. \emph{J.
  Phys. Chem. Lett.} \textbf{2012}, \emph{3}, 145--150\relax
\mciteBstWouldAddEndPuncttrue
\mciteSetBstMidEndSepPunct{\mcitedefaultmidpunct}
{\mcitedefaultendpunct}{\mcitedefaultseppunct}\relax
\EndOfBibitem
\bibitem[Hedley \latin{et~al.}(2017)Hedley, Ruseckas, and Samuel]{photovol2}
Hedley,~G.~J.; Ruseckas,~A.; Samuel,~I. D.~W. Light Harvesting for Organic
  Photovoltaics. \emph{Chem. Rev.} \textbf{2017}, \emph{117}, 796--837\relax
\mciteBstWouldAddEndPuncttrue
\mciteSetBstMidEndSepPunct{\mcitedefaultmidpunct}
{\mcitedefaultendpunct}{\mcitedefaultseppunct}\relax
\EndOfBibitem
\bibitem[Niklas and Poluektov(2017)Niklas, and Poluektov]{photovol3}
Niklas,~J.; Poluektov,~O.~G. Charge Transfer Processes in OPV Materials as
  Revealed by EPR Spectroscopy. \emph{Adv. Energy Mater.} \textbf{2017},
  \emph{7}, 1602226\relax
\mciteBstWouldAddEndPuncttrue
\mciteSetBstMidEndSepPunct{\mcitedefaultmidpunct}
{\mcitedefaultendpunct}{\mcitedefaultseppunct}\relax
\EndOfBibitem
\bibitem[Smith and Michl(2010)Smith, and Michl]{singletfission}
Smith,~M.~B.; Michl,~J. Singlet Fission. \emph{Chem. Rev.} \textbf{2010},
  \emph{110}, 6891--6936\relax
\mciteBstWouldAddEndPuncttrue
\mciteSetBstMidEndSepPunct{\mcitedefaultmidpunct}
{\mcitedefaultendpunct}{\mcitedefaultseppunct}\relax
\EndOfBibitem
\bibitem[Slipchenko and Krylov(2002)Slipchenko, and Krylov]{ch2_krylov}
Slipchenko,~L.~V.; Krylov,~A.~I. Singlet-Triplet Gaps in Diradicals by the
  Spin-Flip Approach: A Benchmark Study. \emph{J. Chem. Phys.} \textbf{2002},
  \emph{117}, 4694--4708\relax
\mciteBstWouldAddEndPuncttrue
\mciteSetBstMidEndSepPunct{\mcitedefaultmidpunct}
{\mcitedefaultendpunct}{\mcitedefaultseppunct}\relax
\EndOfBibitem
\bibitem[W{\l}och \latin{et~al.}(2007)W{\l}och, Gour, and Piecuch]{crccl_jpc}
W{\l}och,~M.; Gour,~J.~R.; Piecuch,~P. Extension of the Renormalized
  Coupled-Cluster Methods Exploiting Left Eigenstates of the
  Similarity-Transformed Hamiltonian to Open-Shell Systems: A Benchmark Study.
  \emph{J. Phys. Chem. A} \textbf{2007}, \emph{111}, 11359--11382\relax
\mciteBstWouldAddEndPuncttrue
\mciteSetBstMidEndSepPunct{\mcitedefaultmidpunct}
{\mcitedefaultendpunct}{\mcitedefaultseppunct}\relax
\EndOfBibitem
\bibitem[Li and Paldus(2008)Li, and Paldus]{ch2-rmrccsdt}
Li,~X.; Paldus,~J. Electronic Structure of Organic Diradicals: Evaluation of
  the Performance of Coupled-Cluster Methods. \emph{J. Chem. Phys.}
  \textbf{2008}, \emph{129}, 174101\relax
\mciteBstWouldAddEndPuncttrue
\mciteSetBstMidEndSepPunct{\mcitedefaultmidpunct}
{\mcitedefaultendpunct}{\mcitedefaultseppunct}\relax
\EndOfBibitem
\bibitem[Demel \latin{et~al.}(2008)Demel, Shamasundar, Kong, and
  Nooijen]{icmrcc6}
Demel,~O.; Shamasundar,~K.~R.; Kong,~L.; Nooijen,~M. Application of Double
  Ionization State-Specific Equation of Motion Coupled Cluster Method to
  Organic Diradicals. \emph{J. Phys. Chem. A} \textbf{2008}, \emph{112},
  11895--11902\relax
\mciteBstWouldAddEndPuncttrue
\mciteSetBstMidEndSepPunct{\mcitedefaultmidpunct}
{\mcitedefaultendpunct}{\mcitedefaultseppunct}\relax
\EndOfBibitem
\bibitem[Saito \latin{et~al.}(2011)Saito, Nishihara, Yamanaka, Kitagawa,
  Kawakami, Yamada, Isobe, Okumura, and Yamaguchi]{saito2011symmetry}
Saito,~T.; Nishihara,~S.; Yamanaka,~S.; Kitagawa,~Y.; Kawakami,~T.; Yamada,~S.;
  Isobe,~H.; Okumura,~M.; Yamaguchi,~K. Symmetry and Broken Symmetry in
  Molecular Orbital Description of Unstable Molecules IV: Comparison Between
  Single- and Multi-Reference Computational Results for Antiaromtic Molecules.
  \emph{Theor. Chem. Acc.} \textbf{2011}, \emph{130}, 749--763\relax
\mciteBstWouldAddEndPuncttrue
\mciteSetBstMidEndSepPunct{\mcitedefaultmidpunct}
{\mcitedefaultendpunct}{\mcitedefaultseppunct}\relax
\EndOfBibitem
\bibitem[Ess \latin{et~al.}(2011)Ess, Johnson, Hu, and Yang]{ess2011ST}
Ess,~D.~H.; Johnson,~E.~R.; Hu,~X.; Yang,~W. Singlet--Triplet Energy Gaps for
  Diradicals from Fractional-Spin Density-Functional Theory. \emph{J. Phys.
  Chem. A} \textbf{2011}, \emph{115}, 76--83\relax
\mciteBstWouldAddEndPuncttrue
\mciteSetBstMidEndSepPunct{\mcitedefaultmidpunct}
{\mcitedefaultendpunct}{\mcitedefaultseppunct}\relax
\EndOfBibitem
\bibitem[Shen and Piecuch(2012)Shen, and Piecuch]{jspp-jctc2012}
Shen,~J.; Piecuch,~P. Merging Active-Space and Renormalized Coupled-Cluster
  Methods via the CC($P$;$Q$) Formalism, with Benchmark Calculations for
  Singlet--Triplet Gaps in Biradical Systems. \emph{J. Chem. Theory Comput.}
  \textbf{2012}, \emph{8}, 4968--4988\relax
\mciteBstWouldAddEndPuncttrue
\mciteSetBstMidEndSepPunct{\mcitedefaultmidpunct}
{\mcitedefaultendpunct}{\mcitedefaultseppunct}\relax
\EndOfBibitem
\bibitem[Abe(2013)]{abe2013diradicals}
Abe,~M. Diradicals. \emph{Chem. Rev.} \textbf{2013}, \emph{113},
  7011--7088\relax
\mciteBstWouldAddEndPuncttrue
\mciteSetBstMidEndSepPunct{\mcitedefaultmidpunct}
{\mcitedefaultendpunct}{\mcitedefaultseppunct}\relax
\EndOfBibitem
\bibitem[Garza \latin{et~al.}(2014)Garza, Jim{\'e}nez-Hoyos, and
  Scuseria]{garza2014electronic}
Garza,~A.~J.; Jim{\'e}nez-Hoyos,~C.~A.; Scuseria,~G.~E. Electronic Correlation
  Without Double Counting via a Combination of Spin Projected Hartree-Fock and
  Density Functional Theories. \emph{J. Chem. Phys.} \textbf{2014}, \emph{140},
  244102\relax
\mciteBstWouldAddEndPuncttrue
\mciteSetBstMidEndSepPunct{\mcitedefaultmidpunct}
{\mcitedefaultendpunct}{\mcitedefaultseppunct}\relax
\EndOfBibitem
\bibitem[Ibeji and Ghosh(2015)Ibeji, and Ghosh]{ibeji2015singlet}
Ibeji,~C.~U.; Ghosh,~D. Singlet--Triplet Gaps in Polyacenes: A Delicate Balance
  Between Dynamic and Static Correlations Investigated by Spin--Flip Methods.
  \emph{Phys. Chem. Chem. Phys.} \textbf{2015}, \emph{17}, 9849--9856\relax
\mciteBstWouldAddEndPuncttrue
\mciteSetBstMidEndSepPunct{\mcitedefaultmidpunct}
{\mcitedefaultendpunct}{\mcitedefaultseppunct}\relax
\EndOfBibitem
\bibitem[Ajala \latin{et~al.}(2017)Ajala, Shen, and Piecuch]{ajala2017}
Ajala,~A.~O.; Shen,~J.; Piecuch,~P. Economical Doubly Electron-Attached
  Equation-of-Motion Coupled-Cluster Methods with an Active-Space Treatment of
  Three-Particle--One-Hole and Four-Particle--Two-Hole Excitations. \emph{J.
  Phys. Chem. A} \textbf{2017}, \emph{121}, 3469--3485\relax
\mciteBstWouldAddEndPuncttrue
\mciteSetBstMidEndSepPunct{\mcitedefaultmidpunct}
{\mcitedefaultendpunct}{\mcitedefaultseppunct}\relax
\EndOfBibitem
\bibitem[Stoneburner \latin{et~al.}(2017)Stoneburner, Shen, Ajala, Piecuch,
  Truhlar, and Gagliardi]{stoneburner2017}
Stoneburner,~S.~J.; Shen,~J.; Ajala,~A.~O.; Piecuch,~P.; Truhlar,~D.~G.;
  Gagliardi,~L. Systematic Design of Active Spaces for Multi-Reference
  Calculations of Singlet--Triplet Gaps of Organic Diradicals, with Benchmarks
  Against Doubly Electron-Attached Coupled-Cluster Data. \emph{J. Chem. Phys.}
  \textbf{2017}, \emph{147}, 164120\relax
\mciteBstWouldAddEndPuncttrue
\mciteSetBstMidEndSepPunct{\mcitedefaultmidpunct}
{\mcitedefaultendpunct}{\mcitedefaultseppunct}\relax
\EndOfBibitem
\bibitem[Zimmerman(2017)]{zimmerman2017}
Zimmerman,~P.~M. Singlet--Triplet Gaps Through Incremental Full Configuration
  Interaction. \emph{J. Phys. Chem. A} \textbf{2017}, \emph{121},
  4712--4720\relax
\mciteBstWouldAddEndPuncttrue
\mciteSetBstMidEndSepPunct{\mcitedefaultmidpunct}
{\mcitedefaultendpunct}{\mcitedefaultseppunct}\relax
\EndOfBibitem
\bibitem[Shen and Piecuch(2021)Shen, and Piecuch]{jspp-dea-2021}
Shen,~J.; Piecuch,~P. Double Electron-Attachment Equation-of-Motion
  Coupled-Cluster Methods with up to 4-Particle--2-Hole Excitations: Improved
  Implementation and Application to Singlet--Triplet Gaps in \textit{ortho}-,
  \textit{meta}-, and \textit{para}-Benzyne Isomers. \emph{Mol. Phys.}
  \textbf{2021}, \emph{119}, e1966534\relax
\mciteBstWouldAddEndPuncttrue
\mciteSetBstMidEndSepPunct{\mcitedefaultmidpunct}
{\mcitedefaultendpunct}{\mcitedefaultseppunct}\relax
\EndOfBibitem
\bibitem[Gulania \latin{et~al.}(2021)Gulania, Kj{\o}nstad, Stanton, Koch, and
  Krylov]{gulania2021EOM}
Gulania,~S.; Kj{\o}nstad,~E.~F.; Stanton,~J.~F.; Koch,~H.; Krylov,~A.~I.
  Equation-of-Motion Coupled-Cluster Method with Double Electron-Attaching
  Operators: Theory, Implementation, and Benchmarks. \emph{J. Chem. Phys.}
  \textbf{2021}, \emph{154}, 114115\relax
\mciteBstWouldAddEndPuncttrue
\mciteSetBstMidEndSepPunct{\mcitedefaultmidpunct}
{\mcitedefaultendpunct}{\mcitedefaultseppunct}\relax
\EndOfBibitem
\bibitem[Boyn and Mazziotti(2021)Boyn, and Mazziotti]{mazziotti2021}
Boyn,~J.-N.; Mazziotti,~D.~A. Accurate Singlet--Triplet Gaps in Biradicals via
  the Spin Averaged Anti-Hermitian Contracted Schr\"{o}dinger Equation.
  \emph{J. Chem. Phys.} \textbf{2021}, \emph{154}, 134103\relax
\mciteBstWouldAddEndPuncttrue
\mciteSetBstMidEndSepPunct{\mcitedefaultmidpunct}
{\mcitedefaultendpunct}{\mcitedefaultseppunct}\relax
\EndOfBibitem
\bibitem[Chakraborty \latin{et~al.}(2022)Chakraborty, Yuwono, Deustua, Shen,
  and Piecuch]{arnab-stgap-2022}
Chakraborty,~A.; Yuwono,~S.~H.; Deustua,~J.~E.; Shen,~J.; Piecuch,~P.
  Benchmarking the Semi-Stochastic CC($P$;$Q$) Approach for Singlet--Triplet
  Gaps in Biradicals. \emph{J. Chem. Phys.} \textbf{2022}, \emph{157},
  134101\relax
\mciteBstWouldAddEndPuncttrue
\mciteSetBstMidEndSepPunct{\mcitedefaultmidpunct}
{\mcitedefaultendpunct}{\mcitedefaultseppunct}\relax
\EndOfBibitem
\bibitem[Demel \latin{et~al.}(2023)Demel, Brandejs, Lang, Brabec, Veis, Legeza,
  and Pittner]{demel-et-al-jcp-cbd-2023}
Demel,~O.; Brandejs,~J.; Lang,~J.; Brabec,~J.; Veis,~L.; Legeza,~{\" O}.;
  Pittner,~J. Hilbert Space Multireference Coupled Cluster Tailored by Matrix
  Product States. \emph{J. Chem. Phys.} \textbf{2023}, \emph{159}, 224115\relax
\mciteBstWouldAddEndPuncttrue
\mciteSetBstMidEndSepPunct{\mcitedefaultmidpunct}
{\mcitedefaultendpunct}{\mcitedefaultseppunct}\relax
\EndOfBibitem
\bibitem[Szabados \latin{et~al.}(2025)Szabados, Mih{\' a}lka, and Surj{\'
  a}n]{szabados-et-al-molphys-2025}
Szabados,~{\' A}.; Mih{\' a}lka,~Z.~{\' E}.; Surj{\' a}n,~P.~R. Orbital
  Optimisation with Spin-Unrestricted and Projected Geminals Reference.
  \emph{Mol. Phys.} \textbf{2025}, \emph{XXX}, e2501778\relax
\mciteBstWouldAddEndPuncttrue
\mciteSetBstMidEndSepPunct{\mcitedefaultmidpunct}
{\mcitedefaultendpunct}{\mcitedefaultseppunct}\relax
\EndOfBibitem
\bibitem[Craig(1950)]{craig1950}
Craig,~D.~P. Electronic Levels in Simple Conjugated Systems, I. Configuration
  Interaction in Cyclobutadiene. \emph{Proc. R. Soc. London A} \textbf{1950},
  \emph{202}, 498--506\relax
\mciteBstWouldAddEndPuncttrue
\mciteSetBstMidEndSepPunct{\mcitedefaultmidpunct}
{\mcitedefaultendpunct}{\mcitedefaultseppunct}\relax
\EndOfBibitem
\bibitem[Buenker and Peyerimhoff(1968)Buenker, and
  Peyerimhoff]{peyerimhoff1968}
Buenker,~R.~J.; Peyerimhoff,~S.~D. \textit{Ab Initio} Study on the Stability
  and Geometry of Cyclobutadiene. \emph{J. Chem. Phys.} \textbf{1968},
  \emph{48}, 354--373\relax
\mciteBstWouldAddEndPuncttrue
\mciteSetBstMidEndSepPunct{\mcitedefaultmidpunct}
{\mcitedefaultendpunct}{\mcitedefaultseppunct}\relax
\EndOfBibitem
\bibitem[Nakamura \latin{et~al.}(1989)Nakamura, Osamura, and Iwata]{iwata1989}
Nakamura,~K.; Osamura,~Y.; Iwata,~S. Second-Order Jahn-Teller Effect of
  Cyclobutadiene in Low-Lying States. An MCSCF Study. \emph{Chem. Phys.}
  \textbf{1989}, \emph{136}, 67--77\relax
\mciteBstWouldAddEndPuncttrue
\mciteSetBstMidEndSepPunct{\mcitedefaultmidpunct}
{\mcitedefaultendpunct}{\mcitedefaultseppunct}\relax
\EndOfBibitem
\bibitem[Balkov\'{a} and Bartlett(1994)Balkov\'{a}, and Bartlett]{balkova1994}
Balkov\'{a},~A.; Bartlett,~R.~J. A Multireference Coupled-Cluster Study of the
  Ground State and Lowest Excited States of Cyclobutadiene. \emph{J. Chem.
  Phys.} \textbf{1994}, \emph{101}, 8972--8987\relax
\mciteBstWouldAddEndPuncttrue
\mciteSetBstMidEndSepPunct{\mcitedefaultmidpunct}
{\mcitedefaultendpunct}{\mcitedefaultseppunct}\relax
\EndOfBibitem
\bibitem[Levchenko and Krylov(2004)Levchenko, and Krylov]{krylov2004}
Levchenko,~S.~V.; Krylov,~A.~I. Equation-of-Motion Spin-Flip Coupled-Cluster
  Model with Single and Double Substitutions: Theory and Application to
  Cyclobutadiene. \emph{J. Chem. Phys.} \textbf{2004}, \emph{120},
  175--185\relax
\mciteBstWouldAddEndPuncttrue
\mciteSetBstMidEndSepPunct{\mcitedefaultmidpunct}
{\mcitedefaultendpunct}{\mcitedefaultseppunct}\relax
\EndOfBibitem
\bibitem[Eckert-Maksi\'{c} \latin{et~al.}(2006)Eckert-Maksi\'{c}, Vazdar,
  Barbatti, Lischka, and Maksi\'{c}]{MR-AQCC}
Eckert-Maksi\'{c},~M.; Vazdar,~M.; Barbatti,~M.; Lischka,~H.; Maksi\'{c},~Z.~B.
  Automerization Reaction of Cyclobutadiene and Its Barrier Height: An
  \textit{Ab Initio} Benchmark Multireference Average-Quadratic Coupled Cluster
  Study. \emph{J. Chem. Phys.} \textbf{2006}, \emph{125}, 064310\relax
\mciteBstWouldAddEndPuncttrue
\mciteSetBstMidEndSepPunct{\mcitedefaultmidpunct}
{\mcitedefaultendpunct}{\mcitedefaultseppunct}\relax
\EndOfBibitem
\bibitem[Monino \latin{et~al.}(2022)Monino, Boggio-Pasqua, Scemama, Jacquemin,
  and Loos]{cbd-loos-2022}
Monino,~E.; Boggio-Pasqua,~M.; Scemama,~A.; Jacquemin,~D.; Loos,~P.-F.
  Reference Energies for Cyclobutadiene: Automerization and Excited States.
  \emph{J. Phys. Chem. A} \textbf{2022}, \emph{126}, 4664--4679\relax
\mciteBstWouldAddEndPuncttrue
\mciteSetBstMidEndSepPunct{\mcitedefaultmidpunct}
{\mcitedefaultendpunct}{\mcitedefaultseppunct}\relax
\EndOfBibitem
\bibitem[Borden(1975)]{borden1975}
Borden,~W.~T. Can a Square or Effectively Square Singlet be the Ground State of
  Cyclobutadiene? \emph{J. Am. Chem. Soc.} \textbf{1975}, \emph{97},
  5968--5970\relax
\mciteBstWouldAddEndPuncttrue
\mciteSetBstMidEndSepPunct{\mcitedefaultmidpunct}
{\mcitedefaultendpunct}{\mcitedefaultseppunct}\relax
\EndOfBibitem
\bibitem[Kollmar and Staemmler(1977)Kollmar, and Staemmler]{kollmar1977}
Kollmar,~H.; Staemmler,~V. A Theoretical Study of the Structure of
  Cyclobutadiene. \emph{J. Am. Chem. Soc.} \textbf{1977}, \emph{99},
  3583--3587\relax
\mciteBstWouldAddEndPuncttrue
\mciteSetBstMidEndSepPunct{\mcitedefaultmidpunct}
{\mcitedefaultendpunct}{\mcitedefaultseppunct}\relax
\EndOfBibitem
\bibitem[Borden \latin{et~al.}(1978)Borden, Davidson, and Hart]{borden1978}
Borden,~W.~T.; Davidson,~E.~R.; Hart,~P. The Potential Surfaces for the Lowest
  Singlet and Triplet States of Cyclobutadiene. \emph{J. Am. Chem. Soc.}
  \textbf{1978}, \emph{100}, 388--392\relax
\mciteBstWouldAddEndPuncttrue
\mciteSetBstMidEndSepPunct{\mcitedefaultmidpunct}
{\mcitedefaultendpunct}{\mcitedefaultseppunct}\relax
\EndOfBibitem
\bibitem[Roos(1987)]{Roos1987}
Roos,~B.~O. The Complete Active Space Self-Consistent Field Method and Its
  Applications in Electronic Structure Calculations. \emph{Adv. Chem. Phys.}
  \textbf{1987}, \emph{69}, 399--445\relax
\mciteBstWouldAddEndPuncttrue
\mciteSetBstMidEndSepPunct{\mcitedefaultmidpunct}
{\mcitedefaultendpunct}{\mcitedefaultseppunct}\relax
\EndOfBibitem
\bibitem[Schmidt and Gordon(1998)Schmidt, and Gordon]{ref:mrmpreview}
Schmidt,~M.~W.; Gordon,~M.~S. The Construction and Interpretation of MCSCF
  Wavefunctions. \emph{Annu. Rev. Phys. Chem.} \textbf{1998}, \emph{49},
  233--266\relax
\mciteBstWouldAddEndPuncttrue
\mciteSetBstMidEndSepPunct{\mcitedefaultmidpunct}
{\mcitedefaultendpunct}{\mcitedefaultseppunct}\relax
\EndOfBibitem
\bibitem[Szalay \latin{et~al.}(2012)Szalay, M{\" u}ller, Gidofalvi, Lischka,
  and Shepard]{chemrev-2012a}
Szalay,~P.~G.; M{\" u}ller,~T.; Gidofalvi,~G.; Lischka,~H.; Shepard,~R.
  Multiconfiguration Self-Consistent Field and Multireference Configuration
  Interaction Methods and Applications. \emph{Chem. Rev.} \textbf{2012},
  \emph{112}, 108--181\relax
\mciteBstWouldAddEndPuncttrue
\mciteSetBstMidEndSepPunct{\mcitedefaultmidpunct}
{\mcitedefaultendpunct}{\mcitedefaultseppunct}\relax
\EndOfBibitem
\bibitem[Roca-Sanju{\' a}n \latin{et~al.}(2012)Roca-Sanju{\' a}n, Aquilante,
  and Lindh]{lindh-review-2012}
Roca-Sanju{\' a}n,~D.; Aquilante,~F.; Lindh,~R. Muticonfiguration Second-Order
  Perturbation Theory Approach to Strong Electron Correlation in Chemistry and
  Photochemistry. \emph{WIREs Comput. Mol. Sci.} \textbf{2012}, \emph{2},
  585--603\relax
\mciteBstWouldAddEndPuncttrue
\mciteSetBstMidEndSepPunct{\mcitedefaultmidpunct}
{\mcitedefaultendpunct}{\mcitedefaultseppunct}\relax
\EndOfBibitem
\bibitem[Chattopadhyay \latin{et~al.}(2016)Chattopadhyay, Chaudhuri, Mahapatra,
  Ghosh, and Ray]{sinha-review-2016}
Chattopadhyay,~S.; Chaudhuri,~R.~K.; Mahapatra,~U.~S.; Ghosh,~A.; Ray,~S.~S.
  State-Specific Multireference Perturbation Theory: Development and Present
  Status. \emph{WIREs Comput. Mol. Sci.} \textbf{2016}, \emph{6},
  266--291\relax
\mciteBstWouldAddEndPuncttrue
\mciteSetBstMidEndSepPunct{\mcitedefaultmidpunct}
{\mcitedefaultendpunct}{\mcitedefaultseppunct}\relax
\EndOfBibitem
\bibitem[Lyakh \latin{et~al.}(2012)Lyakh, Musia{\l}, Lotrich, and
  Bartlett]{chemrev-2012b}
Lyakh,~D.~I.; Musia{\l},~M.; Lotrich,~V.~F.; Bartlett,~R.~J. Multireference
  Nature of Chemistry: The Coupled-Cluster View. \emph{Chem. Rev.}
  \textbf{2012}, \emph{112}, 182--243\relax
\mciteBstWouldAddEndPuncttrue
\mciteSetBstMidEndSepPunct{\mcitedefaultmidpunct}
{\mcitedefaultendpunct}{\mcitedefaultseppunct}\relax
\EndOfBibitem
\bibitem[Piecuch and Kowalski(2002)Piecuch, and Kowalski]{succ5}
Piecuch,~P.; Kowalski,~K. The State-Universal Multi-Reference Coupled-Cluster
  Theory: An Overview of Some Recent Advances. \emph{Int. J. Mol. Sci.}
  \textbf{2002}, \emph{3}, 676--709\relax
\mciteBstWouldAddEndPuncttrue
\mciteSetBstMidEndSepPunct{\mcitedefaultmidpunct}
{\mcitedefaultendpunct}{\mcitedefaultseppunct}\relax
\EndOfBibitem
\bibitem[Evangelista(2018)]{evangelista-perspective-jcp-2018}
Evangelista,~F.~A. Perspective: Multireference Coupled Cluster Theories of
  Dynamical Electron Correlation. \emph{J. Chem. Phys.} \textbf{2018},
  \emph{149}, 030901\relax
\mciteBstWouldAddEndPuncttrue
\mciteSetBstMidEndSepPunct{\mcitedefaultmidpunct}
{\mcitedefaultendpunct}{\mcitedefaultseppunct}\relax
\EndOfBibitem
\bibitem[Coester(1958)]{Coester:1958}
Coester,~F. Bound States of a Many-Particle System. \emph{Nucl. Phys.}
  \textbf{1958}, \emph{7}, 421--424\relax
\mciteBstWouldAddEndPuncttrue
\mciteSetBstMidEndSepPunct{\mcitedefaultmidpunct}
{\mcitedefaultendpunct}{\mcitedefaultseppunct}\relax
\EndOfBibitem
\bibitem[Coester and K{\" u}mmel(1960)Coester, and K{\" u}mmel]{Coester:1960}
Coester,~F.; K{\" u}mmel,~H. Short-Range Correlations in Nuclear Wave
  Functions. \emph{Nucl. Phys.} \textbf{1960}, \emph{17}, 477--485\relax
\mciteBstWouldAddEndPuncttrue
\mciteSetBstMidEndSepPunct{\mcitedefaultmidpunct}
{\mcitedefaultendpunct}{\mcitedefaultseppunct}\relax
\EndOfBibitem
\bibitem[{\v C}{\'\i}{\v z}ek(1966)]{cizek1}
{\v C}{\'\i}{\v z}ek,~J. On the Correlation Problem in Atomic and Molecular
  Systems. Calculation of Wavefunction Components in Ursell-Type Expansion
  Using Quantum-Field Theoretical Methods. \emph{J. Chem. Phys.} \textbf{1966},
  \emph{45}, 4256--4266\relax
\mciteBstWouldAddEndPuncttrue
\mciteSetBstMidEndSepPunct{\mcitedefaultmidpunct}
{\mcitedefaultendpunct}{\mcitedefaultseppunct}\relax
\EndOfBibitem
\bibitem[{\v C}{\'\i}{\v z}ek(1969)]{cizek2}
{\v C}{\'\i}{\v z}ek,~J. On the Use of the Cluster Expansion and the Technique
  of Diagrams in Calculations of Correlation Effects in Atoms and Molecules.
  \emph{Adv. Chem. Phys.} \textbf{1969}, \emph{14}, 35--89\relax
\mciteBstWouldAddEndPuncttrue
\mciteSetBstMidEndSepPunct{\mcitedefaultmidpunct}
{\mcitedefaultendpunct}{\mcitedefaultseppunct}\relax
\EndOfBibitem
\bibitem[Paldus \latin{et~al.}(1972)Paldus, {\v C}{\'\i}{\v z}ek, and
  Shavitt]{cizek4}
Paldus,~J.; {\v C}{\'\i}{\v z}ek,~J.; Shavitt,~I. Correlation Problems in
  Atomic and Molecular Systems. IV. Extended Coupled-Pair Many-Electron Theory
  and Its Application to the BH$_3$ Molecule. \emph{Phys. Rev. A}
  \textbf{1972}, \emph{5}, 50--67\relax
\mciteBstWouldAddEndPuncttrue
\mciteSetBstMidEndSepPunct{\mcitedefaultmidpunct}
{\mcitedefaultendpunct}{\mcitedefaultseppunct}\relax
\EndOfBibitem
\bibitem[Paldus and Li(1999)Paldus, and Li]{paldus-li}
Paldus,~J.; Li,~X. A Critical Assessment of Coupled Cluster Method in Quantum
  Chemistry. \emph{Adv. Chem. Phys.} \textbf{1999}, \emph{110}, 1--175\relax
\mciteBstWouldAddEndPuncttrue
\mciteSetBstMidEndSepPunct{\mcitedefaultmidpunct}
{\mcitedefaultendpunct}{\mcitedefaultseppunct}\relax
\EndOfBibitem
\bibitem[Bartlett and Musia{\l}(2007)Bartlett, and
  Musia{\l}]{bartlett-musial2007}
Bartlett,~R.~J.; Musia{\l},~M. Coupled-Cluster Theory in Quantum Chemistry.
  \emph{Rev. Mod. Phys.} \textbf{2007}, \emph{79}, 291--352\relax
\mciteBstWouldAddEndPuncttrue
\mciteSetBstMidEndSepPunct{\mcitedefaultmidpunct}
{\mcitedefaultendpunct}{\mcitedefaultseppunct}\relax
\EndOfBibitem
\bibitem[Hubbard(1957)]{Hubbard:1957}
Hubbard,~J. The Description of Collective Motions in Terms of Many-Body
  Perturbation Theory. \emph{Proc. R. Soc. London A} \textbf{1957}, \emph{240},
  539--560\relax
\mciteBstWouldAddEndPuncttrue
\mciteSetBstMidEndSepPunct{\mcitedefaultmidpunct}
{\mcitedefaultendpunct}{\mcitedefaultseppunct}\relax
\EndOfBibitem
\bibitem[Hugenholtz(1957)]{Hugenholtz:1957}
Hugenholtz,~N.~M. Perturbation Theory of Large Quantum Systems. \emph{Physica}
  \textbf{1957}, \emph{23}, 481--532\relax
\mciteBstWouldAddEndPuncttrue
\mciteSetBstMidEndSepPunct{\mcitedefaultmidpunct}
{\mcitedefaultendpunct}{\mcitedefaultseppunct}\relax
\EndOfBibitem
\bibitem[Purvis and Bartlett(1982)Purvis, and Bartlett]{ccsd}
Purvis,~G.~D.,~III; Bartlett,~R.~J. A Full Coupled-Cluster Singles and Doubles
  Model: The Inclusion of Disconnected Triples. \emph{J. Chem. Phys.}
  \textbf{1982}, \emph{76}, 1910--1918\relax
\mciteBstWouldAddEndPuncttrue
\mciteSetBstMidEndSepPunct{\mcitedefaultmidpunct}
{\mcitedefaultendpunct}{\mcitedefaultseppunct}\relax
\EndOfBibitem
\bibitem[Cullen and Zerner(1982)Cullen, and Zerner]{ccsd2}
Cullen,~J.~M.; Zerner,~M.~C. The Linked Singles and Doubles Model: An
  Approximate Theory of Electron Correlation Based on the Coupled-Cluster
  Ansatz. \emph{J. Chem. Phys.} \textbf{1982}, \emph{77}, 4088--4109\relax
\mciteBstWouldAddEndPuncttrue
\mciteSetBstMidEndSepPunct{\mcitedefaultmidpunct}
{\mcitedefaultendpunct}{\mcitedefaultseppunct}\relax
\EndOfBibitem
\bibitem[Scuseria \latin{et~al.}(1987)Scuseria, Scheiner, Lee, Rice, and
  Schaefer]{ccsdfritz}
Scuseria,~G.~E.; Scheiner,~A.~C.; Lee,~T.~J.; Rice,~J.~E.; Schaefer,~H.~F.,~III
  The Closed-Shell Coupled Cluster Single and Double Excitation (CCSD) Model
  for the Description of Electron Correlation. A Comparison with Configuration
  Interaction (CISD) Results. \emph{J. Chem. Phys.} \textbf{1987}, \emph{86},
  2881--2890\relax
\mciteBstWouldAddEndPuncttrue
\mciteSetBstMidEndSepPunct{\mcitedefaultmidpunct}
{\mcitedefaultendpunct}{\mcitedefaultseppunct}\relax
\EndOfBibitem
\bibitem[Piecuch and Paldus(1989)Piecuch, and Paldus]{osaccsd}
Piecuch,~P.; Paldus,~J. Orthogonally Spin-Adapted Coupled-Cluster Equations
  Involving Singly and Doubly Excited Clusters. Comparison of Different
  Procedures for Spin-Adaptation. \emph{Int. J. Quantum Chem.} \textbf{1989},
  \emph{36}, 429--453\relax
\mciteBstWouldAddEndPuncttrue
\mciteSetBstMidEndSepPunct{\mcitedefaultmidpunct}
{\mcitedefaultendpunct}{\mcitedefaultseppunct}\relax
\EndOfBibitem
\bibitem[Hoffmann and Schaefer(1986)Hoffmann, and Schaefer]{ccsdt1986}
Hoffmann,~M.~R.; Schaefer,~H.~F.,~III A Full Coupled-Cluster Singles, Doubles,
  and Triples Model for the Description of Electron Correlation. \emph{Adv.
  Quantum Chem.} \textbf{1986}, \emph{18}, 207--279\relax
\mciteBstWouldAddEndPuncttrue
\mciteSetBstMidEndSepPunct{\mcitedefaultmidpunct}
{\mcitedefaultendpunct}{\mcitedefaultseppunct}\relax
\EndOfBibitem
\bibitem[Noga and Bartlett(1987)Noga, and Bartlett]{ccfullt}
Noga,~J.; Bartlett,~R.~J. The Full CCSDT Model for Molecular Electronic
  Structure. \emph{J. Chem. Phys.} \textbf{1987}, \emph{86}, 7041--7050, {\bf
  1988}, {\it 89}, 3401 [Erratum]\relax
\mciteBstWouldAddEndPuncttrue
\mciteSetBstMidEndSepPunct{\mcitedefaultmidpunct}
{\mcitedefaultendpunct}{\mcitedefaultseppunct}\relax
\EndOfBibitem
\bibitem[Scuseria and Schaefer(1988)Scuseria, and Schaefer]{ccfullt2}
Scuseria,~G.~E.; Schaefer,~H.~F.,~III A New Implementation of the Full CCSDT
  Model for Molecular Electronic Structure. \emph{Chem. Phys. Lett.}
  \textbf{1988}, \emph{152}, 382--386\relax
\mciteBstWouldAddEndPuncttrue
\mciteSetBstMidEndSepPunct{\mcitedefaultmidpunct}
{\mcitedefaultendpunct}{\mcitedefaultseppunct}\relax
\EndOfBibitem
\bibitem[Watts and Bartlett(1990)Watts, and Bartlett]{ch2-bartlett2}
Watts,~J.~D.; Bartlett,~R.~J. The Coupled-Cluster Single, Double, and Triple
  Excitation Model for Open-Shell Single Reference Functions. \emph{J. Chem.
  Phys.} \textbf{1990}, \emph{93}, 6104--6105\relax
\mciteBstWouldAddEndPuncttrue
\mciteSetBstMidEndSepPunct{\mcitedefaultmidpunct}
{\mcitedefaultendpunct}{\mcitedefaultseppunct}\relax
\EndOfBibitem
\bibitem[Oliphant and Adamowicz(1991)Oliphant, and Adamowicz]{ccsdtq0}
Oliphant,~N.; Adamowicz,~L. Coupled-Cluster Method Truncated at Quadruples.
  \emph{J. Chem. Phys.} \textbf{1991}, \emph{95}, 6645--6651\relax
\mciteBstWouldAddEndPuncttrue
\mciteSetBstMidEndSepPunct{\mcitedefaultmidpunct}
{\mcitedefaultendpunct}{\mcitedefaultseppunct}\relax
\EndOfBibitem
\bibitem[Kucharski and Bartlett(1991)Kucharski, and Bartlett]{ccsdtq1}
Kucharski,~S.~A.; Bartlett,~R.~J. Recursive Intermediate Factorization and
  Complete Computational Linearization of the Coupled-Cluster Single, Double,
  Triple, and Quadruple Excitation Equations. \emph{Theor. Chim. Acta}
  \textbf{1991}, \emph{80}, 387--405\relax
\mciteBstWouldAddEndPuncttrue
\mciteSetBstMidEndSepPunct{\mcitedefaultmidpunct}
{\mcitedefaultendpunct}{\mcitedefaultseppunct}\relax
\EndOfBibitem
\bibitem[Kucharski and Bartlett(1992)Kucharski, and Bartlett]{ccsdtq2}
Kucharski,~S.~A.; Bartlett,~R.~J. The Coupled-Cluster Single, Double, Triple,
  and Quadruple Excitation Method. \emph{J. Chem. Phys.} \textbf{1992},
  \emph{97}, 4282--4288\relax
\mciteBstWouldAddEndPuncttrue
\mciteSetBstMidEndSepPunct{\mcitedefaultmidpunct}
{\mcitedefaultendpunct}{\mcitedefaultseppunct}\relax
\EndOfBibitem
\bibitem[Piecuch and Adamowicz(1994)Piecuch, and Adamowicz]{ccsdtq3}
Piecuch,~P.; Adamowicz,~L. State-Selective Multireference Coupled-Cluster
  Theory Employing the Single-Reference Formalism: Implementation and
  Application to the H$_8$ Model System. \emph{J. Chem. Phys.} \textbf{1994},
  \emph{100}, 5792--5809\relax
\mciteBstWouldAddEndPuncttrue
\mciteSetBstMidEndSepPunct{\mcitedefaultmidpunct}
{\mcitedefaultendpunct}{\mcitedefaultseppunct}\relax
\EndOfBibitem
\bibitem[Szalay and Bartlett(1993)Szalay, and Bartlett]{mraqcc1}
Szalay,~P.~G.; Bartlett,~R.~J. Multi-Reference Averaged Quadratic
  Coupled-Cluster Method: A Size-Extensive Modification of Multi-Reference CI.
  \emph{Chem. Phys. Lett.} \textbf{1993}, \emph{214}, 481--488\relax
\mciteBstWouldAddEndPuncttrue
\mciteSetBstMidEndSepPunct{\mcitedefaultmidpunct}
{\mcitedefaultendpunct}{\mcitedefaultseppunct}\relax
\EndOfBibitem
\bibitem[Szalay and Bartlett(1995)Szalay, and Bartlett]{mraqcc2}
Szalay,~P.~G.; Bartlett,~R.~J. Approximately Extensive Modifications of the
  Multireference Configuration Interaction Method: A Theoretical and Practical
  Analysis. \emph{J. Chem. Phys.} \textbf{1995}, \emph{103}, 3600--3612\relax
\mciteBstWouldAddEndPuncttrue
\mciteSetBstMidEndSepPunct{\mcitedefaultmidpunct}
{\mcitedefaultendpunct}{\mcitedefaultseppunct}\relax
\EndOfBibitem
\bibitem[Huron \latin{et~al.}(1973)Huron, Malrieu, and Rancurel]{sci_3}
Huron,~B.; Malrieu,~J.~P.; Rancurel,~P. Iterative Perturbation Calculations of
  Ground and Excited State Energies from Multiconfigurational Zeroth-Order
  Wavefunctions. \emph{J. Chem. Phys.} \textbf{1973}, \emph{58},
  5745--5759\relax
\mciteBstWouldAddEndPuncttrue
\mciteSetBstMidEndSepPunct{\mcitedefaultmidpunct}
{\mcitedefaultendpunct}{\mcitedefaultseppunct}\relax
\EndOfBibitem
\bibitem[Garniron \latin{et~al.}(2017)Garniron, Scemama, Loos, and
  Caffarel]{cipsi_1}
Garniron,~Y.; Scemama,~A.; Loos,~P.-F.; Caffarel,~M. Hybrid
  Stochastic-Deterministic Calculation of the Second-Order Perturbative
  Contribution of Multireference Perturbation Theory. \emph{J. Chem. Phys.}
  \textbf{2017}, \emph{147}, 034101\relax
\mciteBstWouldAddEndPuncttrue
\mciteSetBstMidEndSepPunct{\mcitedefaultmidpunct}
{\mcitedefaultendpunct}{\mcitedefaultseppunct}\relax
\EndOfBibitem
\bibitem[Garniron \latin{et~al.}(2019)Garniron, Applencourt, Gasperich, Benali,
  Fert\'{e}, Paquier, Pradines, Assaraf, Reinhardt, Toulouse, Barbaresco,
  Renon, David, Malrieu, V\'{e}ril, Caffarel, Loos, Giner, and
  Scemama]{cipsi_2}
Garniron,~Y.; Applencourt,~T.; Gasperich,~K.; Benali,~A.; Fert\'{e},~A.;
  Paquier,~J.; Pradines,~B.; Assaraf,~R.; Reinhardt,~P.; Toulouse,~J.;
  Barbaresco,~P.; Renon,~N.; David,~G.; Malrieu,~J.-P.; V\'{e}ril,~M.;
  Caffarel,~M.; Loos,~P.-F.; Giner,~E.; Scemama,~A. Quantum Package 2.0: An
  Open-Source Determinant-Driven Suite of Programs. \emph{J. Chem. Theory
  Comput.} \textbf{2019}, \emph{15}, 3591--3609\relax
\mciteBstWouldAddEndPuncttrue
\mciteSetBstMidEndSepPunct{\mcitedefaultmidpunct}
{\mcitedefaultendpunct}{\mcitedefaultseppunct}\relax
\EndOfBibitem
\bibitem[Musia{\l} \latin{et~al.}(2011)Musia{\l}, Perera, and Bartlett]{dipea5}
Musia{\l},~M.; Perera,~A.; Bartlett,~R.~J. Multireference Coupled-Cluster
  Theory: The Easy Way. \emph{J. Chem. Phys.} \textbf{2011}, \emph{134},
  114108\relax
\mciteBstWouldAddEndPuncttrue
\mciteSetBstMidEndSepPunct{\mcitedefaultmidpunct}
{\mcitedefaultendpunct}{\mcitedefaultseppunct}\relax
\EndOfBibitem
\bibitem[Musia{\l} \latin{et~al.}(2011)Musia{\l}, Kucharski, and
  Bartlett]{dipea6}
Musia{\l},~M.; Kucharski,~S.~A.; Bartlett,~R.~J. Multireference Double Electron
  Attached Coupled Cluster Method with Full Inclusion of the Connected Triple
  Excitations: {MR-DA-CCSDT}. \emph{J. Chem. Theory Comput.} \textbf{2011},
  \emph{7}, 3088--3096\relax
\mciteBstWouldAddEndPuncttrue
\mciteSetBstMidEndSepPunct{\mcitedefaultmidpunct}
{\mcitedefaultendpunct}{\mcitedefaultseppunct}\relax
\EndOfBibitem
\bibitem[Shen and Piecuch(2013)Shen, and Piecuch]{jspp-dea-dip-2013}
Shen,~J.; Piecuch,~P. Doubly Electron-Attached and Doubly Ionized
  Equation-of-Motion Coupled-Cluster Methods with 4-Particle--2-Hole and
  4-Hole--2-Particle Excitations and Their Active-Space Extensions. \emph{J.
  Chem. Phys.} \textbf{2013}, \emph{138}, 194102\relax
\mciteBstWouldAddEndPuncttrue
\mciteSetBstMidEndSepPunct{\mcitedefaultmidpunct}
{\mcitedefaultendpunct}{\mcitedefaultseppunct}\relax
\EndOfBibitem
\bibitem[Shen and Piecuch(2014)Shen, and Piecuch]{jspp-dea-dip-2014}
Shen,~J.; Piecuch,~P. Doubly Electron-Attached and Doubly Ionised
  Equation-of-Motion Coupled-Cluster Methods with Full and Active-Space
  Treatments of 4-Particle--2-Hole and 4-Hole--2-Particle Excitations: The Role
  of Orbital Choices. \emph{Mol. Phys.} \textbf{2014}, \emph{112},
  868--885\relax
\mciteBstWouldAddEndPuncttrue
\mciteSetBstMidEndSepPunct{\mcitedefaultmidpunct}
{\mcitedefaultendpunct}{\mcitedefaultseppunct}\relax
\EndOfBibitem
\bibitem[Dunning(1989)]{ccpvnz}
Dunning,~T.~H.,~Jr. Gaussian Basis Sets for Use in Correlated Molecular
  Calculations. I. The Atoms Boron Through Neon and Hydrogen. \emph{J. Chem.
  Phys.} \textbf{1989}, \emph{90}, 1007--1023\relax
\mciteBstWouldAddEndPuncttrue
\mciteSetBstMidEndSepPunct{\mcitedefaultmidpunct}
{\mcitedefaultendpunct}{\mcitedefaultseppunct}\relax
\EndOfBibitem
\bibitem[Raghavachari \latin{et~al.}(1989)Raghavachari, Trucks, Pople, and
  Head-Gordon]{ccsdpt}
Raghavachari,~K.; Trucks,~G.~W.; Pople,~J.~A.; Head-Gordon,~M. A Fifth-Order
  Perturbation Comparison of Electron Correlation Theories. \emph{Chem. Phys.
  Lett.} \textbf{1989}, \emph{157}, 479--483\relax
\mciteBstWouldAddEndPuncttrue
\mciteSetBstMidEndSepPunct{\mcitedefaultmidpunct}
{\mcitedefaultendpunct}{\mcitedefaultseppunct}\relax
\EndOfBibitem
\bibitem[Watts \latin{et~al.}(1993)Watts, Gauss, and
  Bartlett]{watts-gauss-bartlett-1993}
Watts,~J.~D.; Gauss,~J.; Bartlett,~R.~J. Coupled-Cluster Methods with
  Noniterative Triple Excitations for Restricted Open-Shell Hartree--Fock and
  Other General Single Determinant Reference Functions. Energies and Analytical
  Gradients. \emph{J. Chem. Phys.} \textbf{1993}, \emph{98}, 8718--8733\relax
\mciteBstWouldAddEndPuncttrue
\mciteSetBstMidEndSepPunct{\mcitedefaultmidpunct}
{\mcitedefaultendpunct}{\mcitedefaultseppunct}\relax
\EndOfBibitem
\bibitem[Piecuch and W{\l}och(2005)Piecuch, and W{\l}och]{crccl_jcp}
Piecuch,~P.; W{\l}och,~M. Renormalized Coupled-Cluster Methods Exploiting Left
  Eigenstates of the Similarity-Transformed Hamiltonian. \emph{J. Chem. Phys.}
  \textbf{2005}, \emph{123}, 224105\relax
\mciteBstWouldAddEndPuncttrue
\mciteSetBstMidEndSepPunct{\mcitedefaultmidpunct}
{\mcitedefaultendpunct}{\mcitedefaultseppunct}\relax
\EndOfBibitem
\bibitem[Piecuch \latin{et~al.}(2006)Piecuch, W{\l}och, Gour, and
  Kinal]{crccl_cpl}
Piecuch,~P.; W{\l}och,~M.; Gour,~J.~R.; Kinal,~A. Single-Reference,
  Size-Extensive, Non-Iterative Coupled-Cluster Approaches to Bond Breaking and
  Biradicals. \emph{Chem. Phys. Lett.} \textbf{2006}, \emph{418},
  467--474\relax
\mciteBstWouldAddEndPuncttrue
\mciteSetBstMidEndSepPunct{\mcitedefaultmidpunct}
{\mcitedefaultendpunct}{\mcitedefaultseppunct}\relax
\EndOfBibitem
\bibitem[W{\l}och \latin{et~al.}(2006)W{\l}och, Lodriguito, Piecuch, and
  Gour]{crccl_molphys}
W{\l}och,~M.; Lodriguito,~M.~D.; Piecuch,~P.; Gour,~J.~R. Two New Classes of
  Non-Iterative Coupled-Cluster Methods Derived from the Method of Moments of
  Coupled-Cluster Equations. \emph{Mol. Phys.} \textbf{2006}, \emph{104},
  2149--2172, {\bf 2006}, {\it 104}, 2991 [Erratum]\relax
\mciteBstWouldAddEndPuncttrue
\mciteSetBstMidEndSepPunct{\mcitedefaultmidpunct}
{\mcitedefaultendpunct}{\mcitedefaultseppunct}\relax
\EndOfBibitem
\bibitem[Stanton(1997)]{stanton1997}
Stanton,~J.~F. Why CCSD(T) Works: A Different Perspective. \emph{Chem. Phys.
  Lett.} \textbf{1997}, \emph{281}, 130--134\relax
\mciteBstWouldAddEndPuncttrue
\mciteSetBstMidEndSepPunct{\mcitedefaultmidpunct}
{\mcitedefaultendpunct}{\mcitedefaultseppunct}\relax
\EndOfBibitem
\bibitem[Crawford and Stanton(1998)Crawford, and Stanton]{crawford1998}
Crawford,~T.~D.; Stanton,~J.~F. Investigation of an Asymmetric
  Triple-Excitation Correction for Coupled-Cluster Energies. \emph{Int. J.
  Quantum Chem.} \textbf{1998}, \emph{70}, 601--611\relax
\mciteBstWouldAddEndPuncttrue
\mciteSetBstMidEndSepPunct{\mcitedefaultmidpunct}
{\mcitedefaultendpunct}{\mcitedefaultseppunct}\relax
\EndOfBibitem
\bibitem[Kucharski and Bartlett(1998)Kucharski, and Bartlett]{ref:26}
Kucharski,~S.~A.; Bartlett,~R.~J. Noniterative Energy Corrections Through
  Fifth-Order to the Coupled Cluster Singles and Doubles Method. \emph{J. Chem.
  Phys.} \textbf{1998}, \emph{108}, 5243--5254\relax
\mciteBstWouldAddEndPuncttrue
\mciteSetBstMidEndSepPunct{\mcitedefaultmidpunct}
{\mcitedefaultendpunct}{\mcitedefaultseppunct}\relax
\EndOfBibitem
\bibitem[Hirata \latin{et~al.}(2001)Hirata, Nooijen, Grabowski, and
  Bartlett]{eomccpt}
Hirata,~S.; Nooijen,~M.; Grabowski,~I.; Bartlett,~R.~J. Perturbative
  Corrections to Coupled-Cluster and Equation-of-Motion Coupled-Cluster
  Energies: A Determinantal Analysis. \emph{J. Chem. Phys.} \textbf{2001},
  \emph{114}, 3919--3928, {\bf 2001} {\it 115}, 3967 [Erratum]\relax
\mciteBstWouldAddEndPuncttrue
\mciteSetBstMidEndSepPunct{\mcitedefaultmidpunct}
{\mcitedefaultendpunct}{\mcitedefaultseppunct}\relax
\EndOfBibitem
\bibitem[Hirata \latin{et~al.}(2004)Hirata, Fan, Auer, Nooijen, and
  Piecuch]{ccsdpt2}
Hirata,~S.; Fan,~P.-D.; Auer,~A.~A.; Nooijen,~M.; Piecuch,~P. Combined
  Coupled-Cluster and Many-Body Perturbation Theories. \emph{J. Chem. Phys.}
  \textbf{2004}, \emph{121}, 12197--12207\relax
\mciteBstWouldAddEndPuncttrue
\mciteSetBstMidEndSepPunct{\mcitedefaultmidpunct}
{\mcitedefaultendpunct}{\mcitedefaultseppunct}\relax
\EndOfBibitem
\bibitem[Gwaltney and Head-Gordon(2000)Gwaltney, and Head-Gordon]{gwaltney1}
Gwaltney,~S.~R.; Head-Gordon,~M. A Second-Order Correction to Singles and
  Doubles Coupled-Cluster Methods Based on a Perturbative Expansion of a
  Similarity-Transformed Hamiltonian. \emph{Chem. Phys. Lett.} \textbf{2000},
  \emph{323}, 21--28\relax
\mciteBstWouldAddEndPuncttrue
\mciteSetBstMidEndSepPunct{\mcitedefaultmidpunct}
{\mcitedefaultendpunct}{\mcitedefaultseppunct}\relax
\EndOfBibitem
\bibitem[Gwaltney and Head-Gordon(2001)Gwaltney, and Head-Gordon]{gwaltney3}
Gwaltney,~S.~R.; Head-Gordon,~M. A Second-Order Perturbative Correction to the
  Coupled-Cluster Singles and Doubles Method: CCSD(2). \emph{J. Chem. Phys.}
  \textbf{2001}, \emph{115}, 2014--2021\relax
\mciteBstWouldAddEndPuncttrue
\mciteSetBstMidEndSepPunct{\mcitedefaultmidpunct}
{\mcitedefaultendpunct}{\mcitedefaultseppunct}\relax
\EndOfBibitem
\bibitem[Piecuch and Kowalski(2000)Piecuch, and Kowalski]{leszcz}
Piecuch,~P.; Kowalski,~K. In \emph{Computational Chemistry: Reviews of Current
  Trends}; Leszczy{\' n}ski,~J., Ed.; World Scientific: Singapore, 2000;
  Vol.~5; pp 1--104\relax
\mciteBstWouldAddEndPuncttrue
\mciteSetBstMidEndSepPunct{\mcitedefaultmidpunct}
{\mcitedefaultendpunct}{\mcitedefaultseppunct}\relax
\EndOfBibitem
\bibitem[Kowalski and Piecuch(2000)Kowalski, and Piecuch]{ren1}
Kowalski,~K.; Piecuch,~P. The Method of Moments of Coupled-Cluster Equations
  and the Renormalized CCSD[T], CCSD(T), CCSD(TQ), and CCSDT(Q) Approaches.
  \emph{J. Chem. Phys.} \textbf{2000}, \emph{113}, 18--35\relax
\mciteBstWouldAddEndPuncttrue
\mciteSetBstMidEndSepPunct{\mcitedefaultmidpunct}
{\mcitedefaultendpunct}{\mcitedefaultseppunct}\relax
\EndOfBibitem
\bibitem[Piecuch \latin{et~al.}(2002)Piecuch, Kowalski, Pimienta, and
  McGuire]{irpc}
Piecuch,~P.; Kowalski,~K.; Pimienta,~I. S.~O.; McGuire,~M.~J. Recent Advances
  in Electronic Structure Theory: Method of Moments of Coupled-Cluster
  Equations and Renormalized Coupled-Cluster Approaches. \emph{Int. Rev. Phys.
  Chem.} \textbf{2002}, \emph{21}, 527--655\relax
\mciteBstWouldAddEndPuncttrue
\mciteSetBstMidEndSepPunct{\mcitedefaultmidpunct}
{\mcitedefaultendpunct}{\mcitedefaultseppunct}\relax
\EndOfBibitem
\bibitem[Piecuch \latin{et~al.}(2004)Piecuch, Kowalski, Pimienta, Fan,
  Lodriguito, McGuire, Kucharski, Ku{\' s}, and Musia{\l}]{PP:TCA}
Piecuch,~P.; Kowalski,~K.; Pimienta,~I. S.~O.; Fan,~P.-D.; Lodriguito,~M.;
  McGuire,~M.~J.; Kucharski,~S.~A.; Ku{\' s},~T.; Musia{\l},~M. Method of
  Moments of Coupled-Cluster Equations: A New Formalism for Designing Accurate
  Electronic Structure Methods for Ground and Excited States. \emph{Theor.
  Chem. Acc.} \textbf{2004}, \emph{112}, 349--393\relax
\mciteBstWouldAddEndPuncttrue
\mciteSetBstMidEndSepPunct{\mcitedefaultmidpunct}
{\mcitedefaultendpunct}{\mcitedefaultseppunct}\relax
\EndOfBibitem
\bibitem[Kowalski and Piecuch(2005)Kowalski, and Piecuch]{ndcmmcc}
Kowalski,~K.; Piecuch,~P. Extensive Generalization of Renormalized
  Coupled-Cluster Methods. \emph{J. Chem. Phys.} \textbf{2005}, \emph{122},
  074107\relax
\mciteBstWouldAddEndPuncttrue
\mciteSetBstMidEndSepPunct{\mcitedefaultmidpunct}
{\mcitedefaultendpunct}{\mcitedefaultseppunct}\relax
\EndOfBibitem
\bibitem[Taube and Bartlett(2008)Taube, and Bartlett]{bartlett2008a}
Taube,~A.~G.; Bartlett,~R.~J. Improving Upon CCSD(T): $\Lambda$CCSD(T). I.
  Potential Energy Surfaces. \emph{J. Chem. Phys.} \textbf{2008}, \emph{128},
  044110\relax
\mciteBstWouldAddEndPuncttrue
\mciteSetBstMidEndSepPunct{\mcitedefaultmidpunct}
{\mcitedefaultendpunct}{\mcitedefaultseppunct}\relax
\EndOfBibitem
\bibitem[Taube and Bartlett(2008)Taube, and Bartlett]{bartlett2008b}
Taube,~A.~G.; Bartlett,~R.~J. Improving Upon CCSD(T): $\Lambda$CCSD(T). II.
  Stationary Formulation and Derivatives. \emph{J. Chem. Phys.} \textbf{2008},
  \emph{128}, 044111\relax
\mciteBstWouldAddEndPuncttrue
\mciteSetBstMidEndSepPunct{\mcitedefaultmidpunct}
{\mcitedefaultendpunct}{\mcitedefaultseppunct}\relax
\EndOfBibitem
\bibitem[Eriksen \latin{et~al.}(2014)Eriksen, Kristensen, Kj{\ae}rgaard,
  J{\o}rgensen, and Gauss]{eriksen1}
Eriksen,~J.~J.; Kristensen,~K.; Kj{\ae}rgaard,~T.; J{\o}rgensen,~P.; Gauss,~J.
  A Lagrangian Framework for Deriving Triples and Quadruples Corrections to the
  CCSD Energy. \emph{J. Chem. Phys.} \textbf{2014}, \emph{140}, 064108\relax
\mciteBstWouldAddEndPuncttrue
\mciteSetBstMidEndSepPunct{\mcitedefaultmidpunct}
{\mcitedefaultendpunct}{\mcitedefaultseppunct}\relax
\EndOfBibitem
\bibitem[Eriksen \latin{et~al.}(2014)Eriksen, J{\o}rgensen, Olsen, and
  Gauss]{eriksen2}
Eriksen,~J.~J.; J{\o}rgensen,~P.; Olsen,~J.; Gauss,~J. Equation-of-Motion
  Coupled Cluster Perturbation Theory Revisited. \emph{J. Chem. Phys.}
  \textbf{2014}, \emph{140}, 174114\relax
\mciteBstWouldAddEndPuncttrue
\mciteSetBstMidEndSepPunct{\mcitedefaultmidpunct}
{\mcitedefaultendpunct}{\mcitedefaultseppunct}\relax
\EndOfBibitem
\bibitem[Shen and Piecuch(2012)Shen, and Piecuch]{jspp-chemphys2012}
Shen,~J.; Piecuch,~P. Biorthogonal Moment Expansions in Coupled-Cluster Theory:
  Review of Key Concepts and Merging the Renormalized and Active-Space
  Coupled-Cluster Methods. \emph{Chem. Phys.} \textbf{2012}, \emph{401},
  180--202\relax
\mciteBstWouldAddEndPuncttrue
\mciteSetBstMidEndSepPunct{\mcitedefaultmidpunct}
{\mcitedefaultendpunct}{\mcitedefaultseppunct}\relax
\EndOfBibitem
\bibitem[Piecuch \latin{et~al.}(2007)Piecuch, W{\l}och, and Varandas]{ptcp2007}
Piecuch,~P.; W{\l}och,~M.; Varandas,~A. J.~C. In \emph{Topics in the Theory of
  Chemical and Physical Systems}; Lahmar,~S., Maruani,~J., Wilson,~S.,
  Delgado-Barrio,~G., Eds.; Progress in Theoretical Chemistry and Physics;
  Springer: Dordrecht, 2007; Vol.~16; pp 63--121\relax
\mciteBstWouldAddEndPuncttrue
\mciteSetBstMidEndSepPunct{\mcitedefaultmidpunct}
{\mcitedefaultendpunct}{\mcitedefaultseppunct}\relax
\EndOfBibitem
\bibitem[Ge \latin{et~al.}(2007)Ge, Gordon, and Piecuch]{ge1}
Ge,~Y.; Gordon,~M.~S.; Piecuch,~P. Breaking Bonds with the Left Eigenstate
  Completely Renormalized Coupled-Cluster Method. \emph{J. Chem. Phys.}
  \textbf{2007}, \emph{127}, 174106\relax
\mciteBstWouldAddEndPuncttrue
\mciteSetBstMidEndSepPunct{\mcitedefaultmidpunct}
{\mcitedefaultendpunct}{\mcitedefaultseppunct}\relax
\EndOfBibitem
\bibitem[Ge \latin{et~al.}(2008)Ge, Gordon, Piecuch, W{\l}och, and Gour]{ge2}
Ge,~Y.; Gordon,~M.~S.; Piecuch,~P.; W{\l}och,~M.; Gour,~J.~R. Breaking Bonds of
  Open-Shell Species with the Restricted Open-Shell Size Extensive Left
  Eigenstate Completely Renormalized Coupled-Cluster Method. \emph{J. Phys.
  Chem. A} \textbf{2008}, \emph{112}, 11873--11884\relax
\mciteBstWouldAddEndPuncttrue
\mciteSetBstMidEndSepPunct{\mcitedefaultmidpunct}
{\mcitedefaultendpunct}{\mcitedefaultseppunct}\relax
\EndOfBibitem
\bibitem[Shen and Piecuch(2012)Shen, and Piecuch]{jspp-jcp2012}
Shen,~J.; Piecuch,~P. Combining Active-Space Coupled-Cluster Methods with
  Moment Energy Corrections via the CC($P$;$Q$) Methodology, with Benchmark
  Calculations for Biradical Transition States. \emph{J. Chem. Phys.}
  \textbf{2012}, \emph{136}, 144104\relax
\mciteBstWouldAddEndPuncttrue
\mciteSetBstMidEndSepPunct{\mcitedefaultmidpunct}
{\mcitedefaultendpunct}{\mcitedefaultseppunct}\relax
\EndOfBibitem
\bibitem[Piecuch \latin{et~al.}(2006)Piecuch, W{\l}och, Lodriguito, and
  Gour]{ptcp2006}
Piecuch,~P.; W{\l}och,~M.; Lodriguito,~M.; Gour,~J.~R. In \emph{Recent Advances
  in the Theory of Chemical and Physical Systems}; Julien,~J.-P., Maruani,~J.,
  Mayou,~D., Wilson,~S., Delgado-Barrio,~G., Eds.; Progress in Theoretical
  Chemistry and Physics; Springer: Dordrecht, 2006; Vol.~15; pp 45--106\relax
\mciteBstWouldAddEndPuncttrue
\mciteSetBstMidEndSepPunct{\mcitedefaultmidpunct}
{\mcitedefaultendpunct}{\mcitedefaultseppunct}\relax
\EndOfBibitem
\bibitem[Piecuch \latin{et~al.}(2009)Piecuch, Gour, and W{\l}och]{crccl_ijqc2}
Piecuch,~P.; Gour,~J.~R.; W{\l}och,~M. Left-Eigenstate Completely Renormalized
  Equation-of-Motion Coupled-Cluster Methods: Review of Key Concepts, Extension
  to Excited States of Open-Shell Systems, and Comparison with
  Electron-Attached and Ionized Approaches. \emph{Int. J. Quantum Chem.}
  \textbf{2009}, \emph{109}, 3268--3304\relax
\mciteBstWouldAddEndPuncttrue
\mciteSetBstMidEndSepPunct{\mcitedefaultmidpunct}
{\mcitedefaultendpunct}{\mcitedefaultseppunct}\relax
\EndOfBibitem
\bibitem[Fradelos \latin{et~al.}(2011)Fradelos, Lutz, Weso{\l}owski, Piecuch,
  and W{\l}och]{7hq}
Fradelos,~G.; Lutz,~J.~J.; Weso{\l}owski,~T.~A.; Piecuch,~P.; W{\l}och,~M.
  Embedding vs Supermolecular Strategies in Evaluating the
  Hydrogen-Bonding-Induced Shifts of Excitation Energies. \emph{J. Chem. Theory
  Comput.} \textbf{2011}, \emph{7}, 1647--1666\relax
\mciteBstWouldAddEndPuncttrue
\mciteSetBstMidEndSepPunct{\mcitedefaultmidpunct}
{\mcitedefaultendpunct}{\mcitedefaultseppunct}\relax
\EndOfBibitem
\bibitem[Piecuch \latin{et~al.}(2008)Piecuch, W{\l}och, and Varandas]{msg65}
Piecuch,~P.; W{\l}och,~M.; Varandas,~A. J.~C. Application of Renormalized
  Coupled-Cluster Methods to Potential Function of Water. \emph{Theor. Chem.
  Acc.} \textbf{2008}, \emph{120}, 59--78\relax
\mciteBstWouldAddEndPuncttrue
\mciteSetBstMidEndSepPunct{\mcitedefaultmidpunct}
{\mcitedefaultendpunct}{\mcitedefaultseppunct}\relax
\EndOfBibitem
\bibitem[Horoi \latin{et~al.}(2007)Horoi, Gour, W{\l}och, Lodriguito, Brown,
  and Piecuch]{nuclei8}
Horoi,~M.; Gour,~J.~R.; W{\l}och,~M.; Lodriguito,~M.~D.; Brown,~B.~A.;
  Piecuch,~P. Coupled-Cluster and Configuration-Interaction Calculations for
  Heavy Nuclei. \emph{Phys. Rev. Lett.} \textbf{2007}, \emph{98}, 112501\relax
\mciteBstWouldAddEndPuncttrue
\mciteSetBstMidEndSepPunct{\mcitedefaultmidpunct}
{\mcitedefaultendpunct}{\mcitedefaultseppunct}\relax
\EndOfBibitem
\bibitem[Bauman \latin{et~al.}(2017)Bauman, Shen, and
  Piecuch]{nbjspp-molphys2017}
Bauman,~N.~P.; Shen,~J.; Piecuch,~P. Combining Active-Space Coupled-Cluster
  Approaches with Moment Energy Corrections via the CC($P$;$Q$) Methodology:
  Connected Quadruple Excitations. \emph{Mol. Phys.} \textbf{2017}, \emph{115},
  2860--2891\relax
\mciteBstWouldAddEndPuncttrue
\mciteSetBstMidEndSepPunct{\mcitedefaultmidpunct}
{\mcitedefaultendpunct}{\mcitedefaultseppunct}\relax
\EndOfBibitem
\bibitem[Magoulas \latin{et~al.}(2018)Magoulas, Bauman, Shen, and
  Piecuch]{ccpq-be2-jpca-2018}
Magoulas,~I.; Bauman,~N.~P.; Shen,~J.; Piecuch,~P. Application of the
  CC($P$;$Q$) Hierarchy of Coupled-Cluster Methods to the Beryllium Dimer.
  \emph{J. Phys. Chem. A} \textbf{2018}, \emph{122}, 1350--1368\relax
\mciteBstWouldAddEndPuncttrue
\mciteSetBstMidEndSepPunct{\mcitedefaultmidpunct}
{\mcitedefaultendpunct}{\mcitedefaultseppunct}\relax
\EndOfBibitem
\bibitem[Emrich(1981)]{emrich}
Emrich,~K. An Extension of the Coupled Cluster Formalism to Excited States (I).
  \emph{Nucl. Phys. A} \textbf{1981}, \emph{351}, 379--396\relax
\mciteBstWouldAddEndPuncttrue
\mciteSetBstMidEndSepPunct{\mcitedefaultmidpunct}
{\mcitedefaultendpunct}{\mcitedefaultseppunct}\relax
\EndOfBibitem
\bibitem[Emrich(1981)]{emrich2}
Emrich,~K. An Extension of the Coupled Cluster Formalism to Excited States:
  (II). Approximations and Tests. \emph{Nucl. Phys. A} \textbf{1981},
  \emph{351}, 397--438\relax
\mciteBstWouldAddEndPuncttrue
\mciteSetBstMidEndSepPunct{\mcitedefaultmidpunct}
{\mcitedefaultendpunct}{\mcitedefaultseppunct}\relax
\EndOfBibitem
\bibitem[Geertsen \latin{et~al.}(1989)Geertsen, Rittby, and Bartlett]{eomcc1}
Geertsen,~J.; Rittby,~M.; Bartlett,~R.~J. The Equation-of-Motion
  Coupled-Cluster Method: Excitation Energies of {Be} and {CO}. \emph{Chem.
  Phys. Lett.} \textbf{1989}, \emph{164}, 57--62\relax
\mciteBstWouldAddEndPuncttrue
\mciteSetBstMidEndSepPunct{\mcitedefaultmidpunct}
{\mcitedefaultendpunct}{\mcitedefaultseppunct}\relax
\EndOfBibitem
\bibitem[Stanton and Bartlett(1993)Stanton, and Bartlett]{eomcc3}
Stanton,~J.~F.; Bartlett,~R.~J. The Equation of Motion Coupled-Cluster Method.
  {A} Systematic Biorthogonal Approach to Molecular Excitation Energies,
  Transition Probabilities, and Excited State Properties. \emph{J. Chem. Phys.}
  \textbf{1993}, \emph{98}, 7029--7039\relax
\mciteBstWouldAddEndPuncttrue
\mciteSetBstMidEndSepPunct{\mcitedefaultmidpunct}
{\mcitedefaultendpunct}{\mcitedefaultseppunct}\relax
\EndOfBibitem
\bibitem[Yuwono \latin{et~al.}(2019)Yuwono, Magoulas, Shen, and
  Piecuch]{ccpq-mg2-mp-2019}
Yuwono,~S.~H.; Magoulas,~I.; Shen,~J.; Piecuch,~P. Application of the
  Coupled-Cluster CC($P$;$Q$) Approaches to the Magnesium Dimer. \emph{Mol.
  Phys.} \textbf{2019}, \emph{117}, 1486--1506\relax
\mciteBstWouldAddEndPuncttrue
\mciteSetBstMidEndSepPunct{\mcitedefaultmidpunct}
{\mcitedefaultendpunct}{\mcitedefaultseppunct}\relax
\EndOfBibitem
\bibitem[Deustua \latin{et~al.}(2017)Deustua, Shen, and
  Piecuch]{stochastic-ccpq-prl-2017}
Deustua,~J.~E.; Shen,~J.; Piecuch,~P. Converging High-Level Coupled-Cluster
  Energetics by Monte Carlo Sampling and Moment Expansions. \emph{Phys. Rev.
  Lett.} \textbf{2017}, \emph{119}, 223003\relax
\mciteBstWouldAddEndPuncttrue
\mciteSetBstMidEndSepPunct{\mcitedefaultmidpunct}
{\mcitedefaultendpunct}{\mcitedefaultseppunct}\relax
\EndOfBibitem
\bibitem[Yuwono \latin{et~al.}(2020)Yuwono, Chakraborty, Deustua, Shen, and
  Piecuch]{stochastic-ccpq-molphys-2020}
Yuwono,~S.~H.; Chakraborty,~A.; Deustua,~J.~E.; Shen,~J.; Piecuch,~P.
  Accelerating Convergence of Equation-of-Motion Coupled-Cluster Computations
  Using the Semi-Stochastic CC($P$;$Q$) Formalism. \emph{Mol. Phys.}
  \textbf{2020}, \emph{118}, e1817592\relax
\mciteBstWouldAddEndPuncttrue
\mciteSetBstMidEndSepPunct{\mcitedefaultmidpunct}
{\mcitedefaultendpunct}{\mcitedefaultseppunct}\relax
\EndOfBibitem
\bibitem[Deustua \latin{et~al.}(2021)Deustua, Shen, and
  Piecuch]{stochastic-ccpq-jcp-2021}
Deustua,~J.~E.; Shen,~J.; Piecuch,~P. High-Level Coupled-Cluster Energetics by
  Monte Carlo Sampling and Moment Expansions: Further Details and Comparisons.
  \emph{J. Chem. Phys.} \textbf{2021}, \emph{154}, 124103\relax
\mciteBstWouldAddEndPuncttrue
\mciteSetBstMidEndSepPunct{\mcitedefaultmidpunct}
{\mcitedefaultendpunct}{\mcitedefaultseppunct}\relax
\EndOfBibitem
\bibitem[Gururangan \latin{et~al.}(2021)Gururangan, Deustua, Shen, and
  Piecuch]{cipsi-ccpq-2021}
Gururangan,~K.; Deustua,~J.~E.; Shen,~J.; Piecuch,~P. High-Level
  Coupled-Cluster Energetics by Merging Moment Expansions with Selected
  Configuration Interaction. \emph{J. Chem. Phys.} \textbf{2021}, \emph{155},
  174114\relax
\mciteBstWouldAddEndPuncttrue
\mciteSetBstMidEndSepPunct{\mcitedefaultmidpunct}
{\mcitedefaultendpunct}{\mcitedefaultseppunct}\relax
\EndOfBibitem
\bibitem[Gururangan and Piecuch(2023)Gururangan, and Piecuch]{adaptiveccpq2023}
Gururangan,~K.; Piecuch,~P. Converging High-Level Coupled-Cluster Energetics
  via Adaptive Selection of Excitation Manifolds Driven by Moment Expansions.
  \emph{J. Chem. Phys.} \textbf{2023}, \emph{159}, 084108\relax
\mciteBstWouldAddEndPuncttrue
\mciteSetBstMidEndSepPunct{\mcitedefaultmidpunct}
{\mcitedefaultendpunct}{\mcitedefaultseppunct}\relax
\EndOfBibitem
\bibitem[Gururangan \latin{et~al.}(2025)Gururangan, Shen, and
  Piecuch]{adaptive-active-ccpq-2025}
Gururangan,~K.; Shen,~J.; Piecuch,~P. Extension of the Active-Orbital-Based and
  Adaptive CC($P$;$Q$) Approaches to Excited Electronic States: Application to
  Potential Cuts of Water. \emph{Chem. Phys. Lett.} \textbf{2025}, \emph{862},
  141840\relax
\mciteBstWouldAddEndPuncttrue
\mciteSetBstMidEndSepPunct{\mcitedefaultmidpunct}
{\mcitedefaultendpunct}{\mcitedefaultseppunct}\relax
\EndOfBibitem
\bibitem[Oliphant and Adamowicz(1992)Oliphant, and Adamowicz]{semi0b}
Oliphant,~N.; Adamowicz,~L. The Implementation of the Multireference
  Coupled-Cluster Method Based on the Single-Reference Formalism. \emph{J.
  Chem. Phys.} \textbf{1992}, \emph{96}, 3739--3744\relax
\mciteBstWouldAddEndPuncttrue
\mciteSetBstMidEndSepPunct{\mcitedefaultmidpunct}
{\mcitedefaultendpunct}{\mcitedefaultseppunct}\relax
\EndOfBibitem
\bibitem[Oliphant and Adamowicz(1993)Oliphant, and Adamowicz]{semi1}
Oliphant,~N.; Adamowicz,~L. Multireference Coupled Cluster Method for
  Electronic Structure of Molecules. \emph{Int. Rev. Phys. Chem.}
  \textbf{1993}, \emph{12}, 339--362\relax
\mciteBstWouldAddEndPuncttrue
\mciteSetBstMidEndSepPunct{\mcitedefaultmidpunct}
{\mcitedefaultendpunct}{\mcitedefaultseppunct}\relax
\EndOfBibitem
\bibitem[Piecuch \latin{et~al.}(1993)Piecuch, Oliphant, and Adamowicz]{semi2}
Piecuch,~P.; Oliphant,~N.; Adamowicz,~L. A State-Selective Multireference
  Coupled-Cluster Theory Employing the Single-Reference Formalism. \emph{J.
  Chem. Phys.} \textbf{1993}, \emph{99}, 1875--1900\relax
\mciteBstWouldAddEndPuncttrue
\mciteSetBstMidEndSepPunct{\mcitedefaultmidpunct}
{\mcitedefaultendpunct}{\mcitedefaultseppunct}\relax
\EndOfBibitem
\bibitem[Piecuch and Adamowicz(1995)Piecuch, and Adamowicz]{semih2o}
Piecuch,~P.; Adamowicz,~L. Breaking Bonds with the State-Selective
  Multireference Coupled-Cluster Method Employing the Single-Reference
  Formalism. \emph{J. Chem. Phys.} \textbf{1995}, \emph{102}, 898--904\relax
\mciteBstWouldAddEndPuncttrue
\mciteSetBstMidEndSepPunct{\mcitedefaultmidpunct}
{\mcitedefaultendpunct}{\mcitedefaultseppunct}\relax
\EndOfBibitem
\bibitem[Ghose \latin{et~al.}(1995)Ghose, Piecuch, and Adamowicz]{ghose}
Ghose,~K.~B.; Piecuch,~P.; Adamowicz,~L. Improved Computational Strategy for
  the State-Selective Coupled-Cluster Theory with Semi-Internal Triexcited
  Clusters: Potential Energy Surface of the HF Molecule. \emph{J. Chem. Phys.}
  \textbf{1995}, \emph{103}, 9331--9346\relax
\mciteBstWouldAddEndPuncttrue
\mciteSetBstMidEndSepPunct{\mcitedefaultmidpunct}
{\mcitedefaultendpunct}{\mcitedefaultseppunct}\relax
\EndOfBibitem
\bibitem[Adamowicz \latin{et~al.}(1998)Adamowicz, Piecuch, and Ghose]{semi3c}
Adamowicz,~L.; Piecuch,~P.; Ghose,~K.~B. The State-Selective Coupled Cluster
  Method for Quasi-Degenerate Electronic States. \emph{Mol. Phys.}
  \textbf{1998}, \emph{94}, 225--234\relax
\mciteBstWouldAddEndPuncttrue
\mciteSetBstMidEndSepPunct{\mcitedefaultmidpunct}
{\mcitedefaultendpunct}{\mcitedefaultseppunct}\relax
\EndOfBibitem
\bibitem[Piecuch \latin{et~al.}(1999)Piecuch, Kucharski, and Bartlett]{semi4}
Piecuch,~P.; Kucharski,~S.~A.; Bartlett,~R.~J. Coupled-Cluster Methods with
  Internal and Semi-Internal Triply and Quadruply Excited Clusters: {CCSDt} and
  {CCSDtq} Approaches. \emph{J. Chem. Phys.} \textbf{1999}, \emph{110},
  6103--6122\relax
\mciteBstWouldAddEndPuncttrue
\mciteSetBstMidEndSepPunct{\mcitedefaultmidpunct}
{\mcitedefaultendpunct}{\mcitedefaultseppunct}\relax
\EndOfBibitem
\bibitem[Piecuch \latin{et~al.}(1999)Piecuch, Kucharski, and {\v
  S}pirko]{semi4new}
Piecuch,~P.; Kucharski,~S.~A.; {\v S}pirko,~V. Coupled-Cluster Methods with
  Internal and Semi-Internal Triply Excited Clusters: Vibrational Spectrum of
  the {HF} Molecule. \emph{J. Chem. Phys.} \textbf{1999}, \emph{111},
  6679--6692\relax
\mciteBstWouldAddEndPuncttrue
\mciteSetBstMidEndSepPunct{\mcitedefaultmidpunct}
{\mcitedefaultendpunct}{\mcitedefaultseppunct}\relax
\EndOfBibitem
\bibitem[Piecuch(2010)]{piecuch-qtp}
Piecuch,~P. Active-Space Coupled-Cluster Methods. \emph{Mol. Phys.}
  \textbf{2010}, \emph{108}, 2987--3015\relax
\mciteBstWouldAddEndPuncttrue
\mciteSetBstMidEndSepPunct{\mcitedefaultmidpunct}
{\mcitedefaultendpunct}{\mcitedefaultseppunct}\relax
\EndOfBibitem
\bibitem[Booth \latin{et~al.}(2009)Booth, Thom, and Alavi]{Booth2009}
Booth,~G.~H.; Thom,~A. J.~W.; Alavi,~A. Fermion Monte Carlo Without Fixed
  Nodes: A Game of Life, Death, and Annihilation in Slater Determinant Space.
  \emph{J. Chem. Phys.} \textbf{2009}, \emph{131}, 054106\relax
\mciteBstWouldAddEndPuncttrue
\mciteSetBstMidEndSepPunct{\mcitedefaultmidpunct}
{\mcitedefaultendpunct}{\mcitedefaultseppunct}\relax
\EndOfBibitem
\bibitem[Cleland \latin{et~al.}(2010)Cleland, Booth, and Alavi]{Cleland2010}
Cleland,~D.; Booth,~G.~H.; Alavi,~A. Communications: Survival of the Fittest:
  Accelerating Convergence in Full Configuration-Interaction Quantum Monte
  Carlo. \emph{J. Chem. Phys.} \textbf{2010}, \emph{132}, 041103\relax
\mciteBstWouldAddEndPuncttrue
\mciteSetBstMidEndSepPunct{\mcitedefaultmidpunct}
{\mcitedefaultendpunct}{\mcitedefaultseppunct}\relax
\EndOfBibitem
\bibitem[Dobrautz \latin{et~al.}(2019)Dobrautz, Smart, and
  Alavi]{fciqmc-uga-2019}
Dobrautz,~W.; Smart,~S.~D.; Alavi,~A. Efficient Formulation of Full
  Configuration Interaction Quantum Monte Carlo in a Spin Eigenbasis via the
  Graphical Unitary Group Approach. \emph{J. Chem. Phys.} \textbf{2019},
  \emph{151}, 094104\relax
\mciteBstWouldAddEndPuncttrue
\mciteSetBstMidEndSepPunct{\mcitedefaultmidpunct}
{\mcitedefaultendpunct}{\mcitedefaultseppunct}\relax
\EndOfBibitem
\bibitem[Ghanem \latin{et~al.}(2019)Ghanem, Lozovoi, and
  Alavi]{ghanem_alavi_fciqmc_jcp_2019}
Ghanem,~K.; Lozovoi,~A.~Y.; Alavi,~A. Unbiasing the Initiator Approximation in
  Full Configuration Interaction Quantum Monte Carlo. \emph{J. Chem. Phys.}
  \textbf{2019}, \emph{151}, 224108\relax
\mciteBstWouldAddEndPuncttrue
\mciteSetBstMidEndSepPunct{\mcitedefaultmidpunct}
{\mcitedefaultendpunct}{\mcitedefaultseppunct}\relax
\EndOfBibitem
\bibitem[Ghanem \latin{et~al.}(2020)Ghanem, Guther, and
  Alavi]{ghanem_alavi_fciqmc_2020}
Ghanem,~K.; Guther,~K.; Alavi,~A. The Adaptive Shift Method in Full
  Configuration Interaction Quantum Monte Carlo: Development and Applications.
  \emph{J. Chem. Phys.} \textbf{2020}, \emph{153}, 224115\relax
\mciteBstWouldAddEndPuncttrue
\mciteSetBstMidEndSepPunct{\mcitedefaultmidpunct}
{\mcitedefaultendpunct}{\mcitedefaultseppunct}\relax
\EndOfBibitem
\bibitem[Thom(2010)]{Thom2010}
Thom,~A. J.~W. Stochastic Coupled Cluster Theory. \emph{Phys. Rev. Lett.}
  \textbf{2010}, \emph{105}, 263004\relax
\mciteBstWouldAddEndPuncttrue
\mciteSetBstMidEndSepPunct{\mcitedefaultmidpunct}
{\mcitedefaultendpunct}{\mcitedefaultseppunct}\relax
\EndOfBibitem
\bibitem[Franklin \latin{et~al.}(2016)Franklin, Spencer, Zoccante, and
  Thom]{Franklin2016}
Franklin,~R. S.~T.; Spencer,~J.~S.; Zoccante,~A.; Thom,~A. J.~W. Linked Coupled
  Cluster Monte Carlo. \emph{J. Chem. Phys.} \textbf{2016}, \emph{144},
  044111\relax
\mciteBstWouldAddEndPuncttrue
\mciteSetBstMidEndSepPunct{\mcitedefaultmidpunct}
{\mcitedefaultendpunct}{\mcitedefaultseppunct}\relax
\EndOfBibitem
\bibitem[Spencer and Thom(2016)Spencer, and Thom]{Spencer2016}
Spencer,~J.~S.; Thom,~A. J.~W. Developments in Stochastic Coupled Cluster
  Theory: The Initiator Approximation and Application to the Uniform Electron
  Gas. \emph{J. Chem. Phys.} \textbf{2016}, \emph{144}, 084108\relax
\mciteBstWouldAddEndPuncttrue
\mciteSetBstMidEndSepPunct{\mcitedefaultmidpunct}
{\mcitedefaultendpunct}{\mcitedefaultseppunct}\relax
\EndOfBibitem
\bibitem[Scott and Thom(2017)Scott, and Thom]{Scott2017}
Scott,~C. J.~C.; Thom,~A. J.~W. Stochastic Coupled Cluster Theory: Efficient
  Sampling of the Coupled Cluster Expansion. \emph{J. Chem. Phys.}
  \textbf{2017}, \emph{147}, 124105\relax
\mciteBstWouldAddEndPuncttrue
\mciteSetBstMidEndSepPunct{\mcitedefaultmidpunct}
{\mcitedefaultendpunct}{\mcitedefaultseppunct}\relax
\EndOfBibitem
\bibitem[Jankowski \latin{et~al.}(1991)Jankowski, Paldus, and Piecuch]{moments}
Jankowski,~K.; Paldus,~J.; Piecuch,~P. Method of Moments Approach and Coupled
  Cluster Theory. \emph{Theor. Chim. Acta} \textbf{1991}, \emph{80},
  223--243\relax
\mciteBstWouldAddEndPuncttrue
\mciteSetBstMidEndSepPunct{\mcitedefaultmidpunct}
{\mcitedefaultendpunct}{\mcitedefaultseppunct}\relax
\EndOfBibitem
\bibitem[Whitten and Hackmeyer(1969)Whitten, and Hackmeyer]{sci_1}
Whitten,~J.~L.; Hackmeyer,~M. Configuration Interaction Studies of Ground and
  Excited States of Polyatomic Molecules. I. The CI Formulation and Studies of
  Formaldehyde. \emph{J. Chem. Phys.} \textbf{1969}, \emph{51},
  5584--5596\relax
\mciteBstWouldAddEndPuncttrue
\mciteSetBstMidEndSepPunct{\mcitedefaultmidpunct}
{\mcitedefaultendpunct}{\mcitedefaultseppunct}\relax
\EndOfBibitem
\bibitem[Bender and Davidson(1969)Bender, and Davidson]{sci_2}
Bender,~C.~F.; Davidson,~E.~R. Studies in Configuration Interaction: The
  First-Row Diatomic Hydrides. \emph{Phys. Rev.} \textbf{1969}, \emph{183},
  23--30\relax
\mciteBstWouldAddEndPuncttrue
\mciteSetBstMidEndSepPunct{\mcitedefaultmidpunct}
{\mcitedefaultendpunct}{\mcitedefaultseppunct}\relax
\EndOfBibitem
\bibitem[Buenker and Peyerimhoff(1974)Buenker, and Peyerimhoff]{sci_4}
Buenker,~R.~J.; Peyerimhoff,~S.~D. Individualized Configuration Selection in CI
  Calculations with Subsequent Energy Extrapolation. \emph{Theor. Chim. Acta.}
  \textbf{1974}, \emph{35}, 33--58\relax
\mciteBstWouldAddEndPuncttrue
\mciteSetBstMidEndSepPunct{\mcitedefaultmidpunct}
{\mcitedefaultendpunct}{\mcitedefaultseppunct}\relax
\EndOfBibitem
\bibitem[Schriber and Evangelista(2016)Schriber, and
  Evangelista]{adaptive_ci_1}
Schriber,~J.~B.; Evangelista,~F.~A. Communication: An Adaptive Configuration
  Interaction Approach for Strongly Correlated Electrons with Tunable Accuracy.
  \emph{J. Chem. Phys.} \textbf{2016}, \emph{144}, 161106\relax
\mciteBstWouldAddEndPuncttrue
\mciteSetBstMidEndSepPunct{\mcitedefaultmidpunct}
{\mcitedefaultendpunct}{\mcitedefaultseppunct}\relax
\EndOfBibitem
\bibitem[Schriber and Evangelista(2017)Schriber, and
  Evangelista]{adaptive_ci_2}
Schriber,~J.~B.; Evangelista,~F.~A. Adaptive Configuration Interaction for
  Computing Challenging Electronic Excited States with Tunable Accuracy.
  \emph{J. Chem. Theory Comput.} \textbf{2017}, \emph{13}, 5354--5366\relax
\mciteBstWouldAddEndPuncttrue
\mciteSetBstMidEndSepPunct{\mcitedefaultmidpunct}
{\mcitedefaultendpunct}{\mcitedefaultseppunct}\relax
\EndOfBibitem
\bibitem[Tubman \latin{et~al.}(2016)Tubman, Lee, Takeshita, Head-Gordon, and
  Whaley]{asci_1}
Tubman,~N.~M.; Lee,~J.; Takeshita,~T.~Y.; Head-Gordon,~M.; Whaley,~K.~B. A
  Deterministic Alternative to the Full Configuration Interaction Quantum Monte
  Carlo Method. \emph{J. Chem. Phys.} \textbf{2016}, \emph{145}, 044112\relax
\mciteBstWouldAddEndPuncttrue
\mciteSetBstMidEndSepPunct{\mcitedefaultmidpunct}
{\mcitedefaultendpunct}{\mcitedefaultseppunct}\relax
\EndOfBibitem
\bibitem[Tubman \latin{et~al.}(2020)Tubman, Freeman, Levine, Hait, Head-Gordon,
  and Whaley]{asci_2}
Tubman,~N.~M.; Freeman,~C.~D.; Levine,~D.~S.; Hait,~D.; Head-Gordon,~M.;
  Whaley,~K.~B. Modern Approaches to Exact Diagonalization and Selected
  Configuration Interaction with the Adaptive Sampling CI Method. \emph{J.
  Chem. Theory Comput.} \textbf{2020}, \emph{16}, 2139--2159\relax
\mciteBstWouldAddEndPuncttrue
\mciteSetBstMidEndSepPunct{\mcitedefaultmidpunct}
{\mcitedefaultendpunct}{\mcitedefaultseppunct}\relax
\EndOfBibitem
\bibitem[Liu and Hoffmann(2016)Liu, and Hoffmann]{ici_1}
Liu,~W.; Hoffmann,~M.~R. iCI: Iterative CI Toward Full CI. \emph{J. Chem.
  Theory Comput.} \textbf{2016}, \emph{12}, 1169--1178, {\bf 2016}, {\it 12},
  3000 [Erratum]\relax
\mciteBstWouldAddEndPuncttrue
\mciteSetBstMidEndSepPunct{\mcitedefaultmidpunct}
{\mcitedefaultendpunct}{\mcitedefaultseppunct}\relax
\EndOfBibitem
\bibitem[Zhang \latin{et~al.}(2020)Zhang, Liu, and Hoffmann]{ici_2}
Zhang,~N.; Liu,~W.; Hoffmann,~M.~R. Iterative Configuration Interaction with
  Selection. \emph{J. Chem. Theory Comput.} \textbf{2020}, \emph{16},
  2296--2316\relax
\mciteBstWouldAddEndPuncttrue
\mciteSetBstMidEndSepPunct{\mcitedefaultmidpunct}
{\mcitedefaultendpunct}{\mcitedefaultseppunct}\relax
\EndOfBibitem
\bibitem[Holmes \latin{et~al.}(2016)Holmes, Tubman, and Umrigar]{shci_1}
Holmes,~A.~A.; Tubman,~N.~M.; Umrigar,~C.~J. Heat-Bath Configuration
  Interaction: An Efficient Selected Configuration Interaction Algorithm
  Inspired by Heat-Bath Sampling. \emph{J. Chem. Theory Comput.} \textbf{2016},
  \emph{12}, 3674--3680\relax
\mciteBstWouldAddEndPuncttrue
\mciteSetBstMidEndSepPunct{\mcitedefaultmidpunct}
{\mcitedefaultendpunct}{\mcitedefaultseppunct}\relax
\EndOfBibitem
\bibitem[Sharma \latin{et~al.}(2017)Sharma, Holmes, Jeanmairet, Alavi, and
  Umrigar]{shci_2}
Sharma,~S.; Holmes,~A.~A.; Jeanmairet,~G.; Alavi,~A.; Umrigar,~C.~J.
  Semistochastic Heat-Bath Configuration Interaction Method: Selected
  Configuration Interaction with Semistochastic Perturbation Theory. \emph{J.
  Chem. Theory Comput.} \textbf{2017}, \emph{13}, 1595--1604\relax
\mciteBstWouldAddEndPuncttrue
\mciteSetBstMidEndSepPunct{\mcitedefaultmidpunct}
{\mcitedefaultendpunct}{\mcitedefaultseppunct}\relax
\EndOfBibitem
\bibitem[Li \latin{et~al.}(2018)Li, Otten, Holmes, Sharma, and Umrigar]{shci_3}
Li,~J.; Otten,~M.; Holmes,~A.~A.; Sharma,~S.; Umrigar,~C.~J. Fast
  Semistochastic Heat-Bath Configuration Interaction. \emph{J. Chem. Phys.}
  \textbf{2018}, \emph{149}, 214110\relax
\mciteBstWouldAddEndPuncttrue
\mciteSetBstMidEndSepPunct{\mcitedefaultmidpunct}
{\mcitedefaultendpunct}{\mcitedefaultseppunct}\relax
\EndOfBibitem
\bibitem[CCp()]{CCpy-GitHub}
K. Gururangan, J. E. Deustua, and P. Piecuch, ``CCpy: A Coupled-Cluster Package
  Written in Python,'' see https://github.com/piecuch-group/ccpy, last accessed
  October 17, 2025.\relax
\mciteBstWouldAddEndPunctfalse
\mciteSetBstMidEndSepPunct{\mcitedefaultmidpunct}
{}{\mcitedefaultseppunct}\relax
\EndOfBibitem
\bibitem[Schmidt \latin{et~al.}(1993)Schmidt, Baldridge, Boatz, Elbert, Gordon,
  Jensen, Koseki, Matsunaga, Nguyen, Su, Windus, Dupuis, and
  Montgomery]{gamess}
Schmidt,~M.~W.; Baldridge,~K.~K.; Boatz,~J.~A.; Elbert,~S.~T.; Gordon,~M.~S.;
  Jensen,~J.~H.; Koseki,~S.; Matsunaga,~N.; Nguyen,~K.~A.; Su,~S.;
  Windus,~T.~L.; Dupuis,~M.; Montgomery,~J.~A. General Atomic and Molecular
  Electronic Structure System. \emph{J. Comput. Chem.} \textbf{1993},
  \emph{14}, 1347--1363\relax
\mciteBstWouldAddEndPuncttrue
\mciteSetBstMidEndSepPunct{\mcitedefaultmidpunct}
{\mcitedefaultendpunct}{\mcitedefaultseppunct}\relax
\EndOfBibitem
\bibitem[Barca \latin{et~al.}(2020)Barca, Bertoni, Carrington, Datta, De~Silva,
  Deustua, Fedorov, Gour, Gunina, Guidez, Harville, Irle, Magoulas, Mato,
  Mironov, Nakata, Pham, Piecuch, Poole, Pruitt, Rendell, Roskop, Ruedenberg,
  Sattasathuchana, Schmidt, Shen, Slipchenko, Sosonkina, Sundriyal, Tiwari,
  Vallejo, Westheimer, W{\l}och, Xu, Zahariev, and Gordon]{gamess2020}
Barca,~G. M.~J.; Bertoni,~C.; Carrington,~L.; Datta,~D.; De~Silva,~N.;
  Deustua,~J.~E.; Fedorov,~D.~G.; Gour,~J.~R.; Gunina,~A.~O.; Guidez,~E.;
  Harville,~T.; Irle,~S.; Magoulas,~I.; Mato,~J.; Mironov,~V.; Nakata,~H.;
  Pham,~B.~Q.; Piecuch,~P.; Poole,~D.; Pruitt,~S.~R.; Rendell,~A.~P.;
  Roskop,~L.~B.; Ruedenberg,~K.; Sattasathuchana,~T.; Schmidt,~M.~W.; Shen,~J.;
  Slipchenko,~L.; Sosonkina,~M.; Sundriyal,~V.; Tiwari,~A.; Vallejo,~J. L.~G.;
  Westheimer,~B.; W{\l}och,~M.; Xu,~P.; Zahariev,~F.; Gordon,~M.~S. Recent
  Developments in the General Atomic and Molecular Electronic Structure System.
  \emph{J. Chem. Phys.} \textbf{2020}, \emph{152}, 154102\relax
\mciteBstWouldAddEndPuncttrue
\mciteSetBstMidEndSepPunct{\mcitedefaultmidpunct}
{\mcitedefaultendpunct}{\mcitedefaultseppunct}\relax
\EndOfBibitem
\bibitem[Zahariev \latin{et~al.}(2023)Zahariev, Xu, Westheimer, Webb,
  Galvez~Vallejo, Tiwari, Sundriyal, Sosonkina, Shen, Schoendorff, Schlinsog,
  Sattasathuchana, Ruedenberg, Roskop, Rendell, Poole, Piecuch, Pham, Mironov,
  Mato, Leonard, Leang, Ivanic, Hayes, Harville, Gururangan, Guidez, Gerasimov,
  Friedl, Ferreras, Elliott, Datta, Cruz, Carrington, Bertoni, Barca, Alkan,
  and Gordon]{gamess2023}
Zahariev,~F.; Xu,~P.; Westheimer,~B.~M.; Webb,~S.; Galvez~Vallejo,~J.;
  Tiwari,~A.; Sundriyal,~V.; Sosonkina,~M.; Shen,~J.; Schoendorff,~G.;
  Schlinsog,~M.; Sattasathuchana,~T.; Ruedenberg,~K.; Roskop,~L.~B.;
  Rendell,~A.~P.; Poole,~D.; Piecuch,~P.; Pham,~B.~Q.; Mironov,~V.; Mato,~J.;
  Leonard,~S.; Leang,~S.~S.; Ivanic,~J.; Hayes,~J.; Harville,~T.;
  Gururangan,~K.; Guidez,~E.; Gerasimov,~I.~S.; Friedl,~C.; Ferreras,~K.~N.;
  Elliott,~G.; Datta,~D.; Cruz,~D. D.~A.; Carrington,~L.; Bertoni,~C.;
  Barca,~G. M.~J.; Alkan,~M.; Gordon,~M.~S. The General Atomic and Molecular
  Electronic Structure System (GAMESS): Novel Methods on Novel Architectures.
  \emph{J. Chem. Theory Comput.} \textbf{2023}, \emph{19}, 7031--7055\relax
\mciteBstWouldAddEndPuncttrue
\mciteSetBstMidEndSepPunct{\mcitedefaultmidpunct}
{\mcitedefaultendpunct}{\mcitedefaultseppunct}\relax
\EndOfBibitem
\bibitem[Loos \latin{et~al.}(2020)Loos, Damour, and Scemama]{cipsi_benzene}
Loos,~P.-F.; Damour,~Y.; Scemama,~A. The Performance of CIPSI on the Ground
  State Electronic Energy of Benzene. \emph{J. Chem. Phys.} \textbf{2020},
  \emph{153}, 176101\relax
\mciteBstWouldAddEndPuncttrue
\mciteSetBstMidEndSepPunct{\mcitedefaultmidpunct}
{\mcitedefaultendpunct}{\mcitedefaultseppunct}\relax
\EndOfBibitem
\bibitem[Lyakh \latin{et~al.}(2011)Lyakh, Lotrich, and Bartlett]{tailored3}
Lyakh,~D.~I.; Lotrich,~V.~F.; Bartlett,~R.~J. The {`}Tailored{'} CCSD(T)
  Description of the Automerization of Cyclobutadiene. \emph{Chem. Phys. Lett.}
  \textbf{2011}, \emph{501}, 166--171\relax
\mciteBstWouldAddEndPuncttrue
\mciteSetBstMidEndSepPunct{\mcitedefaultmidpunct}
{\mcitedefaultendpunct}{\mcitedefaultseppunct}\relax
\EndOfBibitem
\bibitem[Kowalski and Piecuch(2001)Kowalski, and Piecuch]{eomccsdt1}
Kowalski,~K.; Piecuch,~P. The Active-Space Equation-of-Motion Coupled-Cluster
  Methods for Excited Electronic States: Full EOMCCSDt. \emph{J. Chem. Phys.}
  \textbf{2001}, \emph{115}, 643--651\relax
\mciteBstWouldAddEndPuncttrue
\mciteSetBstMidEndSepPunct{\mcitedefaultmidpunct}
{\mcitedefaultendpunct}{\mcitedefaultseppunct}\relax
\EndOfBibitem
\bibitem[Kowalski and Piecuch(2001)Kowalski, and Piecuch]{eomccsdt2}
Kowalski,~K.; Piecuch,~P. Excited-State Potential Energy Curves of ${\rm
  CH}^{+}$: A Comparison of the EOMCCSDt and Full EOMCCSDT Results. \emph{Chem.
  Phys. Lett.} \textbf{2001}, \emph{347}, 237--246\relax
\mciteBstWouldAddEndPuncttrue
\mciteSetBstMidEndSepPunct{\mcitedefaultmidpunct}
{\mcitedefaultendpunct}{\mcitedefaultseppunct}\relax
\EndOfBibitem
\bibitem[Kucharski \latin{et~al.}(2001)Kucharski, W{\l}och, Musia{\l}, and
  Bartlett]{eomccsdt3}
Kucharski,~S.~A.; W{\l}och,~M.; Musia{\l},~M.; Bartlett,~R.~J. Coupled-Cluster
  Theory for Excited Electronic States: The Full Equation-of-Motion
  Coupled-Cluster Single, Double, and Triple Excitation Method. \emph{J. Chem.
  Phys.} \textbf{2001}, \emph{115}, 8263--8266\relax
\mciteBstWouldAddEndPuncttrue
\mciteSetBstMidEndSepPunct{\mcitedefaultmidpunct}
{\mcitedefaultendpunct}{\mcitedefaultseppunct}\relax
\EndOfBibitem
\end{mcitethebibliography}

\newpage

\clearpage


\onecolumngrid


\squeezetable
\begin{table*}[!ht]
\caption{
Convergence of the CC($P$) and CC($P$;$Q$) Energies of the Lowest Singlet State of Cyclobutadiene,
as Described by the cc-pVDZ Basis Set, Toward CCSDT at Selected Values of Parameter
$\lambda$ Defining the Automerization Coordinate via the Interpolation Formula Given by Eq. (\ref{eq:ell}),
Alongside the Associated Variational and Perturbatively Corrected CIPSI Energies}
\label{tab:table1}
\footnotesize
\begin{ruledtabular}
\begin{tabular}{l p{4.5cm} c c c c c c}
\textrm{$\lambda$}&
\textrm{$N_{\text{det(in)}}$ / $N_{\text{det(out)}}$}&
\textrm{\% of triples}&
\textrm{$E_{\text{var}}$\protect\footnotemark[1]}&
\textrm{$E_{\text{var}}+\Delta E^{(2)}$\footnotemark[1]}&
\textrm{$E_{\text{var}}+\Delta E_{\text{r}}^{(2)}$\protect\footnotemark[1]}&
\textrm{CC$(P)$\footnotemark[2]}&
\textrm{CC$(P;Q)$\footnotemark[2]} \\
\colrule \\[-2.5mm]
    0 & 1/1 & 0 & 596.966\footnotemark[3] & $-84.890$\footnotemark[4] & 119.654 & 26.827\footnotemark[5] & 0.848\footnotemark[6] \\
    & 50,000/55,651         & 0.0  & 120.631  & 25.127(179)  & 27.145(175)  & 25.481 & 0.676 \\
    & 100,000/111,316       & 0.1  & 107.950  & 22.116(146)  & 23.686(143)  & 22.183 & 0.431 \\
    & 250,000/445,296       & 0.6  & 95.932   & 18.756(144)  & 19.974(141)  & 17.706 & 0.278 \\
    & 500,000/890,920       & 1.1  & 91.033   & 17.669(142)  & 18.752(140)  & 16.258 & 0.268 \\
    & 1,000,000/1,781,339   & 2.2  & 86.494   & 16.576(139)  & 17.543(137)  & 14.608 & 0.254 \\
    & 5,000,000/7,127,768   & 7.7  & 75.839   & 14.895(121)  & 15.604(119)  & 10.733 & 0.145 \\
    & 10,000,000/14,258,080 & 15.1 & 67.232   & 13.219(108)  & 13.759(107)  & 7.252  & 0.093 \\
    &                   &      &          &              &              &        &          \\ [-2.5mm]
    0.2 & 1/1 & 0 & 601.559\footnotemark[3] & $-87.141$\footnotemark[4] & 129.221 & 27.964\footnotemark[5] & 1.253\footnotemark[6] \\
    & 50,000/51,630         & 0.0  & 126.605  & 27.112(167)  & 29.321(164)  & 26.641 & 1.012 \\
    & 100,000/103,165       & 0.1  & 111.522  & 24.556(150)  & 26.178(147)  & 23.029 & 0.552 \\
    & 250,000/412,603       & 0.5  & 98.105   & 19.604(152)  & 20.870(149)  & 17.911 & 0.272 \\
    & 500,000/825,242       & 1.1  & 92.595   & 18.791(145)  & 19.889(143)  & 16.130 & 0.270 \\
    & 1,000,000/1,651,057   & 2.0  & 87.874   & 17.728(135)  & 18.703(133)  & 14.653 & 0.264 \\
    & 5,000,000/6,602,235   & 7.3  & 77.378   & 15.889(122)  & 16.612(121)  & 10.783 & 0.148 \\
    & 10,000,000/13,223,732 & 13.9 & 68.937   & 14.333(109)  & 14.887(107)  & 7.524  & 0.101 \\
    &                       &      &          &              &              &        &          \\ [-2.5mm] 
    0.4 & 1/1 & 0 & 605.168\footnotemark[3] & $-91.486$\footnotemark[4] & 141.934 & 29.667\footnotemark[5] & 2.021\footnotemark[6] \\
    & 50,000/50,677         & 0.0  & 129.124  & 28.480(178)  & 30.760(174)  & 28.021 & 1.563 \\
    & 100,000/101,361       & 0.1  & 113.415  & 24.995(166)  & 26.686(163)  & 23.947 & 0.776 \\
    & 250,000/405,591       & 0.5  & 98.290   & 19.122(156)  & 20.416(153)  & 18.165 & 0.271 \\
    & 500,000/811,227       & 1.1  & 92.053   & 17.563(147)  & 18.682(145)  & 15.984 & 0.292 \\
    & 1,000,000/1,621,981   & 2.0  & 87.558   & 17.149(139)  & 18.133(137)  & 14.611 & 0.274 \\
    & 5,000,000/6,488,516   & 7.2  & 76.901   & 15.082(123)  & 15.813(122)  & 10.710 & 0.160 \\
    & 10,000,000/12,976,521 & 10.7 & 73.466   & 14.493(118)  & 15.156(116)  & 10.238 & 0.115 \\
    &                       &      &          &              &              &        &          \\ [-2.5mm]
    0.6 & 1/1 & 0 & 610.659\footnotemark[3] & $-95.640$\footnotemark[4] & 164.690 & 32.473\footnotemark[5] & 3.582\footnotemark[6] \\
    & 50,000/53,206         & 0.0  & 130.859  & 30.784(188)  & 33.041(184)  & 30.198 & 2.679 \\
    & 100,000/106,413       & 0.1  & 115.914  & 26.643(175)  & 28.371(171)  & 25.279 & 1.235 \\
    & 250,000/425,835       & 0.5  & 98.110   & 18.561(152)  & 19.866(150)  & 17.603 & 0.300 \\
    & 500,000/851,740       & 1.1  & 91.503   & 17.269(143)  & 18.381(140)  & 15.781 & 0.304 \\
    & 1,000,000/1,703,867   & 2.0  & 87.073   & 16.711(139)  & 17.693(137)  & 14.282 & 0.284 \\
    & 5,000,000/6,812,598   & 7.5  & 75.664   & 14.470(122)  & 15.183(121)  & 9.967  & 0.167 \\
    & 10,000,000/13,627,034 & 13.9 & 68.381   & 13.315(108)  & 13.878(107)  & 7.311  & 0.109 \\
    &                       &      &          &              &              &        &          \\ [-2.5mm]
    0.8 & 1/1 & 0 & 619.744\footnotemark[3] & $-98.958$\footnotemark[4] & 207.413 & 37.662\footnotemark[5] & 7.008\footnotemark[6] \\
    & 50,000/98,465         & 0.1  & 123.879  & 32.073(170)  & 33.919(167)  & 28.703 & 2.505 \\
    & 100,000/196,965       & 0.2  & 112.359  & 24.572(171)  & 26.222(168)  & 21.842 & 0.628 \\
    & 250,000/394,080       & 0.5  & 100.961  & 19.800(151)  & 21.171(148)  & 17.926 & 0.336 \\
    & 500,000/787,924       & 1.0  & 93.638   & 18.350(149)  & 19.500(147)  & 15.451 & 0.360 \\
    & 1,000,000/1,575,423   & 1.9  & 88.146   & 17.491(137)  & 18.484(135)  & 13.731 & 0.327 \\
    & 5,000,000/6,300,768   & 6.0  & 78.214   & 15.623(125)  & 16.375(123)  & 10.615 & 0.211 \\
    & 10,000,000/12,604,257 & 10.7 & 71.699   & 14.195(115)  & 14.816(113)  & 8.367  & 0.150 \\
    &                       &      &          &              &              &        &          \\ [-2.5mm]
    1 & 1/1 & 0 & 632.766\footnotemark[3] & $-102.757$\footnotemark[4] & 282.305 & 47.979\footnotemark[5] & 14.636\footnotemark[6] \\
    & 50,000/56,219         & 0.0  & 146.883  & 45.172(210)  & 47.519(205)  & 42.119 & 9.569 \\
    & 100,000/112,432       & 0.1  & 130.721  & 36.708(182)  & 38.658(178)  & 32.125 & 3.539 \\
    & 250,000/449,753       & 0.5  & 99.218   & 19.367(152)  & 20.688(150)  & 17.137 & 0.458 \\
    & 500,000/899,464       & 0.9  & 92.482   & 17.938(148)  & 19.059(145)  & 14.685 & 0.435 \\
    & 1,000,000/1,799,702   & 1.7  & 87.614   & 17.261(140)  & 18.243(138)  & 13.223 & 0.375 \\
    & 5,000,000/7,196,961   & 5.5  & 77.242   & 15.348(125)  & 16.078(123)  & 9.969  & 0.246 \\
    & 10,000,000/14,391,011 & 9.6  & 71.571   & 14.183(114)  & 14.800(113)  & 8.486  & 0.167 \\
    &                       &      &          &              &              &        &          \\ [-3.5mm]
\end{tabular}
\end{ruledtabular}

\footnotetext[1]{
\setlength{\baselineskip}{1em}
For each value of $\lambda$,
the $E_{\text{var}}$, $E_{\text{var}}+\Delta E^{(2)}$, and $E_{\text{var}}+\Delta E_\text{r}^{(2)}$
energies are reported as errors, in millihartree, relative to the extrapolated $E_{\text{var}}+\Delta E_{\text{r}}^{(2)}$
energy found using a linear fit based on the last six $E_{\text{var},k}+\Delta E_{\text{r},k}^{(2)}$ values leading to
the largest CIPSI wave function obtained with $N_{\text{det(in)}} = 10,000,000$, plotted against the corresponding
$\Delta E_{\text{r},k}^{(2)}$ corrections, following the procedure described in Refs.\
\onlinecite{cipsi_2,cipsi_benzene,cbd-loos-2022}. The extrapolated
$E_{\text{var}}+\Delta E_{\text{r}}^{(2)}$ energies at $\lambda = 0$, $0.2$, $0.4$, $0.6$, $0.8$, and $1$ are
$-154.248137(398)$, $-154.247883(480)$, $-154.244213(872)$, $-154.239997(642)$, $-154.236928(616)$,
and $-154.235401(1043)$ hartree,
respectively, where the error bounds in parentheses correspond to the uncertainty associated with the linear fit.
The error bounds
for the $E_{\text{var}}+\Delta E^{(2)}$ and $E_{\text{var}}+\Delta E_\text{r}^{(2)}$ energies obtained at the
various values of
$N_{\text{det(in)}}$ reflect on the semi-stochastic design of the $\mathscr{V}_{\text{ext}}^{(k)}$
spaces discussed in the main
text, but they ignore the uncertainties characterizing the reference $E_{\text{var}}+\Delta E_{\text{r}}^{(2)}$ energies
obtained in the above extrapolation procedure.}
\footnotetext[2]{
\setlength{\baselineskip}{1em}
The CC($P$) and CC($P$;$Q$) energies are reported as errors relative to CCSDT, in millihartree.
The total CCSDT energies at $\lambda = 0$, $0.2$, $0.4$, $0.6$, $0.8$, and $1$ are 
$-154.244157$, $-154.242922$, $-154.240027$, $-154.236079$, $-154.232439$, and $-154.232002$ hartree, respectively.}
\footnotetext[3]{
\setlength{\baselineskip}{1em}
Equivalent to RHF.}
\footnotetext[4]{
\setlength{\baselineskip}{1em}
Equivalent to the result obtained with the second-order MBPT approach using the Epstein--Nesbet denominator.}
\footnotetext[5]{
\setlength{\baselineskip}{1em}
Equivalent to CCSD.}
\footnotetext[6]{
\setlength{\baselineskip}{1em}
Equivalent to CR-CC(2,3).}
\end{table*}


\squeezetable
\begin{table*}[!ht]
\caption{
Convergence of the CC($P$) and CC($P$;$Q$) Energies of the Lowest Triplet State of Cyclobutadiene,
as Described by the cc-pVDZ Basis Set, Toward CCSDT at Selected Values of Parameter
$\lambda$ Defining the Automerization Coordinate via the Interpolation Formula Given by Eq. (\ref{eq:ell}),
Alongside the Associated Variational and Perturbatively Corrected CIPSI Energies}
\label{tab:table2}
\footnotesize
\begin{ruledtabular}
\begin{tabular}{l p{4.5cm} c c c c c c}
\textrm{$\lambda$}&
\textrm{$N_{\text{det(in)}}$ / $N_{\text{det(out)}}$}&
\textrm{\% of triples}&
\textrm{$E_{\text{var}}$\protect\footnotemark[1]}&
\textrm{$E_{\text{var}}+\Delta E^{(2)}$\footnotemark[1]}&
\textrm{$E_{\text{var}}+\Delta E_{\text{r}}^{(2)}$\protect\footnotemark[1]}&
\textrm{CC$(P)$\footnotemark[2]}&
\textrm{CC$(P;Q)$\footnotemark[2]} \\
\colrule \\[-2.5mm]
    0 & 1/1 & 0 & 572.232\footnotemark[3] & $-97.167$\footnotemark[4] & 94.195 & 24.646\footnotemark[5] & $-0.033$\footnotemark[6] \\
    & 50,000/84,925         & 0.3  & 126.216	    & 17.881(210)  & 20.609(205)  & 22.589 & 0.016 \\
    & 100,000/169,861       & 0.5  & 100.010	    & 14.692(155)  & 16.283(152)  & 20.446 & 0.028 \\
    & 250,000/339,721       & 0.8  & 90.992	    & 13.111(144)  & 14.389(142)  & 18.341 & 0.100 \\ 
    & 500,000/679,710       & 1.2  & 85.449	    & 12.584(139)  & 13.672(137)  &	16.737 & 0.154 \\
    & 1,000,000/1,359,265   & 1.8  & 81.995	    & 12.345(137)  & 13.324(135)  &	15.567 & 0.173 \\
    & 5,000,000/5,436,202   & 4.6  & 74.316	    & 11.275(126)  & 12.053(124)  & 12.474 & 0.167 \\
    & 10,000,000/10,871,115 & 7.7  & 69.341	    & 10.222(118)  & 10.896(117)  & 10.637 & 0.134 \\
    &                       &      &                &              &              &        &       \\ [-2.5mm]
    0.2 & 1/1 & 0 & 571.775\footnotemark[3] & $-95.504$\footnotemark[4] & 93.218 & 24.424\footnotemark[5] & $-0.043$\footnotemark[6] \\
    & 50,000/95,659         & 0.3  & 119.164	    & 17.033(194)  & 19.415(189)  & 22.170 & 0.015 \\
    & 100,000/191,346       & 0.6  & 97.554	    & 14.541(153)  & 16.027(151)  &	19.833 & 0.072 \\
    & 250,000/382,772       & 0.8  & 90.002	    & 12.999(155)  & 14.237(152)  & 17.991 & 0.100 \\ 
    & 500,000/765,329       & 1.3  & 84.848	    & 12.854(143)  & 13.912(141)  & 16.326 & 0.155\\
    & 1,000,000/1,532,203   & 2.0  & 80.932	    & 12.414(135)  & 13.356(133)  & 14.846 & 0.173 \\
    & 5,000,000/6,122,654   & 5.1  & 72.919	    & 11.032(121)  & 11.776(120)  & 11.898 & 0.152 \\
    & 10,000,000/12,246,843 & 8.5  & 67.669	    & 10.147(114)  & 10.778(113)  & 9.840  & 0.126 \\
    &                       &      &                &              &              &        &       \\ [-2.5mm]
    0.4 & 1/1 & 0 & 572.339\footnotemark[3] & $-93.175$\footnotemark[4] & 93.387 & 24.239\footnotemark[5] & $-0.050$\footnotemark[6] \\
    & 50,000/68,315         & 0.2  & 138.142	    & 20.184(191)  & 23.481(186)  & 22.835 & $-0.028$ \\
    & 100,000/136,635       & 0.4  & 105.157	    & 16.968(158)  & 18.675(155)  & 20.901 & 0.003 \\
    & 250,000/273,285       & 0.7  & 94.039	    & 14.997(147)  & 16.319(145)  & 18.707 & 0.068 \\ 
    & 500,000/546,881       & 1.0  & 88.089	    & 14.172(147)  & 15.295(144)  & 17.028 & 0.130 \\
    & 1,000,000/1,093,480   & 1.6  & 84.427	    & 13.398(141)  & 14.420(139)  & 15.794 & 0.154 \\
    & 5,000,000/8,746,894   & 6.6  & 71.913	    & 12.125(119)  & 12.815(118)  & 10.994 & 0.138 \\
    & 10,000,000/17,483,610 & 12.1 & 62.833	    & 10.294(105)  & 10.813(104)  & 8.130  & 0.084 \\
    &                       &      &                &              &              &        &       \\ [-2.5mm]
    0.6 & 1/1 & 0 & 570.893\footnotemark[3] & $-93.217$\footnotemark[4] & 91.663 & 24.089\footnotemark[5] & $-0.055$\footnotemark[6] \\
    & 50,000/55,070         & 0.2  & 150.423	    & 19.777(414)  & 23.900(401)  & 23.037 & $-0.034$ \\
    & 100,000/110,142       & 0.4  & 110.889	    & 18.391(125)  & 20.292(122)  & 21.542 & $-0.015$ \\
    & 250,000/440,697       & 0.9  & 88.802	    & 13.367(142)  & 14.547(140)  & 17.540 & 0.096 \\ 
    & 500,000/881,321       & 1.3  & 84.455	    & 13.032(131)  & 14.069(129)  & 16.110 & 0.143 \\
    & 1,000,000/1,762,363   & 2.1  & 80.638	    & 12.620(131)  & 13.546(130)  & 14.705 & 0.162 \\
    & 5,000,000/7,051,421   & 5.5  & 73.209	    & 11.431(124)  & 12.173(122)  & 11.869 & 0.132 \\
    & 10,000,000/14,099,214 & 9.1  & 67.881	    & 10.554(114)  & 11.183(113)  & 9.951  & 0.117 \\
    &                       &      &                &              &              &        &       \\ [-2.5mm]
    0.8 & 1/1 & 0 & 570.863\footnotemark[3] & $-92.204$\footnotemark[4] & 91.469 & 23.974\footnotemark[5] & $-0.058$\footnotemark[6] \\
    & 50,000/59,298         & 0.2  & 144.443	    & 20.389(241)  & 24.065(234)  & 22.846 & $-0.040$ \\
    & 100,000/118,602       & 0.4  & 107.656	    & 17.516(184)  & 19.303(180)  & 21.062 & $-0.006$ \\
    & 250,000/474,464       & 0.9  & 88.836	    & 14.143(145)  & 15.297(143)  & 17.489 & 0.094 \\ 
    & 500,000/949,394       & 1.4  & 85.025	    & 13.589(140)  & 14.628(138)  & 16.201 & 0.131 \\
    & 1,000,000/1,898,021   & 2.2  & 81.984	    & 13.095(133)  & 14.049(131)  & 15.224 & 0.131 \\
    & 5,000,000/7,591,707   & 5.5  & 73.852	    & 11.863(123)  & 12.613(122)  & 12.110 & 0.125 \\
    & 10,000,000/15,188,890 & 9.1  & 68.485	    & 11.085(113)  & 11.716(112)  & 10.195 & 0.099 \\
    &                       &      &                &              &              &        &       \\ [-2.5mm]
    1 & 1/1 & 0 & 570.406\footnotemark[3] & $-91.860$\footnotemark[4] & 90.994 & 23.884\footnotemark[5] & $-0.060$\footnotemark[6] \\
    & 50,000/65,391         & 0.2  & 137.892	    & 20.022(218)  & 23.305(212)  & 22.617 & $-0.047$ \\
    & 100,000/130,810       & 0.4  & 103.950	    & 16.288(153)  & 17.965(150)  & 20.624 & $-0.010$ \\
    & 250,000/261,626       & 0.6  & 93.518	    & 14.821(148)  & 16.127(145)  & 18.665 & 0.039 \\ 
    & 500,000/523,285       & 0.9  & 87.775	    & 13.661(137)  & 14.792(134)  & 17.237 & 0.109 \\
    & 1,000,000/1,046,443   & 1.4  & 84.673	    & 13.507(140)  & 14.536(137)  & 16.189 & 0.127 \\
    & 5,000,000/8,373,419   & 5.8  & 74.128	    & 11.775(124)  & 12.535(122)  & 12.371 & 0.117 \\
    & 10,000,000/16,741,696 & 9.3  & 68.611	    & 11.074(115)  & 11.711(114)  & 10.435 & 0.101 \\
    &                       &      &                &              &              &        &       \\ [-3.5mm]
\end{tabular}	
\end{ruledtabular}

\footnotetext[1]{
\setlength{\baselineskip}{1em}
For each value of $\lambda$,
the $E_{\text{var}}$, $E_{\text{var}}+\Delta E^{(2)}$, and $E_{\text{var}}+\Delta E_\text{r}^{(2)}$
energies are reported as errors, in millihartree, relative to the extrapolated $E_{\text{var}}+\Delta E_{\text{r}}^{(2)}$
energy found using a linear fit based on the last six $E_{\text{var},k}+\Delta E_{\text{r},k}^{(2)}$ values leading to
the largest CIPSI wave function obtained with $N_{\text{det(in)}} = 10,000,000$, plotted against the corresponding
$\Delta E_{\text{r},k}^{(2)}$ corrections, following the procedure described in Refs.\
\onlinecite{cipsi_2,cipsi_benzene,cbd-loos-2022}. The extrapolated
$E_{\text{var}}+\Delta E_{\text{r}}^{(2)}$ energies at $\lambda = 0$, $0.2$, $0.4$, $0.6$, $0.8$, and $1$ are 
$-154.195674(1195)$, $-154.206430(1047)$, $-154.215793(862)$, $-154.220754(582)$, $-154.224733(1124)$, and
$-154.225942(668)$ hartree,
respectively, where the error bounds in parentheses correspond to the uncertainty associated with the linear fit.
The error bounds
for the $E_{\text{var}}+\Delta E^{(2)}$ and $E_{\text{var}}+\Delta E_\text{r}^{(2)}$ energies obtained at the
various values of 
$N_{\text{det(in)}}$ reflect on the semi-stochastic design of the $\mathscr{V}_{\text{ext}}^{(k)}$
spaces discussed in the main
text, but they ignore the uncertainties characterizing the reference $E_{\text{var}}+\Delta E_{\text{r}}^{(2)}$ energies
obtained in the above extrapolation procedure.}
\footnotetext[2]{
\setlength{\baselineskip}{1em}
The CC($P$) and CC($P$;$Q$) energies are reported as errors relative to CCSDT, in millihartree.
The total CCSDT energies at $\lambda = 0$, $0.2$, $0.4$, $0.6$, $0.8$, and $1$ are 
$-154.195389$, $-154.205779$, $-154.213867$, $-154.219672$, $-154.223190$, and $-154.224380$ hartree, respectively.}
\footnotetext[3]{
\setlength{\baselineskip}{1em}
Equivalent to ROHF.}
\footnotetext[4]{
\setlength{\baselineskip}{1em}
Equivalent to the result obtained with the second-order MBPT approach using the Epstein--Nesbet denominator.}
\footnotetext[5]{
\setlength{\baselineskip}{1em}
Equivalent to CCSD.}
\footnotetext[6]{
\setlength{\baselineskip}{1em}
Equivalent to CR-CC(2,3).}
\end{table*} 


\squeezetable
\begin{table*}[!ht]
\caption{
Convergence of the CC($P$) and CC($P$;$Q$) Singlet--Triplet Gaps $\Delta E_\text{S--T}= E_\text{S} - E_\text{T}$
Characterizing Cyclobutadiene, as Described by the cc-pVDZ Basis Set, Toward Their CCSDT Parents at Selected Values
of Parameter $\lambda$ Defining the Automerization Coordinate via the Interpolation Formula Given by Eq. (\ref{eq:ell}),
Along With the  $\Delta E_\text{S--T}$ Data Resulting From the Associated Variational and Perturbatively Corrected CIPSI
Computations}
\label{tab:table3}
\footnotesize
\begin{ruledtabular}
\begin{tabular}{l p{4.5cm} c c c c c c}
\textrm{$\lambda$}&
\textrm{$N_{\text{det(in)}}$ / $N_{\text{det(out)}}$}&
\textrm{\% of triples}&
\textrm{$E_{\text{var}}$\protect\footnotemark[1]}&
\textrm{$E_{\text{var}}+\Delta E^{(2)}$\footnotemark[1]}&
\textrm{$E_{\text{var}}+\Delta E_{\text{r}}^{(2)}$\protect\footnotemark[1]}&
\textrm{CC$(P)$\footnotemark[2]}&
\textrm{CC$(P;Q)$\footnotemark[2]} \\
\colrule \\[-2.5mm]
    0 & 1/1; 1 & 0; 0 & 15.521\footnotemark[3] & 7.704\footnotemark[4] & 15.976 & 1.368\footnotemark[5] & 0.553\footnotemark[6] \\
    & 50,000/55,651; 84,925             & 0.0; 0.3  & $-3.504$    & 4.546(173)    & 4.101(169)    & 1.815      & 0.414 \\
    & 100,000/111,316; 169,861          & 0.1; 0.5  & 4.982       & 4.658(133)    & 4.645(131)    & 1.090      & 0.253 \\
    & 250,000/445,296; 339,721          & 0.6; 0.8  & 3.100       & 3.542(128)    & 3.505(126)    & $-0.398$   & 0.112 \\
    & 500,000/890,920; 679,710          & 1.1; 1.2  & 3.504       & 3.191(125)    & 3.188(123)    & $-0.300$   & 0.071 \\
    & 1,000,000/1,781,339; 1,359,265    & 2.2; 1.8  & 2.823       & 2.655(122)    & 2.647(121)    & $-0.602$   & 0.051 \\
    & 5,000,000/7,127,768; 5,436,202    & 7.7; 4.6  & 0.956       & 2.272(110)    & 2.228(108)    & $-1.092$   & $-0.014$ \\
    & 10,000,000/14,258,080; 10,871,115 & 15.1; 7.7 & $-1.324$    & 1.881(100)    & 1.797(099)    & $-2.124$   & $-0.025$ \\
    &                                   &           &             &               &               &            &       \\ [-2.5mm]
    0.2 & 1/1; 1 & 0; 0 & 18.690\footnotemark[3] & 5.248\footnotemark[4] & 22.592 & 2.221\footnotemark[5] & 0.813\footnotemark[6] \\
    & 50,000/51,630; 95,659             & 0.0; 0.3  & 4.670       & 6.325(161)   & 6.216(157)    & 2.805	& 0.626 \\
    & 100,000/103,165; 191,346          & 0.1; 0.6  & 8.765       & 6.284(135)	 & 6.370(132)    & 2.006	& 0.301 \\
    & 250,000/412,603; 382,772          & 0.5; 0.8  & 5.085       & 4.145(136)	 & 4.162(134)	 & $-0.050$	& 0.108 \\
    & 500,000/825,242; 765,329          & 1.1; 1.3  & 4.861       & 3.726(128)	 & 3.751(126)	 & $-0.123$	& 0.072 \\
    & 1,000,000/1,651,057; 1,532,203    & 2.0; 2.0  & 4.357       & 3.334(120)	 & 3.355(118)	 & $-0.121$	& 0.057 \\
    & 5,000,000/6,602,235; 6,122,654    & 7.3; 5.1  & 2.798       & 3.048(108)	 & 3.035(107)	 & $-0.700$	& $-0.003$ \\
    & 10,000,000/13,223,732; 12,246,843 & 13.9; 8.5 & 0.796       & 2.627(099)	 & 2.578(098)	 & $-1.453$	& $-0.015$ \\
    &                                   &           &             &              &               &              &       \\ [-2.5mm]
    0.4 & 1/1; 1 & 0; 0 & 20.600\footnotemark[3] & 1.060\footnotemark[4] & 30.464 & 3.406\footnotemark[5] & 1.300\footnotemark[6] \\
    & 50,000/50,677; 68,315             & 0.0; 0.2   & $-5.659$   & 5.205(164)   & 4.567(160)    & 3.254	& 0.998 \\
    & 100,000/101,361; 136,635          & 0.1; 0.4   & 5.182	  & 5.037(144)	 & 5.027(141)	 & 1.911	& 0.485 \\
    & 250,000/405,591; 273,285          & 0.5; 0.7   & 2.667      & 2.589(135)	 & 2.570(132)	 & $-0.340$	& 0.128 \\
    & 500,000/811,227; 546,881          & 1.1; 1.0   & 2.488	  & 2.128(130)	 & 2.125(128)	 & $-0.655$	& 0.102 \\
    & 1,000,000/1,621,981; 1,093,480    & 2.0; 1.6   & 1.965	  & 2.354(125)	 & 2.330(123)	 & $-0.743$	& 0.076 \\
    & 5,000,000/6,488,516; 8,746,894    & 7.2; 6.6   & 3.130	  & 1.855(108)	 & 1.881(106)	 & $-0.178$	& 0.014 \\
    & 10,000,000/12,976,521; 17,483,610 & 10.7; 12.1 & 6.672	  & 2.635(099)	 & 2.725(098)	 & 1.323        & 0.020 \\
    &                                   &            &            &              &               &              &       \\ [-2.5mm]
    0.6 & 1/1; 1 & 0; 0 & 24.953\footnotemark[3] & $-1.520$\footnotemark[4] & 45.826 & 5.261\footnotemark[5] & 2.282\footnotemark[6] \\
    & 50,000/53,206; 55,070             & 0.0; 0.2   & $-12.277$  & 6.907(285)	 & 5.736(277)	 & 4.493        & 1.703 \\
    & 100,000/106,413; 110,142          & 0.1; 0.4   & 3.154      & 5.179(135)   & 5.070(132)	 & 2.345        & 0.784 \\
    & 250,000/425,835; 440,697          & 0.5; 0.9   & 5.841	  & 3.259(131)	 & 3.338(128)	 & 0.040	& 0.128 \\
    & 500,000/851,740; 881,321          & 1.1; 1.3   & 4.423	  & 2.659(121)	 & 2.706(120)    & $-0.207$	& 0.101 \\
    & 1,000,000/1,703,867; 1,762,363    & 2.0; 2.1   & 4.038	  & 2.567(120)	 & 2.602(119)    & $-0.266$	& 0.077 \\
    & 5,000,000/6,812,598; 7,051,421    & 7.5; 5.5   & 1.541	  & 1.907(109)	 & 1.889(108)	 & $-1.193$	& 0.022 \\
    & 10,000,000/13,627,034; 14,099,214 & 13.9; 9.1  & 0.314      & 1.732(099)	 & 1.691(098)	 & $-1.657$	& $-0.005$ \\
    &                                   &            &            &              &               &              &       \\ [-2.5mm]
    0.8 & 1/1; 1 & 0; 0 & 30.673\footnotemark[3] & $-4.238$\footnotemark[4] & 72.756 & 8.589\footnotemark[5] & 4.434\footnotemark[6] \\
    & 50,000/98,465; 59,298             & 0.1; 0.2   & $-12.905$  & 7.332(151)	 & 6.184(180)    & 3.676        & 1.597 \\
    & 100,000/196,965; 118,602          & 0.2; 0.4   & 2.951	  & 4.428(152)	 & 4.342(154)	 & 0.490	& 0.398 \\
    & 250,000/394,080; 474,464          & 0.5; 0.9   & 7.609	  & 3.550(134)   & 3.685(129)	 & 0.274	& 0.152 \\
    & 500,000/787,924; 949,394          & 1.0; 1.4   & 5.404	  & 2.987(132)	 & 3.057(126)	 & $-0.471$	& 0.143 \\
    & 1,000,000/1,575,423; 1,898,021    & 1.9; 2.2   & 3.867	  & 2.759(121)	 & 2.783(118)	 & $-0.937$	& 0.123 \\
    & 5,000,000/6,300,768; 7,591,707    & 6.0; 5.5   & 2.737	  & 2.359(111)	 & 2.360(109)	 & $-0.938$	& 0.054 \\
    & 10,000,000/12,604,257; 15,188,890 & 10.7; 9.1  & 2.016	  & 1.952(102)	 & 1.945(100)	 & $-1.147$	& 0.032 \\
    &                                   &            &            &              &               &              &       \\ [-2.5mm]
    1 & 1/1; 1 & 0; 0 & 39.131\footnotemark[3] & $-6.838$\footnotemark[4] & 120.049 & 15.120\footnotemark[5] & 9.222\footnotemark[6] \\
    & 50,000/56,219; 65,391             & 0.0; 0.2   & 5.642	   & 15.782(190) & 15.195(185)	 & 12.238	& 6.035 \\
    & 100,000/112,432; 130,810          & 0.1; 0.4   & 16.799	   & 12.814(149) & 12.985(146)	 & 7.217        & 2.227 \\
    & 250,000/449,753; 261,626          & 0.5; 0.6   & 3.577	   & 2.853(133)	 & 2.862(131)    & $-0.959$	& 0.263 \\
    & 500,000/899,464; 523,285          & 0.9; 0.9   & 2.954	   & 2.683(126)	 & 2.678(124)	 & $-1.601$	& 0.205 \\
    & 1,000,000/1,799,702; 1,046,443    & 1.7; 1.4   & 1.845	   & 2.356(124)	 & 2.326(122)	 & $-1.861$	& 0.156 \\
    & 5,000,000/7,196,961; 8,373,419    & 5.5; 5.8   & 1.954	   & 2.243(110)	 & 2.223(109)	 & $-1.507$	& 0.081 \\
    & 10,000,000/14,391,011; 16,741,696 & 9.6; 9.3   & 1.857	   & 1.951(101)	 & 1.938(100)	 & $-1.223$	& 0.042 \\
    &                                   &            &             &             &               &          &       \\ [-2.5mm]
\end{tabular}
\end{ruledtabular}

\footnotetext[1]{
\setlength{\baselineskip}{1em}
For each value of $\lambda$,
the $E_{\text{var}}$, $E_{\text{var}}+\Delta E^{(2)}$, and $E_{\text{var}}+\Delta E_\text{r}^{(2)}$
singlet--triplet gaps are reported as errors, in kcal/mol, relative to the parent CIPSI data obtained by
forming the differences between the extrapolated $E_{\text{var}}+\Delta E_{\text{r}}^{(2)}$ energies of
the lowest singlet and triplet states given in footnotes `a' of Tables \ref{tab:table1} and \ref{tab:table2}.
The resulting reference $E_{\text{var}}+\Delta E_{\text{r}}^{(2)}$ singlet--triplet gap values at
$\lambda = 0$, $0.2$, $0.4$, $0.6$, $0.8$, and $1$ are
$-32.921(790)$, $-26.013(723)$, $-17.833(769)$, $-12.076(544)$, $-7.653(804)$, and $-5.936(777)$ kcal/mol, respectively.}
\footnotetext[2]{
\setlength{\baselineskip}{1em}
The CC($P$) and CC($P$;$Q$) singlet--triplet gaps are reported as errors relative to CCSDT, in kcal/mol.
The CCSDT singlet--triplet gap values at $\lambda = 0$, $0.2$, $0.4$, $0.6$, $0.8$, and $1$ are 
$-30.603$, $-23.308$, $-16.416$, $-10.295$, $-5.804$, and $-4.783$ kcal/mol, respectively.}
\footnotetext[3]{
\setlength{\baselineskip}{1em}
Equivalent to RHF/ROHF.}
\footnotetext[4]{
\setlength{\baselineskip}{1em}
Equivalent to the result obtained with the second-order MBPT approach using the Epstein--Nesbet denominator.}
\footnotetext[5]{
\setlength{\baselineskip}{1em}
Equivalent to CCSD.}
\footnotetext[6]{
\setlength{\baselineskip}{1em}
Equivalent to CR-CC(2,3).}
\end{table*}

\squeezetable
\begin{table*}[!ht]
\caption{
Convergence of the CC($P$) and CC($P$;$Q$) Energies of the Lowest Singlet State of Cyclobutadiene,
as Described by the cc-pVTZ Basis Set, Toward CCSDT at the R ($\lambda = 0$) and TS ($\lambda = 1$) Geometries,
Alongside the Associated Variational and Perturbatively Corrected CIPSI Energies
}
\label{tab:table4}
\footnotesize
\begin{ruledtabular}
\begin{tabular}{l p{4.5cm} c c c c c c}
\textrm{$\lambda$}&
\textrm{$N_{\text{det(in)}}$ / $N_{\text{det(out)}}$}&
\textrm{\% of triples}&
\textrm{$E_{\text{var}}$\protect\footnotemark[1]}&
\textrm{$E_{\text{var}}+\Delta E^{(2)}$\footnotemark[1]}&
\textrm{$E_{\text{var}}+\Delta E_{\text{r}}^{(2)}$\protect\footnotemark[1]}&
\textrm{CC$(P)$\footnotemark[2]}&
\textrm{CC$(P;Q)$\footnotemark[2]} \\
\colrule \\[-1.0mm]
    0 & 1/1 & 0.0 & 702.920\footnotemark[3] & $-76.833$\footnotemark[4] & 151.271 & 36.016\footnotemark[5] & 0.941\footnotemark[6] \\
    & 50,000/88,980         & 0.0  & 204.608  & 39.243(330)  & 44.233(320)  & 35.977 & 0.934 \\
    & 100,000/177,965       & 0.0  & 162.217  & 36.025(252)  & 38.853(246)  & 35.450 & 0.873 \\
    & 250,000/355,932       & 0.0  & 139.878  & 34.161(203)  & 36.131(200)  & 32.851 & 0.637 \\
    & 500,000/711,877       & 0.0  & 129.692  & 31.659(193)  & 33.310(190)  & 29.254 & 0.439 \\
    & 1,000,000/1,423,810   & 0.1  & 122.284  & 28.864(186)  & 30.323(183)  & 26.113 & 0.359 \\
    & 5,000,000/5,695,067   & 0.5  & 108.430  & 25.507(166)  & 26.600(163)  & 20.099 & 0.291 \\
    & 10,000,000/11,390,227 & 0.9  & 101.366  & 24.119(154)  & 25.041(152)  & 16.896 & 0.236 \\
    &                   &      &          &              &              &        &          \\ [-2.5mm]
    1 & 1/1 & 0.0 & 743.761\footnotemark[3] & $-84.810$\footnotemark[4] & 326.149 & 55.205\footnotemark[5] & 13.793\footnotemark[6] \\
    & 50,000/83,877         & 0.0  & 239.272  & 68.057(304)  & 73.536(294)  & 53.912 & 12.098 \\
    & 100,000/169,536       & 0.0  & 196.632  & 64.661(261)  & 67.837(254)  & 52.060 & 10.916 \\
    & 250,000/339,078       & 0.0  & 172.961  & 60.875(205)  & 63.128(201)  & 45.898 & 7.126 \\
    & 500,000/678,157       & 0.0  & 158.270  & 53.134(210)  & 55.059(206)  & 35.931 & 2.296 \\
    & 1,000,000/1,342,908   & 0.1  & 139.683  & 39.101(201)  & 40.795(197)  & 26.039 & 0.591 \\
    & 5,000,000/5,425,949   & 0.3  & 116.366  & 32.766(167)  & 33.884(165)  & 18.615 & 0.515 \\
    & 10,000,000/10,744,113 & 0.6  & 109.897  & 31.233(157)  & 32.198(155)  & 16.177 & 0.418 \\
    &                       &      &          &              &              &        &          \\ [-2.0mm]
\end{tabular}
\end{ruledtabular}

\footnotetext[1]{
\setlength{\baselineskip}{1em}
For each value of $\lambda$, the $E_{\text{var}}$, $E_{\text{var}}+\Delta E^{(2)}$, and $E_{\text{var}}+\Delta E_\text{r}^{(2)}$
energies are reported as errors, in millihartree, relative to the extrapolated $E_{\text{var}}+\Delta E_{\text{r}}^{(2)}$
energy found using a linear fit based on the last four $E_{\text{var},k}+\Delta E_{\text{r},k}^{(2)}$ values leading to
the largest CIPSI wave function obtained with $N_{\text{det(in)}} = 10,000,000$, plotted against the corresponding
$\Delta E_{\text{r},k}^{(2)}$ corrections, following the procedure described in Refs.\
\onlinecite{cipsi_2,cipsi_benzene,cbd-loos-2022}.
The extrapolated $E_{\text{var}}+\Delta E_{\text{r}}^{(2)}$ energies at $\lambda = 0$ and 1 are
$-154.397265(1917)$ and $-154.388862(2773)$ hartree, respectively,
where the error bounds in parentheses correspond to the uncertainty associated with the linear fit.
The error bounds for the $E_{\text{var}}+\Delta E^{(2)}$ and $E_{\text{var}}+\Delta E_\text{r}^{(2)}$ energies obtained at the
various values of $N_{\text{det(in)}}$ reflect on the semi-stochastic design of the $\mathscr{V}_{\text{ext}}^{(k)}$
spaces discussed in the main text, but they ignore the uncertainties characterizing the reference
$E_{\text{var}}+\Delta E_{\text{r}}^{(2)}$ energies obtained in the above extrapolation procedure.}
\footnotetext[2]{
\setlength{\baselineskip}{1em}
The CC($P$) and CC($P$;$Q$) energies are reported as errors relative to CCSDT, in millihartree.
The total CCSDT energies at $\lambda = 0$ and 1 are $-154.390763$ and $-154.373902$ hartree, respectively.}
\footnotetext[3]{
\setlength{\baselineskip}{1em}
Equivalent to RHF.}
\footnotetext[4]{
\setlength{\baselineskip}{1em}
Equivalent to the result obtained with the second-order MBPT approach using the Epstein--Nesbet denominator.}
\footnotetext[5]{
\setlength{\baselineskip}{1em}
Equivalent to CCSD.}
\footnotetext[6]{
\setlength{\baselineskip}{1em}
Equivalent to CR-CC(2,3).}
\end{table*}

\squeezetable
\begin{table*}[!ht]
\caption{
Convergence of the CC($P$) and CC($P$;$Q$) Energies of the Lowest Triplet State of Cyclobutadiene,
as Described by the cc-pVTZ Basis Set, Toward CCSDT at the R ($\lambda = 0$) and TS ($\lambda = 1$) Geometries,
Alongside the Associated Variational and Perturbatively Corrected CIPSI Energies
}
\label{tab:table5}
\footnotesize
\begin{ruledtabular}
\begin{tabular}{l p{4.5cm} c c c c c c}
\textrm{$\lambda$}&
\textrm{$N_{\text{det(in)}}$ / $N_{\text{det(out)}}$}&
\textrm{\% of triples}&
\textrm{$E_{\text{var}}$\protect\footnotemark[1]}&
\textrm{$E_{\text{var}}+\Delta E^{(2)}$\footnotemark[1]}&
\textrm{$E_{\text{var}}+\Delta E_{\text{r}}^{(2)}$\protect\footnotemark[1]}&
\textrm{CC$(P)$\footnotemark[2]}&
\textrm{CC$(P;Q)$\footnotemark[2]} \\
\colrule \\[-1.0mm]
    0 & 1/1 & 0 & 669.778\footnotemark[3] & $-108.330$\footnotemark[4] & 118.961 & 33.952\footnotemark[5] & $-0.023$\footnotemark[6] \\
    & 50,000/85,613         & 0.0  & 248.570  & 22.169(331)  & 32.314(316)  & 33.124 & $-0.012$ \\
    & 100,000/167,642       & 0.0  & 191.558  & 19.197(329)  & 24.688(319)  & 32.708 & $-0.003$ \\
    & 250,000/341,366       & 0.1  & 144.738  & 16.373(252)  & 19.315(246)  & 31.619 & $-0.006$ \\
    & 500,000/684,969       & 0.1  & 116.979  & 13.807(205)  & 15.681(202)  & 29.243 & 0.036 \\
    & 1,000,000/1,369,977   & 0.2  & 104.356  & 11.401(186)  & 12.878(183)  & 25.752 & 0.130 \\
    & 5,000,000/5,463,192   & 0.5  & 91.508   & 9.527(159)   & 10.617(157)  & 20.217 & 0.183 \\
    & 10,000,000/10,729,824 & 0.9  & 86.061   & 9.824(147)   & 10.758(145)  & 17.620 & 0.169 \\
    &                       &      &          &              &              &        &       \\ [-2.5mm]
    1 & 1/1 & 0.0 & 677.397\footnotemark[3] & $-93.025$\footnotemark[4] & 124.713 & 33.145\footnotemark[5] & $-0.047$\footnotemark[6] \\
    & 50,000/50,010         & 0.0  & 300.824  & 35.294(413)  & 50.017(390)  & 32.563 & $-0.044$ \\
    & 100,000/100,035       & 0.0  & 237.746  & 31.657(406)  & 39.836(390)  & 32.398 & $-0.027$ \\
    & 250,000/400,180       & 0.1  & 142.976  & 26.006(234)  & 28.422(229)  & 30.664 & $-0.017$ \\
    & 500,000/850,568       & 0.2  & 120.112  & 23.424(190)  & 25.046(187)  & 27.514 & 0.056 \\
    & 1,000,000/1,600,766   & 0.2  & 111.708  & 21.860(176)  & 23.221(173)  & 24.619 & 0.138 \\
    & 5,000,000/6,403,314   & 0.6  & 100.023  & 19.912(159)  & 20.942(157)  & 19.293 & 0.178 \\
    & 10,000,000/12,806,196 & 1.0  & 94.485   & 18.841(151)  & 19.737(149)  & 16.670 & 0.162 \\
    &                       &      &          &              &              &        &       \\ [-2.0mm]
\end{tabular}
\end{ruledtabular}

\footnotetext[1]{
\setlength{\baselineskip}{1em}
For each value of $\lambda$, the $E_{\text{var}}$, $E_{\text{var}}+\Delta E^{(2)}$, and $E_{\text{var}}+\Delta E_\text{r}^{(2)}$
energies are reported as errors, in millihartree, relative to the extrapolated $E_{\text{var}}+\Delta E_{\text{r}}^{(2)}$
energy found using a linear fit based on the last four $E_{\text{var},k}+\Delta E_{\text{r},k}^{(2)}$ values leading to
the largest CIPSI wave function obtained with $N_{\text{det(in)}} = 10,000,000$, plotted against the corresponding
$\Delta E_{\text{r},k}^{(2)}$ corrections, following the procedure described in Refs.\
\onlinecite{cipsi_2,cipsi_benzene,cbd-loos-2022}.
The extrapolated $E_{\text{var}}+\Delta E_{\text{r}}^{(2)}$ energies at $\lambda = 0$ and $1$ are
$-154.331596(4818)$ and $-154.373574(1906)$ hartree, respectively,
where the error bounds in parentheses correspond to the uncertainty associated with the linear fit.
The error bounds for the $E_{\text{var}}+\Delta E^{(2)}$ and $E_{\text{var}}+\Delta E_\text{r}^{(2)}$ energies obtained at the
various values of $N_{\text{det(in)}}$ reflect on the semi-stochastic design of the $\mathscr{V}_{\text{ext}}^{(k)}$
spaces discussed in the main text, but they ignore the uncertainties characterizing the reference
$E_{\text{var}}+\Delta E_{\text{r}}^{(2)}$ energies obtained in the above extrapolation procedure.}
\footnotetext[2]{
\setlength{\baselineskip}{1em}
The CC($P$) and CC($P$;$Q$) energies are reported as errors relative to CCSDT, in millihartree.
The total CCSDT energies at $\lambda = 0$ and 1 are $-154.339738$ and $-154.370744$ hartree, respectively.}
\footnotetext[3]{
\setlength{\baselineskip}{1em}
Equivalent to ROHF.}
\footnotetext[4]{
\setlength{\baselineskip}{1em}
Equivalent to the result obtained with the second-order MBPT approach using the Epstein--Nesbet denominator.}
\footnotetext[5]{
\setlength{\baselineskip}{1em}
Equivalent to CCSD.}
\footnotetext[6]{
\setlength{\baselineskip}{1em}
Equivalent to CR-CC(2,3).}
\end{table*}

\squeezetable
\begin{table*}[!ht]
\caption{
Convergence of the CC($P$) and CC($P$;$Q$) Singlet--Triplet Gaps $\Delta E_\text{S--T}= E_\text{S} - E_\text{T}$
Characterizing Cyclobutadiene, as Described by the cc-pVTZ Basis Set, Toward Their CCSDT Parents at
the R ($\lambda = 0$) and TS ($\lambda = 1$) Geometries,
Along With the $\Delta E_\text{S--T}$ Data Resulting From the Associated Variational and Perturbatively Corrected CIPSI
Computations
}
\label{tab:table6}
\footnotesize
\begin{ruledtabular}
\begin{tabular}{l p{4.5cm} c c c c c c}
\textrm{$\lambda$}&
\textrm{$N_{\text{det(in)}}$ / $N_{\text{det(out)}}$}&
\textrm{\% of triples}&
\textrm{$E_{\text{var}}$\protect\footnotemark[1]}&
\textrm{$E_{\text{var}}+\Delta E^{(2)}$\footnotemark[1]}&
\textrm{$E_{\text{var}}+\Delta E_{\text{r}}^{(2)}$\protect\footnotemark[1]}&
\textrm{CC$(P)$\footnotemark[2]}&
\textrm{CC$(P;Q)$\footnotemark[2]} \\
\colrule \\[-1.0mm]
    0 & 1/1; 1 & 0.0; 0.0 & 20.797\footnotemark[3] & 19.764\footnotemark[4] & 20.275 & 1.295\footnotemark[5] & 0.605\footnotemark[6] \\
    & 50,000/88,980; 85,613             & 0.0; 0.0   & $-27.586$  & 10.714(294)  & 7.480(283)   & 1.790    & 0.593 \\
    & 100,000/177,965; 167,642          & 0.0; 0.0   & $-18.412$  & 10.560(260)  & 8.888(253)  & 1.721    & 0.549 \\
    & 250,000/355,932; 341,366          & 0.0; 0.1   & $-3.050$   & 11.162(203)  & 10.552(199)  & 0.773    & 0.403 \\
    & 500,000/711,877; 684,969          & 0.0; 0.1   & 7.978      & 11.202(177)  & 11.062(174)  & 0.007    & 0.253 \\
    & 1,000,000/1,423,810; 1,369,977    & 0.1; 0.2   & 11.250     & 10.958(165)  & 10.946(162)  & 0.227    & 0.144 \\
    & 5,000,000/5,695,067; 5,463,192    & 0.5; 0.5   & 10.619     & 10.028(144)  & 10.030(142)  & $-0.074$ & 0.068 \\
    & 10,000,000/11,390,227; 10,729,824 & 0.9; 0.9   & 9.604      & 8.970(134)   & 8.962(132)  & $-0.454$ & 0.042 \\
    &                                   &            &            &              &              &          &       \\ [-2.5mm]
    1 & 1/1; 1 & 0.0; 0.0 & 41.644\footnotemark[3] & 5.155\footnotemark[4] & 126.403 & 13.843\footnotemark[5] & 8.685\footnotemark[6] \\
    & 50,000/83,877; 50,010             & 0.0; 0.0   & $-38.625$  & 20.559(322)  & 14.758(307)  & 13.397    & 7.619 \\
    & 100,000/169,536; 100,035          & 0.0; 0.0   & $-25.800$  & 20.710(303)  & 17.571(292)  & 12.338    & 6.866 \\
    & 250,000/339,078; 400,180          & 0.0; 0.1   & 18.815     & 21.880(195)  & 21.778(191)  & 9.559     & 4.482 \\
    & 500,000/678,157; 850,568          & 0.0; 0.2   & 23.944     & 18.643(178)  & 18.834(175)  & 5.282     & 1.406 \\
    & 1,000,000/1,342,908; 1,600,766    & 0.1; 0.2   & 17.555     & 10.819168)   & 11.028(165)   & 0.891     & 0.285 \\
    & 5,000,000/5,425,949; 6,403,314    & 0.3; 0.6   & 10.255     & 8.066(145)   & 8.121(143)   & $-0.425$  & 0.212 \\
    & 10,000,000/10,744,113; 12,806,196 & 0.6; 1.0   & 9.671      & 7.776(137)   & 7.819(135)   & $-0.309$  & 0.161 \\
    &                                   &            &            &              &              &           &       \\ [-2.5mm]
\end{tabular}
\end{ruledtabular}

\footnotetext[1]{
\setlength{\baselineskip}{1em}
For each value of $\lambda$, the $E_{\text{var}}$, $E_{\text{var}}+\Delta E^{(2)}$, and $E_{\text{var}}+\Delta E_\text{r}^{(2)}$
singlet--triplet gaps are reported as errors, in kcal/mol, relative to the parent CIPSI data obtained by
forming the differences between the extrapolated $E_{\text{var}}+\Delta E_{\text{r}}^{(2)}$ energies of
the lowest singlet and triplet states given in footnotes `a' of Tables \ref{tab:table4} and \ref{tab:table5}.
The resulting reference $E_{\text{var}}+\Delta E_{\text{r}}^{(2)}$ singlet--triplet gap values at $\lambda = 0$ and 1 are
$-41.208(3.254)$ and $-9.593(2.111)$ kcal/mol, respectively.}
\footnotetext[2]{
\setlength{\baselineskip}{1em}
The CC($P$) and CC($P$;$Q$) singlet--triplet gaps are reported as errors relative to CCSDT, in kcal/mol.
The CCSDT singlet--triplet gap values at $\lambda = 0$ and 1 are $-32.019$ and $-1.981$ kcal/mol, respectively.}
\footnotetext[3]{
\setlength{\baselineskip}{1em}
Equivalent to RHF/ROHF.}
\footnotetext[4]{
\setlength{\baselineskip}{1em}
Equivalent to the result obtained with the second-order MBPT approach using the Epstein--Nesbet denominator.}
\footnotetext[5]{
\setlength{\baselineskip}{1em}
Equivalent to CCSD.}
\footnotetext[6]{
\setlength{\baselineskip}{1em}
Equivalent to CR-CC(2,3).}
\end{table*}

\newpage
\clearpage
\setcounter{table}{0}
\renewcommand{\thetable}{A.\arabic{table}}

\squeezetable
\begin{table*}[!ht]
\caption{
Programmable Expressions for the Matrix Elements of $\overline{H}_{N}^{(2)}$
in the ST, DT, and TT Sectors Entering Term (II) of Eq. (\ref{A:ampeqs3}),
Excluding the Contributions in the TT Block Due to the
Three-Body Component of the $\overline{H}_{N}^{(2)}$ Operator,
Organized According
to the Possible $\mu p$-$\nu h$-Differences Between Bra and Ket Determinants, as Defined in the Appendix.
}
\label{A:table1}
\footnotesize
\begin{ruledtabular}
\begin{tabular}{p{2.5cm} p{3.0cm} p{7.5cm} p{3.0cm}}
\textrm{Matrix Element}&
\textrm{$\mu p$-$\nu h$ Difference}&
\textrm{Expression\footnotemark[1]}&
\textrm{Index Constraint\footnotemark[2]}\\
\colrule \\[-2.5mm]
%
%
$\displaystyle \langle \Phi_{i}^{a} | \overline{H}_{N}^{(2)} | \Phi_{lmn}^{def} \rangle$\footnotemark[3] &
2$p$-2$h$ &
$\displaystyle \mathscr{A}_{l/mn}\mathscr{A}^{d/ef} \bar{h}_{mn}^{ef}\delta^{i}_{l}\delta_{a}^{d}$ &
$\displaystyle |S_h| = |S_p| = 1$ \vspace{1em} \\
%
%
$\displaystyle \langle \Phi_{ij}^{ab} | \overline{H}_{N}^{(2)} | \Phi_{lmn}^{def} \rangle$\footnotemark[4] &
1$p$-1$h$ &
$\displaystyle \mathscr{A}_{n/lm}\mathscr{A}^{f/de} \bar{h}_{n}^{f}\delta^{i}_{l}\delta^{j}_{m}\delta_{a}^{d}\delta_{b}^{e}$ \vspace{0.2em} &
$\displaystyle |S_h| = |S_p| = 2$ \\
&
1$p$-2$h$ &
$\displaystyle -\mathscr{A}^{ij}\mathscr{A}_{l/mn}\mathscr{A}^{f/de}\bar{h}_{mn}^{jf}\delta^{i}_{l}\delta_{a}^{d}\delta_{b}^{e}$ &
$\displaystyle |S_h| = 1, \; |S_p| = 2$ \vspace{0.2em} \\
&
2$p$-1$h$ &
$\displaystyle \mathscr{A}_{ab}\mathscr{A}^{d/ef}\mathscr{A}_{n/lm}\bar{h}_{bn}^{ef}\delta^{i}_{l}\delta_{a}^{d}\delta^{j}_{m}$ &
$\displaystyle |S_h| = 2, \; |S_p| = 1$ \vspace{1em} \\
%
%
$\displaystyle \langle \Phi_{ijk}^{abc} | \overline{H}_{N}^{(2)} | \Phi_{lmn}^{def} \rangle$\footnotemark[5] &
0$p$-1$h$ &
$\displaystyle -\mathscr{A}^{k/ij}\mathscr{A}_{n/lm}\bar{h}_{n}^{k}\delta_{a}^{d}\delta_{b}^{e}\delta_{c}^{f}\delta^{i}_{l}\delta^{j}_{m}$ &
$\displaystyle |S_h| = 2, \; |S_p| = 3$ \vspace{0.2em} \\
&
1$p$-0$h$ &
$\displaystyle \mathscr{A}_{c/ab}\mathscr{A}^{f/de}\bar{h}_{c}^{f}\delta_{a}^{d}\delta_{b}^{e}\delta^{i}_{l}\delta^{j}_{m}\delta^{k}_{n}$ &
$\displaystyle |S_h| = 3, \; |S_p| = 2$ \vspace{0.2em} \\
&
1$p$-1$h$ &
$\displaystyle \mathscr{A}^{ijk}\mathscr{A}_{abc}\mathscr{A}_{l/mn}\mathscr{A}^{d/ef} \bar{h}_{al}^{id} \delta^{k}_{n}\delta^{j}_{m}\delta_{b}^{e}\delta_{c}^{f}$ &
$\displaystyle |S_h| = |S_p| = 2$ \vspace{0.2em} \\
&
0$p$-2$h$ &
$\displaystyle \mathscr{A}^{k/ij}\mathscr{A}_{n/lm} \bar{h}_{lm}^{ij} \delta^{k}_{n}\delta_{a}^{d}\delta_{b}^{e}\delta_{c}^{f}$ &
$\displaystyle |S_h| = 1, \; |S_p| = 3$ \vspace{0.2em} \\
&
2$p$-0$h$ &
$\displaystyle \mathscr{A}_{c/ab}\mathscr{A}^{f/de} \bar{h}_{ab}^{de} \delta^{k}_{n}\delta^{i}_{l}\delta^{j}_{m}\delta_{c}^{f}$ &
$|S_h| = 3, \; |S_p| = 1$ \\
\end{tabular}
\end{ruledtabular}
\footnotetext[1]{
\setlength{\baselineskip}{1em}
In the expressions reported in this column, $\delta_{p}^{q}$ is the Kronecker delta and
$\displaystyle \mathscr{A}_{pq} = \mathscr{A}^{pq} = 1 - (pq)$,
$\displaystyle \mathscr{A}_{p/qr} = \mathscr{A}^{p/qr} = 1 - (pq) - (pr)$, and
$\displaystyle \mathscr{A}_{pqr} = \mathscr{A}^{pqr}
= \mathscr{A}_{qr}\mathscr{A}_{p/qr} = \mathscr{A}^{qr}\mathscr{A}^{p/qr}
= 1 - (pq) - (pr) - (qr) + (pqr) + (prq)$
are index antisymmetrizers, with $(pq)$ designating the transposition of $p$ and $q$.}
\footnotetext[2]{
\setlength{\baselineskip}{1em}
The relevant set intersections describing the hole and particle indices common to the bra and ket determinants
are defined as $S_h = \{i,\ldots\} \cap \{l,\ldots\}$ and $S_p = \{a,\ldots\} \cap \{d,\ldots\}$, respectively,
with $|S_{h}|$ and $|S_{p}|$ representing the cardinal numbers of $S_{h}$ and $S_{p}$.}
\footnotetext[3]{
\setlength{\baselineskip}{1em}
The triply excited determinant specifying the ket assumes that the spin-orbital hole and particle
indices are ordered such that $l<m<n$ and $d<e<f$.}
\footnotetext[4]{
\setlength{\baselineskip}{1em}
The doubly and triply excited determinants specifying the bra and ket assume that the spin-orbital hole and particle
indices are ordered such that $i<j$, $a<b$, $l<m<n$, and $d<e<f$.}
\footnotetext[5]{
\setlength{\baselineskip}{1em}
The triply excited determinants specifying the bra and ket assume that the spin-orbital hole and particle
indices are ordered such that $i<j<k$, $a<b<c$, $l<m<n$, and $d<e<f$.
Only the contributions due to the one- and two-body components of $\overline{H}_{N}^{(2)}$ are considered.
The contributions due to the three-body component of $\overline{H}_{N}^{(2)}$ are embedded in the
$\tilde{{\mathfrak{M}}}_{abc}^{ijk}(2)$ quantity defined in Eq. (\ref{A:moment3augmented})
(see the Appendix for further details).}
\end{table*}

\newpage

\clearpage

\end{document}